\documentclass{lmcs} 


\usepackage{hyperref, bookmark}

\usepackage{amsmath, amssymb}
\usepackage{algorithm}
\usepackage{algpseudocode}
\usepackage{mathtools}
\usepackage{thmtools}
\usepackage{thm-restate}
\usepackage{stmaryrd}
\usepackage{enumitem}
\usepackage{tikz}
\usetikzlibrary{automata, arrows.meta, positioning}
\usepackage{caption}
\usepackage{subcaption}
\usepackage{comment}

\newcommand{\incl}{\subseteq}
\newcommand{\Rpos}{\mathbb{R}_{\geq 0}}
\newcommand{\xra}[1]{\xrightarrow{#1}}
\newcommand{\lleq}{\triangleleft}
\newcommand{\N}{\mathbb{N}}
\newcommand{\Z}{\mathbb{Z}}
\newcommand{\fract}{\operatorname{fract}}

\newcommand{\locsign}{\operatorname{loc-sign}}
\newcommand{\U}{\operatorname{U}}
\newcommand{\X}{\operatorname{X}}
\newcommand{\Val}{\operatorname{Val}}

\newcommand{\true}{\operatorname{\textbf{true}}}
\newcommand{\false}{\operatorname{\textbf{false}}}

\newcommand{\Aa}{\mathcal{A}}
\newcommand{\Bb}{\mathcal{B}}
\newcommand{\Ii}{\mathcal{I}}

\newcommand{\inact}[1]{\overline{#1}}
\newcommand{\mc}{\mathsf{M}}
\newcommand{\mca}[1]{\mathsf{M_{#1}}}

\newcommand{\target}{\operatorname{target}}
\newcommand{\Inact}{\operatorname{IA}} 
\newcommand{\F}{\mathbf{F}}
\newcommand{\loc}{\operatorname{loc}}
\newcommand{\range}{\operatorname{range}}
\newcommand{\msat}{\textsc{MONOTONE-3-SAT}}
\newcommand{\etmt}{\textsc{1-ATA-Zone-Non-Entailment}}
\newcommand{\NP}{\mathsf{NP}}
\renewcommand{\flat}{\operatorname{flat}}

\let\entails\sqsubseteq
\let\regeq\simeq
\let\sat\models
\let\dual\widetilde

\def\val{\textit{val}}
\def\Var{\textit{Var}}
\def\bd{\textit{bd}}
\def\ubd{\textit{ubd}}

\begin{document}
\bibliographystyle{alphaurl}
\title[Model-checking real-time systems: revisiting the 1-ATA route]{Model-checking real-time systems: revisiting the alternating automaton route}

\author{Patricia Bouyer}[a]
\author{B Srivathsan}[b,c]
\author{Vaishnavi Vishwanath}[b]

\address{Universit\'e Paris-Saclay, CNRS, ENS Paris-Saclay, LMF, Gif-sur-Yvette, France}	
\email{bouyer@lmf.cnrs.fr}  

\address{Chennai Mathematical Institute, India}	
\email{sri@cmi.ac.in,vaishnaviv@cmi.ac.in}  

\address{CNRS, ReLaX, IRL 2000, Siruseri, India}

\begin{abstract}
  Alternating timed automata (ATA) are an extension of timed automata where the effect of each transition is described by a positive boolean combination of (state-reset) pairs. Unlike classical timed automata, ATA are closed under complementation and are therefore suitable for logic-to-automata translations. Several timed logics, including Metric Temporal Logic (MTL), can be converted to equivalent 1-clock ATAs (1-ATAs). Satisfiability of an MTL formula reduces to checking emptiness of a 1-ATA. A straightforward modification of the 1-ATA emptiness algorithm can be applied for model-checking timed automata models against 1-ATA specifications.

  Existing emptiness algorithms for 1-ATA proceed by a region construction, and are not satisfactory for implementations. On the other hand, emptiness for classical timed automata proceeds by zone-based methods that have been extensively studied with well developed implementations. Therefore, logic-to-automata tools typically perform a sophisticated conversion of the formula into a network of timed automata, sometimes, via a 1-ATA as an intermediate step.

  Our goal in this work is to initiate the study of zone-based methods directly for 1-ATAs. The challenge here, as opposed to timed automata, is the fact that the zone graph may generate an unbounded number of variables. We first introduce a \emph{deactivation operation} on the 1-ATA syntax to allow an explicit deactivation of the clock in  transitions. Using the deactivation operation, we improve the existing MTL-to-1-ATA conversion and present a fragment of MTL for which the equivalent 1-ATA generate a bounded number of variables.  Secondly, we develop the idea of zones for 1-ATA and present an emptiness algorithm which explores a corresponding zone graph. For termination, a special entailment check between zones is necessary. Our main technical contributions are: (1) an algorithm for the entailment check using simple zone operations and (2) an $\NP$-hardness for the entailment check in the general case. Finally, we adapt our methods to the problem of model-checking timed automata models against 1-ATA specifications. We observe that when the timed automaton is strongly non-Zeno or when the 1-ATA generates a bounded number of variables, a modified entailment check with quadratic complexity can be applied.
\end{abstract}

\maketitle

\section{Introduction}\label{sec:introduction}
\paragraph*{Model-checking real-time systems.} The task of verifying if a model $A$ (typically a network of automata) satisfies a specification $\varphi$ (a formula in a logic) can be cast as the emptiness of an automaton $A \times B_{\neg \varphi}$, where $B_{\neg \varphi}$ is an automaton recognizing the set of behaviours that violate the specification -- in other words, an automaton corresponding to $\neg \varphi$. This is the well-known model-checking paradigm. In the context of real-time systems, \emph{timed automata}~\cite{alur1994theory} are the de-facto choice for the automaton model. For the logic part, metric temporal logic (MTL) is a natural extension of the widely used linear temporal logic (LTL), incorporating timing constraints in the modalities. However, timed logics cannot be easily converted to timed automata, since they are not closed under complementation.

A popular choice for logic-to-automata translations for timed logics, is the model of \emph{1-clock Alternating Timed Automata (1-ATAs)}~\cite{ouaknine2007decidability,lasota2005alternating}. 1-ATAs are closed under complementation, and are incomparable in expressive power to (multi-clock) timed automata. The power of alternation enables resetting fresh copies of the single clock at different positions along the word. Therefore, while reading a timed word, a 1-ATA can keep track of the time elapsed from multiple positions in the word. This feature is quite convenient to handle timed modalities like $\U_I$, which is an Until operator with an interval constraint $I$. Therefore, the model-checking problem reduces to checking the emptiness of $A \times B_{\neg \varphi}$ where $A$ is a timed automaton and $B_{\neg \varphi}$ is a
1-ATA. This is a rather natural formulation of the model-checking problem in the real-time setting. Surprisingly, there is no tool that implements an algorithm to this problem, to the best of our knowledge. Decidability of this question follows from \cite{ouaknine2004language} and \cite{lasota2005alternating}. However it is based on extended regions and not suitable for practical implementations.

On the other hand, emptiness for classical timed automata has been extensively studied over the last three decades, and there are mature and well developed tools like UPPAAL~\cite{Larsen:1997:UPPAAL}, PAT~\cite{PAT}, LTSMin~\cite{LTSMin}, TChecker~\cite{TChecker}, Theta~\cite{Theta}. Algorithms for classical timed automata are based on \emph{zones}. These are convenient data structures that symbolically capture reachable configurations of a timed automaton. Our goal is to eventually adapt the best known methods from timed automata literature to solve the emptiness of $A \times B_{\neg \varphi}$. The crucial missing piece in this picture is a zone-based algorithm to handle 1-ATAs, like $B_{\neg \varphi}$. This is the subject of this paper.

\paragraph*{The challenge.} Undoubtedly, the biggest challenge in developing a zone-based procedure for 1-ATA is the unbounded size of the configurations that it generates (illustrated in Fig.~\ref{fig:overview-1}). As said earlier, a 1-ATA maintains several copies of its clock, each of them storing the time since a specific position. This not only adds technical difficulty in describing zone graphs, but it also makes some of the operations used in the computation algorithmically hard. In this work, we pinpoint the challenges and provide some efficient algorithmic solutions for some restricted cases of 1-ATA which arise from fragments of MTL.

\paragraph*{Contributions.} As a first contribution, we introduce a new \emph{deactivation} operation for the clock, on transitions. When a clock is deactivated, the branch of the 1-ATA that passes through the transition maintains only the control state, and not the clock value (until the clock gets reactivated again through a reset). This explicit deactivation helps reducing the number of active clock copies maintained by the 1-ATA. To substantiate this idea, we improve the MTL-to-1-ATA construction of \cite{ouaknine2007decidability} using the deactivation operator and identify a fragment that induces 1-ATAs with \emph{bounded width} -- these are automata which contain a bounded number of active clock copies in any reachable configuration.

Next, we present a definition of a zone graph for the enhanced 1-ATAs with the deactivation operation, and provide an algorithm to compute the zone successors. For termination of the zone graph computation, we adapt the entailment relation between zones, studied in \cite{abdulla2007zone}, to our setting. Our main technical contribution is a close study of this entailment check.
\begin{itemize}
  \item We provide an algorithm for the entailment check using standard zone operations known from the timed automata literature.
  \item We prove that deciding non-entailment  is $\NP$-hard, by reducing the monotone 3-SAT problem to the non-entailment check.
  \item Then, we present a modified entailment check, with quadratic complexity, for 1-ATAs with bounded width.
\end{itemize}

A shorter version of this work, reporting on all the contributions mentioned above, appeared in~\cite{BouyerSV:Fossacs25}. In this extended version, we firstly include all the missing proofs and present a new elaborate section on extending the zone graph algorithm to the product $A \times B_{\neg \varphi}$ in order to solve the model-checking problem. As a new contribution, we focus on the case when $A$ is \emph{strongly non-Zeno}, that is, there is a number $k$ such that in every $k$ actions of $A$, at least $1$ time unit elapses. We prove that when $A$ is strongly non-Zeno, the number of ``useful configurations'' generated by the product $A \times B_{\neg \varphi}$ is bounded. Hence, the modified entailment check with quadratic complexity can be applied to the zone graph of $A \times B_{\neg \varphi}$. This concept of strong non-Zenoness appears frequently in timed automata literature, for instance~\cite{Tripakis:progress:99,ASARIN1998447,baier2009timed}. Models that are not strongly non-Zeno can have \emph{timelocks} -- an infinite number of events happen in a finite time interval. Such models are not realistic and there is a body of work on detecting timelock freedom in timed automata models~\cite{Tripakis:progress:99,DBLP:journals/fac/BowmanG06}. 

Our new observation indicates that for model-checking the practically useful class of strongly non-Zeno automata against specifications given by 1-ATAs, our approach is promising, since a bounded number of variables are generated during the computation, and furthermore all zone operations have the same efficient complexity as in the classical timed automata setting. We believe this is encouraging, since it paves the way for model-checking the full Metric Temporal Logic over realistic timed automata models, as long as we consider finite words. We remark that in the untimed setting, Linear Temporal Logic over finite words (LTLf) has recently gained substantial attention due to its practical applicability (see~\cite{Bansal:ATVA23} for an exposition).

\paragraph*{Related work.} Recall the casting of the model-checking question as an emptiness of $A \times B_{\neg \varphi}$ where $B_{\neg \varphi}$ is a 1-ATA. Since timed automata tools are mature, some of the existing techniques convert $B_{\neg \varphi}$ into a timed automaton. MightyL~\cite{DBLP:conf/cav/BrihayeGHM17} is a tool that converts a fragment of MTL called Metric Interval Temporal Logic (MITL) into a network of timed automata, by observing special properties of the 1-ATA obtained from MITL~\cite{brihaye2013mitl}. A conversion of MITL into a network of generalized timed automata~\cite{DBLP:conf/cav/AkshayGGJS23} has been developed~\cite{DBLP:conf/concur/0001G0S24}, and a tool incorporating MITL model-checking is available~\cite{Tempora}. Both these works deal with the MITL fragment. To the best of our knowledge, the only available logic-to-automaton procedure for the full MTL fragment is the MTL-to-1-ATA conversion of \cite{ouaknine2007decidability}. Hence, our work would be the first attempt at an implementable algorithm for the full MTL model-checking (over finite timed words). 

The core zone based algorithm for timed automata reachability was proposed in~\cite{DBLP:conf/tacas/DawsT98}. Termination of the zone enumeration has been an intricate topic since then~\cite{DBLP:journals/fmsd/Bouyer04,DBLP:journals/sttt/BehrmannBLP06,herbreteau2016better,gastin_et_al2018diag}. We refer the reader to the surveys \cite{DBLP:conf/formats/BouyerGHSS22,DBLP:journals/siglog/Srivathsan22} for a more detailed exposition on this topic. An important ingredient in the zone enumeration is a simulation operation between zones (similar in spirit to the entailment check as said above). For extended models of timed automata -- like timed automata with diagonal constraints~\cite{DBLP:journals/fmsd/Bouyer04} and event-clock automata~\cite{alur1999event}, there is a translation to classical timed automata. Therefore emptiness for these extended automata can be reduced to emptiness of a bigger timed automaton. However, previous works have observed that zone-based algorithms which work directly on the extended automata perform better than the zone-based algorithms that run on equivalent bigger timed automata: \cite{gastin_et_al2018diag} shows it for timed automata with diagonal constraints; \cite{DBLP:conf/cav/AkshayGGJS23} introduces \emph{generalized timed automata} which can model event-clock automata and provides a comparison of the direct method versus the conversion to timed automata. The main reason is that the conversion to timed automata adds many control states which cannot be handled well by the current zone-based algorithms. Algorithms which work directly on the extended automata encode the extra information using zones and manage the explosion using zone simulation techniques. For timed automata with diagonal constraints, the simulation check is $\NP$-hard. Nevertheless an algorithm using basic zone operations has been proposed \cite{gastin_et_al2018diag} and has been shown to work well on examples. This history of successful zone-based methods on extensions of timed automata encourages us to look for direct zone based methods for 1-ATAs as well.

Alternating Timed Automata appeared in~\cite{ouaknine2007decidability,lasota2005alternating}. With two clocks, emptiness is undecidable, simply because universality for 2-clock timed automata is undecidable~\cite{alur1994theory}. Emptiness for 1-ATA is known to be decidable with a non-primitive recursive complexity -- one potential reason why 1-ATAs have been ignored for direct manipulations. However, several timed logics have been compiled into 1-ATA~\cite{brihaye2013mitl,DBLP:conf/concur/KrishnaMP18} and multiple restrictions of 1-ATAs with better complexity, like very weak 1-ATA~\cite{DBLP:conf/cav/BrihayeGHM17}, 1-ATA with reset-free loops~\cite{DBLP:conf/concur/KrishnaMP18} have been independently considered in the literature. 

In the untimed setting, emptiness of Alternating Finite Automata (AFA) can be performed by a \emph{de-alternation} that computes an equivalent NFA on-the-fly. To deal with large state-spaces, techniques based on antichains have been extensively explored in the literature~\cite{DBLP:conf/cav/WulfDHR06,DBLP:journals/corr/abs-0902-3958,DBLP:conf/tacas/DoyenR10}. Essentially, the algorithms maintain an antichain of the state space, induced by various simulation preorders. Several heuristics based on preprocessing of the automaton~\cite{DBLP:conf/aplas/VargovcikH21} and making use of SAT solvers~\cite{DBLP:conf/sat/HolikV24} have been investigated. A comparative study of various AFA emptiness algorithms can be found in \cite{DBLP:conf/cade/FiedorHHRSV23}. 

In the timed setting, the closest to our work is \cite{abdulla2007zone}, which presents a zone-based approach for answering universality of 1-clock timed automata. Our zone construction for 1-ATA builds on this approach. The entailment check between zones that we use has been proposed in \cite{abdulla2007zone}. However, there is no algorithm given for the entailment check. An approximation of the check is encoded as an SMT formula. In this work, we make a comprehensive study of this entailment check in terms of algorithms and complexity.

\paragraph*{Organization.} Section~\ref{sec:preliminaries} defines the syntax and semantics of 1-ATA with the deactivation operation, and defines 1-ATAs with bounded width. Section~\ref{sec:improving-mtl} presents an improvement of the MTL-to-1-ATA construction and identifies a fragment of MTL which generates bounded width 1-ATAs through the new construction. Section~\ref{sec:zone-graph-1} provides a comprehensive description of zone graphs for ATAs. In Section~\ref{sec:entailment}, we delve into an in-depth study of the entailment relation that is essential for termination of the zone graph computation. 
Finally, Section~\ref{sec:zg-mc} defines the zone graph and entailment check for the model checking problem, and studies the strongly non-Zeno fragment.


\section{1-ATA with a deactivation operator}\label{sec:preliminaries}

We present the syntax and semantics of 1-ATA as in~\cite{ouaknine2007decidability}, but with the novel \emph{deactivation operator} that we introduce. 
In what follows, we use $\N,\Z$, and $\Rpos$ to represent the set of natural numbers, integers and non-negative real numbers respectively. For $d \in \Rpos$, we use $\lfloor d \rfloor$ to denote the integral part and $\fract(d)$ to denote the fractional part of $d$.  We will refer to the single clock of the alternating timed automaton as $x$. We define intervals as $I := [a,b] \mid [a,b) \mid (a,b] \mid (a,b) \mid [a, \infty) \mid (a, \infty) $ where $a,b \in \N$. We write $\Ii$ for the set of all such intervals. For a finite set $S$, we define $\Phi(S)$ as the set of formulas generated by the following grammar, where $s \in S$ and $I \in \Ii$:
\[\varphi = \true \mid \false \mid s \mid I \mid \varphi \land \varphi \mid \varphi \lor \varphi \mid x.\varphi \mid \inact{x}.\varphi \]
Here, $I$ denotes clock interval guards, $x.\varphi$ corresponds to resetting the clock (followed by applying $\varphi)$, and $\inact{x}.\varphi$ corresponds to making the clock \emph{inactive} and then applying $\varphi$. 

A \emph{timed word} over a finite alphabet $\Sigma$ is a finite sequence of the form $(d_{1},a_{1}) \dots (d_{n},a_{n})$, where $a_{1},\dots,a_{n} \in \Sigma$ are events and $d_{1},\dots,d_{n} \in \Rpos$ are time delays. For example, $d_2$ is the time between $a_1$ and $a_2$, $d_3$ is between $a_2$ and $a_3$, and so on.  Alternating timed automata were introduced independently in \cite{DBLP:conf/lics/OuaknineW05,ouaknine2007decidability} and \cite{lasota2005alternating,DBLP:journals/tocl/LasotaW08}. In this paper, we consider the model as defined in \cite{DBLP:conf/lics/OuaknineW05,ouaknine2007decidability}, extended with the  deactivation operator.

\begin{defi}[1-ATA]\label{def:1-ata}
A one-clock alternating timed automaton (1-ATA) is given by $\Aa = (Q, \Sigma, q_0, F, \delta)$ where $Q$ is a finite set of (control) locations, $\Sigma$ is a finite alphabet, $q_0 \in Q$ is the initial location, $F \incl Q$ is a set of accepting locations, and $\delta: Q \times \Sigma \to \Phi(Q)$ is a partial function describing the transitions. We assume $\delta(q,a)$ is a formula in disjunctive normal form for $q \in Q$ and $a \in \Sigma$.
\end{defi}

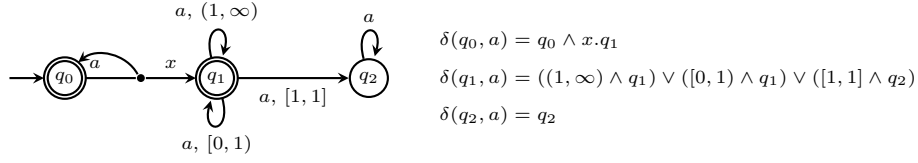
\begin{figure}[t]
\centering
\begin{tikzpicture}[state/.style={draw, circle, inner sep=2pt, thick},node distance = 2cm and 1cm, thick, ->,>=stealth, auto]
  \node (q0) [state,initial,initial text=,accepting] at (-2,0) {\scriptsize $q_{0}$}; 
  \node (q1) [state,accepting] at (0,0) {\scriptsize $q_{1}$}; 
  \node (q2) [state] at (2,0) {\scriptsize $q_{2}$}; 
  \node [circle, fill, inner sep=0pt](t) [] at (-1,0) {.}; 
  \path[-] (q0) edge [above left] node {\scriptsize $a$} (t);
  \path[->] (t) edge [above left,bend right = 50] node {} (q0);
  \path[->] (t) edge [above] node {\scriptsize $x$} (q1); 
  \path[->] (q1) edge [loop above] node {\scriptsize $a,\,(1,\infty)$} (q0); 
  \path[->] (q1) edge [loop below] node {\scriptsize $a,\,[0,1)$} (q0); 
  \path[->] (q1) edge [below] node {\scriptsize $a,\,[1,1]$} (q2); 
  \path[->] (q2) edge [loop above] node {\scriptsize $a$} (q2);
  \begin{scope}[xshift = 2.8cm] 
    \node [right] at (0,0.5) {\scriptsize $\delta(q_0, a) = q_0 \land x. q_1$};
    \node [right] at (0,0){\scriptsize $\delta(q_1, a) = ((1, \infty) \land q_1) \lor ([0, 1) \land q_1) \lor ([1,1] \land q_2)$}; 
    \node [right] at (0, -0.5) {\scriptsize $\delta(q_2, a) = q_2$ };
  \end{scope}
\end{tikzpicture}

\caption{1-ATA $\Aa_1$. The transition function $\delta$ is given on the right. The transition $\delta(q_0, a) = q_0 \land x. q_1$ is depicted as two branches starting from $q_0$, one that leads back to $q_0$ and one that goes onto $q_1$. The edge to $q_1$ contains an $x$ to say that the clock is set to $0$ on this branch.}
\label{fig:ata}
\end{figure}
Before we explain the formal semantics, we give two 1-ATA examples and an informal explanation of their mechanics. 
\begin{exa}
Fig. \ref{fig:ata} shows 1-ATA $\Aa_1$ with $\Sigma = \{a\}$. It accepts all timed words such that no two $a$s are at distance $1$ apart. The initial location is $q_0$. On reading the first $a$, $\Aa_1$ opens two branches, one that comes back to $q_0$ and the other that resets clock $x$ and goes to $q_1$. Similarly, for each of the next $a$s that are read, the $q_0$ branch spawns into two. The purpose of the $q_1$ branch is to check if there is an $a$ at distance $1$ from the $a$ that created the branch. If yes, the branch goes to a non-accepting state $q_2$. Fig.~\ref{fig:ata-a-implies-two-properties} illustrates 1-ATA $\Aa_2$ with $\Sigma = \{a, b, c\}$ that accepts timed words satisfying the property: for every $a$ in the word, there is a $c$ at a later position (with no timing constraints), and a $b$ at a later position which is at $1$ time unit away from $a$. Intuitively, at location $q_a$, whenever an $a$ is read, two obligations are generated in the form of locations $q_c$ and $q_b$. Since the only timing constraints are on $q_b$, we deactivate the clock while going to $q_c$. This is reflected by $\bar{x}. q_c$ in the transition $\delta(q_a, a)$. Location $q_c$ waits for a $c$, and on seeing a $c$, the obligation is discharged (using $\delta(q_c, c) = \true$, the obligation $q_c$ ``disappears''). Location $q_b$ waits for a $b$ at time $1$ from the $a$ where the obligation appeared. The only accepting location is $q_a$. Therefore a configuration containing $q_b$ or $q_c$ will denote unfulfilled obligations, and hence will be non-accepting. 
\end{exa}

\begin{figure}[t]
  \centering \footnotesize \renewcommand{\arraystretch}{1.2}
  \begin{tabular}{c|c|c|c}
    & $a$ & $b$ & $c$ \\
    \hline
    $q_a$ & $q_a \land x. q_b \land \bar{x}.q_c$ & $q_a$ & $q_a$ \\
    \hline
    $q_b$ & $q_b$ & $( [1,1]) \lor ((0,1) \land q_b) \lor ((1, \infty) \land q_b)$ & $q_b$\\
    \hline 
    $q_c$ & $q_c$ & $q_c$ & $\true$ \\
    \hline
  \end{tabular}
  \caption{Transition table of 1-ATA $\Aa_2$. Location $q_a$ is initial and accepting.  }
  \label{fig:ata-a-implies-two-properties}
\end{figure}

\paragraph*{Semantics of 1-ATA} We now present the formal semantics. We fix a 1-ATA $\Aa = (Q, \Sigma, q_0, F, \delta)$ for the rest of this section.
We define the set of valuations, or values of the clock $x$, as $\Val = \Rpos \cup \{\bot\}$, where $\bot$ means the clock is inactive. 
A \emph{state} of $\Aa$ is a pair $(q,v)$, where $q \in Q$ and $v \in \Val$. The state $(q, v)$ is said to be \emph{active} if $v \in \Rpos$ and inactive if $v = \bot$. We say $\val((q,v))=v$. We call $(q_{0},0)$ the initial state. We define $S = Q \times \Val$ as the set of all states of $\Aa$. Since transitions in a 1-ATA are formulas, potentially with conjunctions, each state $(q, v)$ could result in a set of states after a transition. For example, if there is a transition $t:= q_0 \land x. q_2$ from $q$, then $(q, v)$ on $t$ results in $\{(q_0, v), (q_2, 0)\}$. To formalize the idea, we first recursively define when a formula $\varphi \in \Phi(Q)$ is satisfied by a set of states $M \incl S$ on a value $v$, or $M \sat_{v} \varphi$:
\begin{align*}
  & M \not\sat_{v} \false \qquad  && M \sat_{v} \true \\
  & \text{$M \sat_{v} q$ if $(q,v) \in M$} \qquad &&  \text{$M \sat_{v} I$ if $v = \bot$ or $v \in I$} \\
  & \text{$M \sat_{v} \varphi_{1} \land \varphi_{2}$ if $M \sat_{v} \varphi_{1}$ and $M \sat_{v} \varphi_{2}$} && \text{$M \sat_{v} \varphi_{1} \lor \varphi_{2}$ if $M \sat_{v} \varphi_{1}$ or $M \sat_{v} \varphi_{2}$} \\
  & \text{$M \sat_{v} x.\varphi$ if $M \sat_{0} \varphi$} && \text{$M \sat_{v} \inact{x}.\varphi$ if $M \sat_{\bot} \varphi$}
\end{align*}

If $M \sat_v \varphi$, we call $M$ to be a \emph{model} of $\varphi$ on $v$. If $M \sat_{v} \varphi$ and for all $M' \subsetneq M$, $M' \not\sat_{v} \varphi$, then $M$ is called a \emph{minimal model} of $\varphi$ on $v$.

\begin{exa}
Let $\varphi = q_{0} \land x.q_{1}$, $M_{1} = \{(q_{0},1),(q_{1},0),(q_{2},1)\}$, $M_{2} = \{(q_{0},1),(q_{1},0)\}$, and $v=1$ both $M_{1},M_{2} \sat_{v}\varphi$, and $M_{2}$ is a minimal model on $v$ for $\varphi$. Let $\varphi = [1, 2] \land x. q_1$, and $v = 0.2$. Since $v \notin [1,2]$, there is no model for $\varphi$ on $v$. Let $\varphi = x.([0,1] \land q_1)$ and $v = 2.5$. Then $\{(q_1, 0)\}$ is a minimal model for $\varphi$ on $v$.
\end{exa}

A \emph{configuration} $\gamma$ of $\Aa$ is a finite set of states $\{(q_{1},v_{1}),\dots,(q_{n},v_{n})\} \incl S$.  We call $\gamma_{0} = \{(q_{0},0)\}$ the initial configuration. We say that a configuration $\gamma$ is accepting if for \emph{every} $(q,v)\in \gamma$, we have $q \in F$. The point is that each state with a non-accepting location is an \emph{obligation} that is required to be discharged. Therefore, an accepting configuration is one containing no obligations.  Observe that by definition of an accepting configuration, the \emph{empty configuration} $\{ \}$ consisting of no states is accepting. We define the location signature of a configuration $\gamma = \{(q_{1},v_{1}),\dots,(q_{n},v_{n})\}$ as the multiset $\locsign(\gamma) = \{q_{1},\dots,q_{n}\}$. 

Next we look at definition of the transitions. We define two kinds of transitions from $\gamma$:

\paragraph*{Timed transitions.} for $d \in \Rpos$, we define $\gamma + d = \{(q, v + d) \mid (q, v) \in \gamma,\,v \in \Rpos\} \cup \{(q,\bot)\mid (q,\bot) \in \gamma\}$. We add an edge $\gamma \xra{d} \gamma + d$ for all $d \in \Rpos$. 

\paragraph*{Discrete transitions.} let $\gamma = \{(q_{1},v_{1}),\dots,(q_{n},v_{n})\}$ and let $a \in \Sigma$; recall that $\delta(q_i, a)$ is a formula in disjunctive normal form for all $1 \le i \le n$; for each combination $C=(C_{1},\dots,C_{n})$ such that $C_{i}$ is one disjunct of $\delta(q_{i},a)$, we add an edge $\gamma \xra{a,C} \gamma'$ if $\gamma' = \bigcup_{i=1}^{n} M_{i}$ where each $M_{i}$ is a minimal model of $C_{i}$ on $v_{i}$, for all $1 \leq i \leq n$.
  
\begin{exa}
Consider the 1-ATA $\Aa_1$ of Fig.~\ref{fig:ata}. Let $\gamma = \{ (q_0, 0.4), (q_1, 0.2) \}$. We have $\delta(q_0, a)$ to be a single clause and $\delta(q_1, a)$ to be a disjunction of three clauses. So, there are three possible combinations, out of which only $([0, 1) \land q_1)$ has an extension: $\gamma \xra{a, (q_0 \land x.q_1, ~[0,1) \land q_1)} \gamma'$ with $\gamma'= \{(q_0, 0.4), (q_1, 0), (q_1, 0.2)\}$.
\end{exa}

We use $\gamma \xra{d,a,C} \gamma'$ as shorthand to mean there is some $\gamma''$ such that $\gamma \xra{d} \gamma'' \xra{a,C} \gamma'$. We define a run from $\gamma$ as a sequence of transitions: 
\[\gamma \xra{d_{1}}\gamma' \xra{a_{1},C_{1}} \gamma_{1} \xra{d_{2}}\gamma'_1 \xra{a_{2}, C_2} \gamma_{2} \dots \gamma_{n-1}\xra{d_{n}} \gamma'_{n-1} \xra{a_{n},C_{n-1}} \gamma_{n}\]
The run is accepting if $\gamma_{n}$ is accepting. A timed word $(d_1, a_1) \dots (d_n, a_n)$ is accepted by $\Aa$ if there is an accepting run of the above form from $\gamma_{0}$. We define the language of $\Aa$, or $L(\Aa)$ to be the set of words accepted by $\Aa$.

\begin{exa}\label{eg:run-of-A1}
  For the 1-ATA of Fig \ref{fig:ata}, $\{(q_{0},0)\}$ is the initial configuration. For timed word $w = (0.5,a)(0.7,a)$, an accepting run is
  \[\{(q_{0},0)\} \xra{0.5} \{(q_{0},0.5)\} \xra{a,(q_{0} \land x.q_{1})} \{(q_{0},0.5),(q_{1},0)\} \xra{0.7} \{(q_{0},1.2),(q_{1},0.7)\} \]
  \[\xra{a,((q_{0} \land x.q_{1}),([0,1) \land q_{1}))} \{(q_{0},1.2),(q_{1},0),(q_{1},0.7)\}\]
\end{exa}

Given a 1-ATA $\Aa$, the \emph{emptiness problem} asks whether $L(\Aa)$ is empty or not. To decide the emptiness problem, it is enough to find if there is some accepting run from $\gamma_{0}$ in the transition system above. This problem has been proven decidable with non-primitive recursive complexity in \cite{lasota2005alternating}. The decidability result relies on the construction of extended regions, on which a well-quasi order can be defined. In this paper, following standard techniques for timed automata~\cite{DBLP:conf/tacas/DawsT98,DBLP:conf/formats/BouyerGHSS22}, we design a symbolic zone-based approach to decide the emptiness problem for 1-ATA. Before we delve into it, we consider a special restriction of 1-ATAs.

\paragraph*{1-ATA with bounded width}
We define the \emph{width} of a configuration $\gamma$ to be the number of active states present in $\gamma$. For instance, $\{(q_{0},1.2),(q_{1},0),(q_{1},\bot)\}$ has width $2$. From Example~\ref{eg:run-of-A1}, we notice that the run can be extended to generate configurations of larger and larger width. Each transition out of $q_0$ generates two more states, and the existing states with $q_1$ remain. 
\begin{defi}
  A 1-ATA $\Aa$ has width $k$ if every \emph{reachable} configuration starting from the initial configuration of $\Aa$ has width $\le k$. A 1-ATA has \emph{bounded width} if there exists a $k$ such that the 1-ATA has width $k$. 
\end{defi}
It can be checked that $\Aa_1, \Aa_2$ of Fig.~\ref{fig:ata} and \ref{fig:ata-a-implies-two-properties} have unbounded width.
We will see that this is in fact a useful subclass of 1-ATA, because this leads to a global bound on the number of different active clock values that need to be tracked during the emptiness check. 


\section{Improving the MTL-to-1-ATA construction}
\label{sec:improving-mtl}

In this section, we revisit the MTL-to-1-ATA construction of \cite{DBLP:conf/lics/OuaknineW05,ouaknine2007decidability} and propose a modified construction incorporating the deactivation operation. Subsequently, we identify a fragment of MTL which yields bounded width 1-ATAs with the proposed modification. The syntax of MTL is described using the grammar: 
\begin{align*}
\varphi:= a \mid \lnot a \mid \varphi \land \varphi \mid \varphi \lor \varphi \mid \X_{I} \varphi \mid \dual{\X}_{I} \varphi \mid \varphi \U_{I} \varphi \mid \varphi \dual{\U}_{1} \varphi
\end{align*}
where $a \in \Sigma$ and $I$ is an interval. For simplicity of exposition, we have considered a syntax where all formulas are in negation normal form, that is, the negation appears only at the atomic level. In general, by considering dual versions of $\X_I$ and $\U_I$ as we do, all formulas can be converted into negation normal form~\cite{ouaknine2007decidability}. 

Given a timed word $w = (d_{1},a_{1}),\dots,(d_{n},a_{n})$, a position $1 \leq i \leq n$ and an MTL formula $\varphi$, we define $(w,i) \sat \varphi$ as follows:
\begin{itemize}
\item $(w, i) \sat a$ if $a_{i} = a$, and $(w,i) \sat \lnot a$ if $a_{i} \neq a$
\item $(w,i) \sat \varphi_{1} \land \varphi_{2}$ if $(w,i) \sat \varphi_{1}$ and $(w,i) \sat \varphi_{2}$
\item $(w,i) \sat \varphi_{1} \lor \varphi_{2}$ if $(w,i) \sat \varphi_{1}$ or $(w,i) \sat \varphi_{2}$ 
\item $(w,i) \sat \X_{I} \varphi$ if $(w,i+1) \sat \varphi$ and $d_{i+1} \in I$
\item $(w,i) \sat \dual{\X}_I \varphi$ if $i=n$, or $d_{i+1} \not\in I$, or $(w,i+1) \sat \varphi$
\item $(w,i) \sat \varphi_1 \U_{I} \varphi_{2}$ if there is some $i \leq k \leq n$ such that $(w,k) \sat \varphi_{2}$, $\sum_{c=i+1}^{c=k} d_{c} \in I$  
and for all $i \leq j < k$, $(w,j) \sat \varphi_1$
\item $(w,i) \sat \varphi_1 \dual{\U}_I \varphi_2$ if for all $j$ s.t. $i \le j \le n$ and $\sum_{c=i+1}^{c=j} \in I$, either $(w,j) \sat \varphi_{2}$, or there is some $k$ with $i \le k < j$ and $(w,k) \sat \varphi_{1}$
\end{itemize}
We say a timed word $w$ satisfies formula $\varphi$, denoted as $w \sat \varphi$, if $(w, 1) \sat \varphi$. Observe that the time stamp of the first letter is not relevant for the satisfaction of the formula and therefore the semantics is translation invariant. 

We define $L(\varphi)$ as the set of timed words that satisfy $\varphi$.

\subsection{MTL to 1-ATA construction of \cite{ouaknine2007decidability}.} Given a formula $\varphi$, \cite{ouaknine2007decidability} gives a method to construct a 1-ATA $\Aa_\varphi$ whose language is $L(\varphi)$. 
Locations of $\Aa_\varphi$ include a special start location $\varphi_{init}$, all subformulas of the form $\psi_{1} \U_{I} \psi_{2}$ and $\psi_1 \dual{\U}_I \psi_2$ of $\varphi$, and locations $(X_I \psi)^r$ and $(\dual{\X}_I \psi)^r$ for every subformula $X_I \psi$ and $\dual{\X}_I \psi$ of $\varphi$ respectively. The only accepting locations are subformulas of the form $(\dual{\X}_I \psi)^r$ or $\psi_1 \dual{\U}_I \psi_2$. The transition relation $\delta$ is designed so that the following invariant is satisfied~\cite{ouaknine2007decidability}: on a word $w = (d_1, a_1) \dots (d_n, a_n)$ such that $\Aa_\varphi$ has an accepting run, the presence of state $(\psi, 0)$ in the configuration occuring after reading $a_j$, ensures that $(w, j) \sat \psi$. With this in mind, the transition relation $\delta$ is given as follows. It is defined for all subformulas of $\varphi$ and for each individual letter $a,b \in \Sigma$. 

\begin{align*}
\delta(\varphi_{init}, a) & ~=~ x. \delta(\varphi, a) \\
\delta(\psi_1 \lor \psi_2, a) & ~=~ \delta(\psi_1, a) \lor \delta(\psi_2,a) \\
\delta(\psi_1 \land \psi_2, a) & ~=~ \delta(\psi_1, a) \land \delta(\psi_2, a) \\
\delta(\X_I \psi, a) & ~=~ x. (\X_I \psi)^r \\
\delta((\X_I \psi)^r, a) &~=~ I \land x.\delta(\psi, a) \\
\delta(\dual{\X}_I \psi,a) & ~=~ x.(\dual{\X}_I \psi,a)^r\\
\delta((\dual{\X}_I \psi)^r, a) & ~=~ \lnot I \lor x.\delta(\psi, a) \\
\delta(\psi_1 \U_I \psi_2, a) & ~=~ (x. \delta(\psi_2, a) \land I) \lor
  (x. \delta(\psi_1,a) \land (\psi_1 \U_I \psi_2)) \\
\delta(\psi_1 \dual{\U}_I \psi_2, a) & ~=~ (x. \delta(\psi_2, a) \lor I) \land (x. \delta(\psi_1,a) \lor (\psi_1 \U_I \psi_2)) \\
\delta(b, a) & ~=~ \true \text{ if $b = a$, and $\false$ otherwise } \\
\delta(\neg b, a) & ~=~ \true \text{ if $b \neq a$ and $\false$ otherwise }
\end{align*}

\subsection{Proposed modification.} The idea is that if $\psi$ is a pure LTL formula, then in the state $(\psi, v)$ generated to check $\psi$, the value $v$ of the clock $x$ is not utilized. Hence, we can deactivate the clock whenever we generate an obligation for a pure LTL formula. More precisely, we generate the state $(\psi, \bot)$ whenever $\psi$ is a pure LTL formula. The modified transition rules are as follows.
\begin{itemize}
\item In the $\delta$ function defined earlier, replace all occurrences of $x. \delta(\psi, a)$ and $\delta(\psi, a)$ with 
\begin{align*}
 \bar{x}. \delta(\psi, a) \quad  &  \quad \text{ when $\psi$ is a pure LTL formula} \\
\end{align*}
\end{itemize}

Call the resulting transition function as $\delta'$ and the 1-ATA as $\Aa'_\varphi$. 
We can then say the following:
\begin{lem}\label{lem:constcorr}
  For every MTL formula $\varphi$, we have $L(\varphi) = L(\Aa'_{\varphi})$.
\end{lem}
\begin{proof}
  Let $w = (d_1, a_1) (d_2, a_2) \dots (d_n, a_n)$ be a timed word. It has been shown in ~\cite{ouaknine2007decidability} that for every formula $\varphi$ and position $1 \le i \le n$, $(w, i) \sat \varphi$ iff $\Aa_\varphi$ has an accepting run on $(d_i, a_i) \dots (d_n, a_n)$ starting from $(\psi,0)$. To prove our modified construction, it is sufficient to prove that for a purely LTL formula $\psi$ and a position $1 \le i \le n$, $(w, i) \sat \psi$ iff $\Aa'_\psi$ has an accepting run on $(d_i, a_i) \dots (d_{n}, a_n)$ starting from $\{(\psi, \bot)\}$. 

  Observe that if $\psi$ is a purely LTL formula, every subformula of $\psi$ is also purely LTL. Therefore, every new state generated in the run starting from $(\psi, \bot)$ continues to have $\bot$ as the clock value. The correctness of the construction follows from the fact that~\cite{ouaknine2007decidability} is correct also for LTL formulas. Since no interval appears inside, all timing constraints are vacuously satisfied.
\end{proof}

\begin{figure}[t]
  \centering
  \begin{tikzpicture}[state/.style={draw, rectangle, rounded corners, inner sep=3pt}]
    \begin{scope}[every node/.style={state}]
      \node (0) at (0,0) {\footnotesize $\varphi$}; \node (1) at (0,-2) {\footnotesize $\F a$};
    \end{scope}
    \node (2) at (2,0) {\footnotesize $\checkmark$}; \node [circle, fill, inner sep=1pt] (mb) at (0, -1) {}; \node [circle, fill, inner sep=1pt] (ma) at (0, 1) {};
    \draw [thick,auto] (0) to node [right] {\footnotesize $a$} (ma);
    \draw [thick,auto] (0) to node {\footnotesize $b$} (mb);
    \begin{scope}[thick, ->,>=stealth, auto]
      \draw (0) to node [above] {\footnotesize $~c, [1,2]$} (2); 
      \draw (ma) to [bend right] node {} (0); 
      \draw (ma) to [bend left] node {} (2); 
      \draw (mb) to node {\footnotesize \color{red}{$\bar{x}$}} (1); 
      \draw (mb) to [bend left] node {} (0); 
      \draw (1) to node [below] {\footnotesize $a$} (2); 
      \draw (1) to [loop left] node {\footnotesize $b,c$} (1);
    \end{scope}

    \begin{scope}[xshift=8cm,scale=.8]
      \node [blue] at (-2.2, 1) {\footnotesize $(0.3, b)$}; 
      \node [blue] at (-2.2, 0) {\footnotesize $(0.2, b)$}; 
      \node [blue] at (-2.2, -1) {\footnotesize $(0.6, c)$}; \node [blue] at (-2.2,-2.2) {\footnotesize $(1, a)$}; 
      \node (0) at (0,1.8) {\footnotesize $\{(\varphi, 0)\}$}; 
      \node (11) at (-1,0.5) {\footnotesize $\{ (\varphi, 0.3)$}; 
      \node (12) at (1,0.5) {\footnotesize $(\F a, \bot) \}$}; 
      \node (21) at (-1,-0.5) {\footnotesize $\{ (\varphi, 0.5) $}; 
      \node (22) at (1, -0.5) {\footnotesize $ (\F a, \bot) \}$};
      \node (32) at (0, -1.5) {\footnotesize $\{(\F a, \bot) \}$};
      \node (4) at (0, -2.8) {\footnotesize $\{\}$};
    \end{scope}

    \begin{scope}[gray, ->, >=stealth]
      \draw (0) to (11); 
      \draw (0) to (12); 
      \draw (11) to (21); 
      \draw (11) to (22); 
      \draw (12) to (22);
      \draw (21) to (32); 
      \draw (22) to (32);
      \draw (32) to (4);
    \end{scope}
  \end{tikzpicture}
  \caption{Left: 1-ATA $\Aa'_\varphi$ corresponding to the MTL formula $\varphi := (\F a) \U_{[1,2]} c$. The construction of \cite{ouaknine2007decidability} gives the 1-ATA obtained by replacing the $\bar{x}$ above with the reset operation $x$. The initial location $\varphi_{init}$ is not depicted for clarity. The $\checkmark$ is a placeholder for transitions going to $\true$. Missing transitions are assumed to be $\false$. Transitions to $\true$ ($\checkmark$) deactivate the clock -- $\bar{x}$ is not depicted for clarity. Right: The run of $\Aa'_\varphi$ on the word shown in blue, read from top to bottom. It leads to an accepting configuration $\{ \}$.}\label{fig:example-mtl-improvement}
\end{figure}
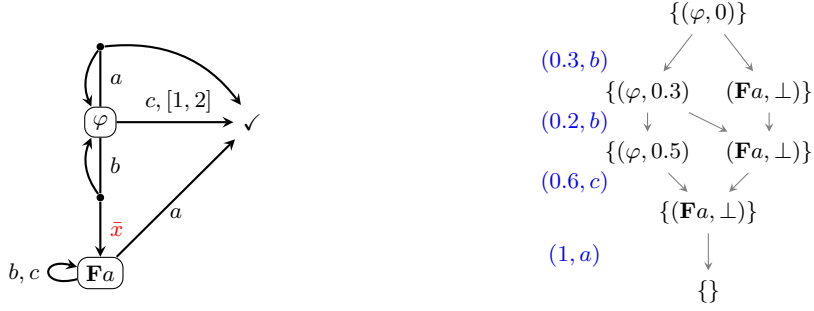

\begin{exa}
Fig.~\ref{fig:example-mtl-improvement} gives an example of the proposed construction for the formula $(\F a) \U_{[1, 2]} c$ where $\F a$, a short form for $\true \U a$, is a pure LTL formula. In the figure, the transition system on the right depicts a run. In the original construction, each $b$ generates a new copy of location $\F a$ with an active clock, and makes the 1-ATA unbounded width. On the other hand, with the deactivated clock, the resulting 1-ATA becomes bounded width, in fact, it has width $1$ -- the only active state checks for an occurrence of $c$ within the interval $[1,2]$ starting from the initial position. 
\end{exa}

Motivated by this observation, we define a fragment of MTL for which our proposed construction results in 1-ATAs with bounded width. This fragment is inspired by the flat-MTL fragment defined in \cite{bouyer2007cost}.

\begin{defi}[One-sided MTL]
  The syntax of one-sided MTL is defined as:
    \[\varphi:= a \mid \lnot a \mid \varphi \land \varphi \mid \varphi \lor \varphi \mid \X_{I} \varphi \mid \dual{\X}_{I} \varphi \mid \psi \U_{I} \varphi \mid \varphi \dual{\U}_{1} \psi\]
  where $a \in \Sigma$, $I \in \Ii$, and $\psi$ is a purely LTL-formula.
\end{defi}

The key idea is that in the $\U_I$ modality, the left branch is an LTL formula, and in the $\dual{\U}_I$ modality the right branch is an LTL formula. For example, the formula $(\F a) \U_{[1,2]} c$, is a one-sided MTL formula. Notice that in the Flat-MTL fragment \cite{bouyer2007cost}, the Until formulas either have bounded intervals, or have pure LTL formulas on the left branch. Here, we only allow the latter kind of formulas. Also, this fragment is different from MITL \cite{alur1996benefits} as we do not forbid singular intervals.
Here is the main result of this section:
\begin{lem}\label{lem:mtltobddcon}
  For every one-sided MTL formula $\varphi$, there is a bound $k_{\varphi}$ such that the 1-ATA $\Aa'_\varphi$ has width $k_{\varphi}$.
\end{lem}
\begin{proof}
  We first define $k_\varphi$ inductively:
  \begin{itemize}
  \item[--] if $\varphi$ is a pure LTL formula, then $k_\varphi = 1$,
  \item[--] if $\varphi = \varphi_1 \lor \varphi_2$, then $k_\varphi = \max(k_{\varphi_1}, k_{\varphi_2})$, 
  \item[--] if $\varphi = \varphi_1 \land \varphi_2$, then $k_\varphi = k_{\varphi_1} + k_{\varphi_2}$,
  \item[--] if $\varphi = \X_{I} \varphi_1$ or $\dual{\X}_I \varphi_{1}$, then $k_\varphi = k_{\varphi_1}$,
  \item[--] if $\varphi = \psi \U_I \varphi_2$ with $\psi$ an LTL formula, then $k_\varphi = k_{\varphi_2}$,
  \item[--] if $\varphi = \varphi_{1} \dual{\U}_I \psi$ with $\psi$ an LTL formula, then $k_\varphi$ = 1 + $k_{\varphi_1}$.
  \end{itemize}
  We will prove that the number of active states present in every reachable configuration of $\Aa'_\varphi$, starting from the initial configuration $\{(\varphi_{init},0)\}$ is at most $k_\varphi$. 
  
  We first observe that because of the way we deactivate clocks in $\Aa'_{\varphi}$ for every transition $\delta'(\psi,a)$ on an LTL subformula $\psi$, once some transition deactivates the clock $x$ in $\Aa'_{\varphi}$, the automaton never reactivates $x$. This means if $\gamma$ is a minimal model of some transition $\inact{x}.\delta'(\psi,a)$ on a valuation $v$, any configuration reachable from $\gamma$ will have no active states. 

  We now look at the case when $\varphi$ is an LTL formula. Initially, we have the configuration $\{(\varphi_{init}, 0)\}$ containing one active state. By definition, $\delta'(\varphi_{init}, a) = \bar{x}. \delta'(\varphi, a)$. As observed above, this means except for the initial configuration which has a single active state, every reachable configuration has no active states. 
  
  Next, when $\varphi$ is not a pure LTL formula, we will prove the required statement by proving that for any subformula $\varphi'$ of $\varphi$, if a configuration $\gamma$ is a minimal model of $\delta'(\psi,a)$ for some $a$ on some valuation $v$, then any configuration reachable from $\gamma$ in $\Aa'_{\varphi}$ has at most $k_{\psi}$ active states. We can see that because $\delta'(\varphi_{init},a) = x.\delta'(\varphi,a)$, any reachable configuration from $\{(\varphi_{init},0)\}$ will be a configuration reachable from some minimal model of $\delta'(\varphi,a)$ on $0$, and so proving this claim will give us the required proof. We now prove our claim by structural induction on $\varphi'$: 

  \paragraph*{Case $\varphi' = a$ or $\lnot a$.} The only transitions from $\delta'(a,a)$ or $\delta'(\lnot a,a)$ are $\true$ or $\false$, meaning if there exists a minimal model of $\delta'(a,a)$ or $\delta'(\lnot a,a)$, it is the empty configuration, hence the claim holds here.

  \paragraph*{Case $\varphi' = \varphi_1 \lor \varphi_2$ or $\varphi' = \varphi_1 \land \varphi_2$.} Looking at the case where $\varphi_1$ and $\varphi_2$ are both not pure LTL formulas, the transition for disjunction is $\delta'(\varphi_1 \lor \varphi_2,a) = \delta'(\varphi_1,a) \lor \delta'(\varphi_2,a)$. Meaning any configuration that is a minimal model of $\delta'(\varphi',a)$ on $v$ is either a minimal model of $\delta'(\varphi_1,a)$ on $v$, or of $\delta'(\varphi_2,a)$ on $v$. Similarly, for the case of conjunction, $\delta'(\varphi_1 \land \varphi_2,a) = \delta'(\varphi_1,a) \land \delta'(\varphi_2,a)$, so any minimal model of $\delta'(\varphi',a)$ on $v$ will be a minimal model of both $\delta'(\varphi_1,a)$ and $\delta'(\varphi_2,a)$ on $v$. Using induction hypothesis on $\varphi_1$ and $\varphi_2$, we can prove the claim for this case. The cases where only one of $\varphi_1$ or $\varphi_2$ is purely LTL can be proved similarly. When both $\varphi_1$ and $\varphi_2$ are pure LTL formulas, we notice that the transitions are now $\delta'(\varphi',a) = \inact{x}.\delta'(\varphi_1,a) \lor \inact{x}.\delta'(\varphi_2,a)$ and $\delta'(\varphi,a) = \inact{x}.\delta'(\varphi_1,a) \land \inact{x}.\delta'(\varphi_2,a)$ respectively. Similar to earlier observation, we see that any configuration that is a minimal model of $\delta'(\varphi',a)$ will have no active variables, and because the clock will not get activated in any configuration reachable from a minimal model of $\delta'(\varphi_1,a)$ or $\delta'(\varphi_2,a)$, the claim will hold.

  \paragraph*{Case $\varphi' = \X_{I} \varphi_1$.} We have $\delta'(\varphi',a)$ to be either $\inact{x}.(\X_{I}\varphi_1)^r$ or $x.(\X_{I}\varphi_1)^r$. Meaning any minimal model of $\delta'(\varphi',a)$ on some $v$ will be of the form $\{((\X_{I}\varphi_1)^r,0)\}$ or $\{((\X_{I}\varphi_1)^r,\bot)\}$. From there, as $\delta'((\X_{I}\varphi_1)^r,a)$ is either $I \land \inact{x}.\delta(\varphi_1,a)$ or $I \land x.\delta(\varphi_1,a)$, the next configuration will be a minimal model of $\delta(\varphi_1,a)$ for either $0$ or $\bot$, or will be $\emptyset$. By induction hypothesis, any configuration reachable from it will have atmost $k_{\varphi_1}$ active states. The proof will follow similarly for $\varphi' = \dual{\X}_I \varphi_1$.

  \paragraph*{Case $\varphi' = \psi \U_I \varphi_2$} When $\varphi_2$ is not an LTL formula, we have 
  \[ \delta'(\psi \U_I \varphi_2,a) = (x. \delta'(\varphi_2, a) \land I) \lor (\bar{x}. \delta'(\psi, a) \land \varphi')\]
  When the left branch of the $\lor$ is considered, the minimal model on $v$ is either $\emptyset$ (when $v \notin I$) or it is a minimal model of $\delta'(\varphi_2,a)$ on $v$. By induction hypothesis, any configuration reachable from it will have atmost $k_{\varphi_2}$ active states. When the right branch of $\lor$ is considered, any minimal model on $v$ is a configuration containing $\{(\varphi',v)\}$ that is also a minimal model of $\delta'(\psi,a)$ on $\bot$. By earlier argument, the configurations reachable from it will have at most 1 active variable in the form of $(\varphi',v')$. This means in general, any configuration reachable from a minimal model of $\delta'(\varphi',a)$ on some $v$ will have at most $k_{\varphi_2}$ active states. When $\varphi_2$ is an LTL formula, the argument is similar.

  \paragraph*{Case $\varphi' = \varphi_1 \dual{\U}_I \psi$}When $\varphi_1$ is not an LTL formula, we have 
  \[\delta'(\varphi_1 \dual{\U}_I \psi,a) = (\inact{x}.\delta'(\psi, a) \lor I) \land (x.\delta'(\varphi_1, a) \lor \varphi')\]
  The minimal model of $\delta'(\varphi',a)$ is a minimal model of both the $\land$ branches. Looking at the left branch of $\land$, any minimal model on some $v$ will be either a minimal model of $\delta'(\psi,a)$ on $\bot$ or be $\emptyset$ (when $v \not\in I$). Meaning any configuration reachable from it will have no active states. Similarly, for the right branch of $\land$, any minimal model on $v$ will be a minimal model of $\delta'(\varphi_{1},a)$ on $0$ or be the configuration $\{(\varphi',v)\}$. By induction hypothesis, any configuration reachable from it will have at most $1 + k_{\varphi_1}$ active states. This means in general, any configuration reachable from a minimal model of $\delta'(\varphi',a)$ on some $v$ will have at most $1 + k_{\varphi_1}$ active states. When $\varphi_1$ is an LTL formula, we have 
  \[\delta'(\varphi_1 \dual{\U}_I \psi,a) = (\inact{x}.\delta'(\psi, a) \lor I) \land (\inact{x}.\delta'(\varphi_1, a) \lor \varphi)\]
  The left $\land$ branch is unchanged, and for the right $\land$ branch, we use the observation earlier to see that any configuration reachable from a minimal model will have at most 1 active state of the form $(\varphi',v')$, meaning any configuration reachable from a minimal model of $\delta'(\varphi',a)$ on some $v$ will have at most $1$ active state.

\end{proof}


\section{A zone graph for 1-ATAs}\label{sec:zone-graph-1}

So far, we have seen 1-ATAs with the deactivation operator and an improved MTL to 1-ATA conversion. We will now move on to solving the emptiness problem for 1-ATAs using zone based techniques. Algorithms based on zones are well known in the timed automata literature. We begin this section with two examples that provide an overview of our zone graphs for 1-ATAs.

\begin{figure}[t]
  \centering 
  \begin{tikzpicture}[scale=0.8]

    \node at (3.5, 10) {\scriptsize $\{(q_0, 0)\}$};
    \node at (3.5, 8.5) {\scriptsize $\{ (q_0, 0.2),\qquad (q_1, 0) \}$};
    \node at (3.5, 7) {\scriptsize $\{ (q_0, 0.5), \qquad (q_1, 0), \qquad (q_1, 0.3) \}$};
    \node at (3.5, 5.5) {\scriptsize $\{(q_0, 1.2), \qquad (q_1, 0), \qquad (q_1, 0.7), \qquad (q_2, 1) \} $};

    \begin{scope}[every node/.style={rectangle, rounded corners, fill=gray!20}]
    \node (z0) at (12, 10) {\tiny $x_{q_0, 1} = 0$};
    \node (z1) at (12, 8.5) {\tiny $x_{q_0, 1} \ge 0 ~\land~ x_{q_1,1} = 0$};
    \node (z2) at (12, 7) {\tiny $x_{q_0, 1} \ge 0 ~\land~ x_{q_1, 1} = 0 ~\land~ x_{q_1, 2} \in [0, 1) ~\land~ x_{q_1,2} \le x_{q_0,1}$};
    \node (z3) at (12, 5.5) {\tiny $x_{q_0,1} \ge 1 ~ \land ~ x_{q_1, 1} = 0 ~ \land ~ x_{q_1, 2} \in [0, 1) ~ \land x_{q_2, 1} = 1$ };
  \end{scope}
  
    \node [left, blue] at (0,9.25) {\tiny $(0.2, a)$};
    \node [left, blue] at (0,7.75) {\tiny $(0.3, a)$};
    \node [left, blue] at (0,6.25) {\tiny $(0.7, a)$};

    \draw [gray] (3.2, 9.7) to (2.6, 8.8);
    \draw [gray] (3.8, 9.7) to (4.4, 8.8);
    \draw [gray] (2.3, 8.2) to (1.6, 7.3);
    \draw [gray] (2.7, 8.2) to (3.5, 7.3);
    \draw [gray] (4.7, 8.2) to (5.3, 7.3);
    \draw [gray] (1.3, 6.7) to (0.8, 5.7);
    \draw [gray] (1.7, 6.7) to (2.6, 5.7);
    \draw [gray] (3.9, 6.7) to (4.5, 5.7);
    \draw [gray] (5.6, 6.7) to (6.3, 5.7);

    \begin{scope}[red]
      \node at (3.5, 10.4) {\tiny $x_{q_0,1}$};
      \node at (2.1, 8.9) {\tiny $x_{q_0,1}$};
      \node at (4.8, 8.9) {\tiny $x_{q_1, 1}$};
      \node at (1.2, 7.4) {\tiny $x_{q_0,1}$};
      \node at (3.8, 7.4) {\tiny $x_{q_1, 1}$};
      \node at (5.8, 7.4) {\tiny $x_{q_1,2}$};
      \node at (0.5, 5.9) {\tiny $x_{q_0,1}$};
      \node at (2.8, 5.9) {\tiny $x_{q_1,1}$};
      \node at (4.8, 5.9) {\tiny $x_{q_1,2}$};
      \node at (6.6, 5.9) {\tiny $x_{q_2, 1}$};
    \end{scope}

    \begin{scope}[->, >=stealth, thick]
      \draw (z0) to node [rectangle, fill=white] {\tiny $q_0 \land x. q_1$} (z1); 
      \draw (z1) to node [rectangle, fill=white] {\tiny $(q_0 \land x.q_1, ~[0, 1) \land q_1)$ } (z2);
      \draw (z2) to node [rectangle, fill=white] {\tiny $(q_0 \land x.q_1, ~[0,1) \land q_1, ~[1,1] \land q_2 )$} (z3);
    \end{scope}
      
  \end{tikzpicture}
  \caption{A timed word $(0.2, a) (0.3, a) (0.7, a)$ shown in blue; the run of the 1-ATA $\Aa_1$ of Fig.~\ref{fig:ata} in the middle; part of the zone graph shown on the right. Variable names corresponding to states in a configuration are shown in red.}\label{fig:overview-1}
\end{figure}
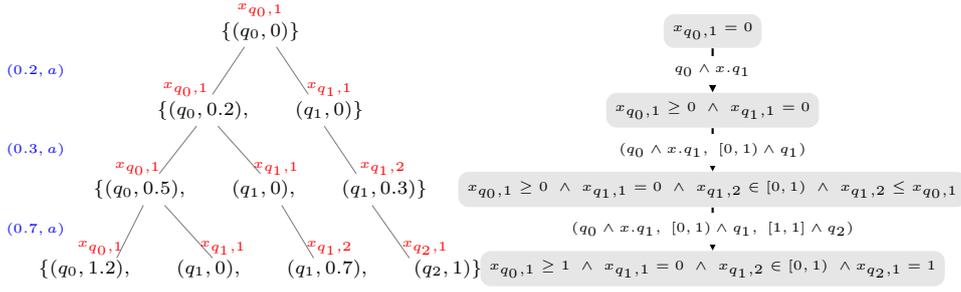

\begin{exa}
  Consider the 1-ATA $\Aa_1$ of Fig.~\ref{fig:ata}. Fig.~\ref{fig:overview-1} illustrates a run of $\Aa_1$ on a timed word $(0.2, a) (0.3, a) (0.7, a)$. The picture on the left (without the red annotations) gives the run. Observe that the size of the configurations keeps growing as we see more and more letters. We need a naming convention to represent each state of a configuration. The annotations in red give the variable names. For each location $q$ of the 1-ATA, we have variables $x_{q, 1}, x_{q,2}, \dots$. At each configuration, we make use of a fresh variable as and when needed. For example, to compute the successor of configuration $\gamma := \{(q_0, 0.5), (q_1, 0), (q_1, 0.3)\}$, we pick one state at a time, pick one clause in an outgoing transition on $a$, and compute the minimal model. In the figure, we first pick $(q_0, 0.5)$ and compute minimal model w.r.t. $q_0 \land x. q_1$. This results in states, corresponding to variable names $x_{q_0,1}$ and $x_{q_1,1}$. Next, we pick $(q_1, 0)$ from $\gamma$, and the transition $[0, 1) \land q_1$. This gives the state $(q_1, 0.7)$ with variable associated being $x_{q_1, 2}$. Finally, we pick $(q_1, 0.3)$ from $\gamma$, and the transition $[1,1] \wedge q_2$. This gives the state $(q_2,1)$ with variable associated being $x_{q_2,1}$. On the right of Fig.~\ref{fig:overview-1} is the zone graph built with this naming convention. Each zone collects the set of all configurations obtained by following the sequence of transitions given alongside the arrow. For example, the initial zone $x_{q_0,1} = 0$ says that the initial configuration is $(q_0, 0)$; the zone $x_{q_0, 1} \ge 0 \land x_{q_1, 1} = 0$ contains all configurations $\{(q_0, \theta), (q_1, 1)\}$ where $\theta \ge 0$.
\end{exa}

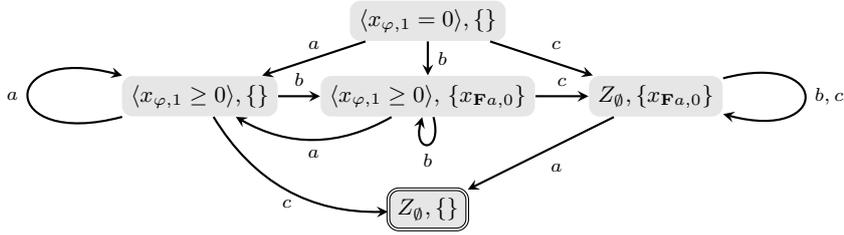
\begin{figure}[t]
  \centering
  \begin{tikzpicture}
    \begin{scope}[every node/.style={rectangle, rounded corners, fill=gray!20}]
    \node (0) at (0,0) {\footnotesize $\langle x_{\varphi,1} = 0 \rangle, \{ \}$};
    \node (1) at (0,-1) {\footnotesize $\langle x_{\varphi,1} \ge 0 \rangle$, $\{x_{\F a, 0}\}$};
    \node (2) at (-3, -1) {\footnotesize $\langle x_{\varphi,1} \ge 0 \rangle, \{ \}$};
    \node (3) at (3, -1) {\footnotesize $Z_\emptyset, \{ x_{\F a, 0} \}$};
    \node [draw, double] (4) at (0, -2.5) {\footnotesize $Z_\emptyset,\{ \} $};
  \end{scope}
  \begin{scope}[->, >=stealth, thick, auto]
    \draw (0) to node {\scriptsize $b$} (1);
    \draw (0) to node [above] {\scriptsize $a$} (2);
    \draw (0) to node {\scriptsize $c$} (3);
    \draw (2) to [loop left] node {\scriptsize $a$} (2);
    \draw (2) to node {\scriptsize $b$} (1);
    \draw (1) to [bend left] node {\scriptsize $a$} (2);
    \draw (1) to node {\scriptsize $c$} (3);
    \draw (2) to [bend right] node [below] {\scriptsize $c$} (4);
    \draw (3) to [loop right] node {\scriptsize $b, c$} (3);
    \draw (3) to node {\scriptsize $a$} (4);
    \draw (1) to [loop below] node {\scriptsize $b$} (1);
  \end{scope}
\end{tikzpicture}
\caption{Zone graph for the 1-ATA $\Aa'_\varphi$ illustrated in Fig.~\ref{fig:example-mtl-improvement}. Node with location $\varphi_{init}$ is not depicted for clarity. $Z_\emptyset$ denotes an empty zone with $\llbracket Z_\emptyset \rrbracket = \emptyset$.}\label{fig:zone-graph-fa-until}
\end{figure}

\begin{exa}
  As a second example, consider the automaton $\Aa'_\varphi$ as depicted in Fig.~\ref{fig:example-mtl-improvement}, for $\varphi = (\F a) \U_{[1,2]} c$. This example shows how to deal with inactive states. Fig.~\ref{fig:zone-graph-fa-until} depicts the zone graph for $\Aa'_\varphi$. There are two remarks. Firstly, the zone graph maintains a zone over the active states, and a set of inactive variables. Successors are computed for all of them, but zone constraints are maintained only for the active states. Secondly, we explicitly keep a node for the ``empty zone''. This is important since it shows that all obligations have been discharged.
  In fact, for $\Aa'_\varphi$, the empty configuration is the only accepting configuration, and therefore the node corresponding to the empty configuration in the zone graph is the only accepting node (c.f. Fig.~\ref{fig:zone-graph-fa-until}).
\end{exa}

We now move on to a formal definition of the zone graph and the procedure to compute successors. Fix a 1-ATA $\Aa = (Q, \Sigma, q_0, F, \delta)$ for the rest of this section. We assume an infinite supply of variable names $x_{q,0}, x_{q,1}, \dots$ and $x'_{q,0},x'_{q,1}, \dots$ for each location $q$ of $\Aa$. For variables $x_{q,i}$ and $x'_{q,i}$, we define $\loc(x_{q,i}) = q$, $\loc(x'_{q,i}) = q$. For a set of variables $X$, we define its \emph{location signature}, $\locsign(X)$ to be the multi-set $\{\loc(y) \mid y \in X \}$. 

\subsection{Zones and nodes.} A \emph{zone} is a conjunction of constraints of the form $(y \sim k)$ or $(y - x \sim k)$, where ${\sim} \in \{<,\leq,>,\geq\}$, $k \in \Z$ and $x, y$ are variables. We write $\Var(Z)$ for the set of variables used in $Z$. We define $\locsign(Z) := \locsign(\Var(Z))$. To deal with the inactive valuations, we define the nodes of a zone graph to be of the form $(Z,\Inact)$ where $Z$ is a zone and $\Inact \incl \{ x_{q,0} \mid q \text{ is a location of } \Aa\}$ is a set of variables that are currently inactive. Notice that we always use variables with index $0$ in $\Inact$, and hence there are finitely many choices for the $\Inact$ component of nodes. We say $\Var(Z,\Inact) = \Var(Z) \cup \Inact$ and $\locsign(Z,\Inact) = \locsign(Z) \cup \locsign(\Inact)$. We will call the pair $(Z,\Inact)$ a \emph{node}. 

We say that a configuration $\gamma$ satisfies a node $(Z,\Inact)$, denoted as $\gamma \sat (Z,\Inact)$, if there is a surjection $h:\Var(Z,\Inact) \mapsto \gamma$ such that (i) $\loc(y) = \loc(h(y))$ for every $y \in \Var(Z, \Inact)$, (ii) for every $x_{q,0} \in \Inact$, $val(h(x_{q,0})) = \bot$ and (iii) replacing every variable $y \in \Var(Z)$ with $val(h(y))$ satisfies all the zone constraints.  We can now define $\llbracket (Z,\Inact) \rrbracket = \{\gamma \mid \gamma \sat (Z,\Inact)\}$.
This means we can look at zones as a representation of a set of configurations. For convenience, we will simply use $(Z,\Inact)$ to refer to both the zone and the set $\llbracket (Z,\Inact) \rrbracket$ of configurations. For instance, we will use $\gamma \in (Z,\Inact)$ to mean $\gamma \in \llbracket (Z,\Inact) \rrbracket$. We illustrate these notions on an example.

\begin{exa}\label{eg:config-zones}
  For the 1-ATA in Fig. \ref{fig:ata}, $(Z,\Inact) = ((x_{q_{1},1} \geq 0) \land (x_{q_{0},1} \geq 0) \land (x_{q_{0},1} - x_{q_{1},1} \geq 0),\emptyset)$ is a zone such that for $\gamma = \{(q_{0},2),(q_{1},1)\}$, $\gamma \in (Z,\Inact)$ by the mapping $h$ with $h(x_{q_{0},1}) = (q_{0},2)$ and $h(x_{q_{1},1}) = (q_{1},1)$. For a zone $(Z',\Inact') = ((x_{q_{0},1} \geq 1), \{x_{q_{1},0}\})$ and a configuration $\gamma' = \{(q_{0},2),(q_{1},\bot)\}$, $\gamma' \in (Z',\Inact')$ by mapping $h'$ with $h'(x_{q_{1},0}) = (q_{1},\bot)$ and $h(x_{q_{0},1}) = (q_{0},2)$. For a zone $(Z_1, \Inact_1) = 0 \le x_{q,0} \le x_{q,1}$ and $\gamma_1 = \{(q, 0.5)\}$, we have $\gamma_1 \in (Z_1,\Inact_1)$ via the mapping $h(x_{q,0}) = (q, 0.5), h(x_{q,1}) = (q,0.5)$. This example illustrates why we require $h$ to be a surjection and not a bijection. Associating the value $0.5$ to both $x_{q_0}$ and $x_{q_1}$ satisfies the constraints of the zone, whereas the resulting configuration (which is a set of states, and not a multi-set) is $\{(q, 0.5)\}$.
\end{exa}

Zones can be represented using Difference-Bound-Matrices (DBMs)~\cite{DBLP:conf/avmfss/Dill89}. We refer the reader to~\cite{DBLP:conf/sfm/BehrmannDL04,DBLP:conf/formats/BouyerGHSS22} for an exposition of algorithms for some of the standard operations on zones.

\subsection{Computing successors.} Given a node $(Z, \Inact)$ and a letter $a \in \Sigma$, we need to first pick an outgoing transition for each variable -- more precisely, for the location corresponding to the variable. For instance, in Fig.~\ref{fig:zone-graph-fa-until}, consider the node $\langle x_{\varphi,1} \ge 0 \rangle, \{x_{\F a,0}\}$. On $b$, location $\varphi$ has transition $\varphi \land \bar{x}. (\F a)$ and location $\F a$ has a loop back to it. Hence the successor is computed on the tuple $(\varphi \land \bar{x}. (\F a), \F a)$. In Fig.~\ref{fig:overview-1}, consider node $\langle x_{q_0, 1} \ge 0 \land x_{q_1, 1} = 0 \rangle$. From $q_0$ there is a unique outgoing transition, whereas from $q_1$, there are three possible transitions. So the targets out of this node are $(q_0 \land x. q_1, (1, \infty) \land q_1)$, $(q_0 \land x. q_1, [0, 1) \land q_1)$ and $(q_0 \land x. q_1, [1,1] \land q_2)$. In Fig.~\ref{fig:overview-1} we depict the only successor on $(q_0 \land x. q_1, [0, 1) \land q_1)$. 

Assume $(Z, \Inact)$ is a non-empty node with $\Var(Z) = \{x_{q_1, i_1}, x_{q_2, i_2}, \dots, x_{q_k, i_k} \}$ (indices $1$ to $k$) and $\Inact = \{ x_{q_{k+1}, 0}, \dots, x_{q_m, 0} \}$ (indices $k+1$ to $m$). For an $a \in \Sigma$, we define: 
\[\target((Z,\Inact), a) = \{ (C_1, \dots, C_m) \mid  C_{j} \text{ is a disjunct in }\delta(q_j, a) \text{ for } 1 \leq j \leq m \} \]
Pick $(C_1, \dots, C_m)$. Our goal is to compute the successor node $(Z, \Inact) \xra{a, (C_1, \dots, C_m)} (Z', \Inact')$. If some $C_j = \false$ then according to our definition, there is no model for it, w.r.t. to any valuation.  We discard such targets. Let us assume none of the $C_j$ is $\false$. We do the following sequence of operations:

\paragraph*{Time elapse.} Compute node $(Z_1, \Inact_1)$ where $\llbracket (Z_1, \Inact_1) \rrbracket = \{ \gamma + \delta \mid \delta \ge 0 \}$ and $\Inact_1 = \Inact$. The zone $Z_1$ represents the closure of $Z$ w.r.t. time successors. It can be computed using a standard DBM technique as in~\cite{DBLP:conf/sfm/BehrmannDL04}.

\paragraph*{Guard intersection.} For every active variable $j$ ranging from $1$ to $k$, and for every interval $I_j \in C_j$, add the constraints corresponding to $x_{q_j,i_j} \in I_j$. After adding all the constraints, we tighten the constraints using an operation known as \emph{canonicalization} in the timed automata literature~\cite{DBLP:conf/sfm/BehrmannDL04}. For instance if $x' = x, y' = y$ and $x = y$, we derive the constraint $x' = y'$. Let the resulting node be $(Z_2, \Inact_2)$ with $\Inact_2 = \Inact_1$. We have $\llbracket (Z_2, \Inact_2) \rrbracket := \{\gamma \in (Z_1, \Inact_1) \mid \gamma \text{ satisfies all clock constraints in the transition}\}$.
  
\paragraph*{Reset and move to new variables.} Since we have dealt with intervals already, we can now assume that each $C_j$ is a conjunction of atoms of the form $q$, $x.q$ and $\bar{x}.q$, or $C_j = \true$. If $C_j = \true$ for all $1 \le j \le m$, the successor of $(Z, \Inact)$ is a special empty node $(Z_\emptyset, \{\})$ where $Z_\emptyset$ denotes a zone with $\llbracket Z_\emptyset \rrbracket = \emptyset$ (see Fig.~\ref{fig:zone-graph-fa-until}, the successor of $(\langle x_{\varphi, 1} \ge 0 \rangle, \{\} )$ for instance). Otherwise, we will iteratively compute a new set of constraints $\Phi$, and a new set of variables $\Inact_3$. Initially, $\Phi := \true$, $\Inact_3 = \emptyset$. Pick each variable $x_{q_j, i_j}$, with $j$ ranging from $1$ to $m$ in some order, and consider all the atoms in the conjunction $C_j$:
\begin{enumerate}
  \item if $q$ is an atom of $C_j$: when $1 \le j \le k$ (active variable),  add $x'_{q, \ell} = x_{q_j,i_j}$ to $\Phi$, where $\ell \ge 2$ is the smallest index (greater than $2$) such that $x'_{q, \ell}$ is not used in $\Phi$; when $k+1 \le j \le m$ (inactive variable), add $x'_{q,0}$ to $\Inact_3$,
  \item if $x. q$ is an atom of $C_j$, add $x'_{q, 1} = 0$ to $\Phi$,
  \item if $\bar{x}.q$ is an atom of $C_j$, add $x'_{q,0}$ to $\Inact_3$.
\end{enumerate}
Define $Z_3 = Z_2 \land \Phi$.

\paragraph*{Remove old variables.} Tighten all the constraints of $Z_3$ by the canonicalization procedure. After canonicalization, remove all the old unprimed variables, and remove the primes from the newly introduced variables, i.e., $x'_{q,1}$ becomes $x_{q,1}$ and so on. This new node is the required $(Z', \Inact')$.

\subsection{Zone graph of a 1-ATA} The initial node is $(Z_0,\Inact_0)$ where $Z_0 := (x_{q_0, 1} = 0)$ (with $q_0$ being the initial state) and $\Inact_0 = \emptyset$. There is a special node $(Z_\emptyset, \{\})$ denoting the empty configuration, with $\llbracket Z_\emptyset \rrbracket = \emptyset$.  Successors are systematically computed by enumerating over all the outgoing targets, and performing the successor computation as explained earlier: we add an edge $(Z, \Inact) \xra{(a, C_1, \dots, C_m)} (Z', \Inact')$ when $(Z', \Inact')$ is the result of applying $(C_1, \dots, C_m)$ on $(Z, \Inact)$, as above. The resulting graph that is computed is called the zone graph of the 1-ATA. A node $(Z, \Inact)$ is said to be accepting if for every $x \in \Var(Z) \cup \Inact$, we have $\loc(x)$ to be accepting. In particular, the special node $(Z_{\emptyset}, \{\})$ is accepting. We now proves that the zone graph is sound and complete using the following lemma which shows that the successor computation as described above is correct. 

Here are some remarks about the statement of the lemma. For each transition $(Z, \Inact) \xra{(a, C_1, \dots, C_m)} (Z', \Inact')$, we wish to say two things: each configuration in the target node $(Z', \Inact')$ is the result of picking some configuration in the source node, performing a delay and then applying the update; and conversely, every successor of a configuration in the source $(Z, \Inact)$ via a delay and a transition given by $(C_1, \dots, C_m)$ appears in the target. However, we need to care for an extra detail: recall that a configuration $\gamma$ belongs to $(Z, \Inact)$ if there exists a surjection $h_\gamma$ that maps variables in $(Z, \Inact)$ to states of $\gamma$. In particular, the number of states $\ell$ in $\gamma$ could be smaller than the number of variables $m$ in $(Z, \Inact)$ since $h_\gamma$ might map multiple variables to the same state of $\gamma$. Hence we cannot simply write $(C_1, \dots, C_m)$ as the outgoing action from $\gamma$ and instead we need to find the action corresponding to each state of $\gamma$ via the mapping $h_\gamma$. This is exactly what the statement of the next lemma makes precise. We assume that $(Z, \Inact)$ is a node with active variables $\{x_{q_1, i_1}, x_{q_2, i_2}, \dots, x_{q_k, i_k} \}$ and inactive variables $\{x_{q_{k+1},i_{k+1}}, \dots, x_{q_m,i_m} \}$ where $i_{k+1} = \cdots = i_m = 0$.

\begin{lem}\label{lem:succ-zone}
  For the successor computation $(Z, \Inact) \xra{a, (C_1, \dots, C_m)} (Z', \Inact')$, we have $(Z',\Inact')$ to be the set of all  $\gamma'$ s.t. there exists: (1) a $\gamma \in (Z, \Inact)$ with $\gamma = \{ (q_1, v_1), \dots, (q_\ell, v_\ell)\}$, (2) a mapping $h_\gamma: \Var(Z,\Inact) \to \gamma$, and (3) a delay $d \in \Rpos$,  satisfying the following: 
  \[  \gamma \xra{~d, a, (C'_1, C'_2, \dots, C'_{\ell})~} \gamma'\] 
  where for all $1 \le p \le \ell$, we have $C'_p = C_j$ when $(q_p, v_p) = h_\gamma(x_{q_j, i_j})$.

\end{lem}
\begin{proof}
  Let $S$ be the set of all $\gamma'$ as defined above. We will show that $S = (Z', \Inact')$. The successor computation proceeds in the following steps:
  \begin{align*}
  (Z, \Inact) \xra{\text{ time elapse }} (Z_1, \Inact_1) \xra{ \text{ guard } } (Z_2, \Inact_2) \xra{ \text{ update }} (Z_3, \Inact_3) \xra{ \text{remove, rename} } (Z', \Inact')
  \end{align*}

  \paragraph*{To prove $S \incl (Z', \Inact') $} 
  Pick $\gamma' \in S$. We will prove that $\gamma' \in (Z', \Inact')$. Since $\gamma' \in S$, we know that there exists some $\gamma \in (Z,\Inact)$, $h_{\gamma}$, and $d$ that satisfy the criteria stated in the lemma. 
  Let $\gamma_{1} = \{(q_{1},v'_{1}),\dots,(q_{\ell},v'_{\ell})\}$ where for all $1 \le p \le \ell, v'_{p} = v_{p}+d$. This is the configuration obtained by elapsing $d$ units from $\gamma$. Hence we have:
  \[ \gamma \xra{~d~} \gamma_1 \xra{a, (C'_1, \dots, C'_\ell)} \gamma'\]

  Firstly, we deduce that $\gamma_{1} \in (Z_{1},\Inact_{1})$, due to the correctness of the time successor computation in classical zones~\cite{DBLP:conf/sfm/BehrmannDL04} and by using the mapping $h_{\gamma_1} = h_\gamma$ to witness the configuration $\gamma_1$ being in $(Z_1, \Inact_1)$. Secondly, each $v'_p$ satisfies all interval constraints present in $C'_p$. Hence, $\gamma_1 \in (Z_2, \Inact_2)$ by the correctness of guard intersection operation on DBMs, and once again using the mapping $h_{\gamma_1}$. It remains to show that the reset and generation of new variables is correct. 
  
  For each $(q_p, v'_p) \in \gamma_1$, let $M_p$ be the minimal model of $C'_p$ w.r.t. $v'_p$. We will now give a map $h'$ from the new variables generated in $Z_3$ to $\bigcup_{p=1}^{\ell} M_{p}$. In the items below, let $x_{q_j, i_j}$ be a variable such that $h_\gamma(x_{q_j, i_j}) = (q_p, v_p)$. Notice that if there exist distinct $j_1$ and $j_2$ such that $h_{\gamma}(x_{q_{j_1}, i_{j_1}}) = h_\gamma(x_{q_{j_2}, i_{j_2}}) = (q_p, v_p)$, then $C_{j_1} = C_{j_2} = C'_p$. Therefore, the proof below works for every choice of $j$ taken in the previous sentence. 
  \begin{enumerate}
  \item For an atom $q$ in $C'_p$, there is a state $(q, v'_p)$ in $M_p$ and a variable $x'_{q, i}$ in $Z_3$, with constraint $x'_{q, i} = x_{q_j, i_j}$. Let $h'(x'_{q,i}) = (q, v'_p)$.
  \item For an atom $x.q$ in $C'_p$, there is a state $(q, 0)$ in $M_p$ and a variable $x'_{q,1}$ with constraint $x'_{q,1} = 0$ added to $Z_3$. Let $h'(x'_{q,1}) = (q, 0)$.
  \item For an atom $\bar{x}.q$ in $C'_p$, there is a state $(q, \bot)$ in $M_p$ and a variable $x'_{q,0}$ added to $\Inact_3$. Let $h'(x'_{q,0}) = (q, \bot)$.
  \end{enumerate}
  By the way the constraints are added, the configuration consisting of $\gamma_1 \cup \bigcup_{p=1}^{p=\ell} M_p$ belongs to $Z_3$ via the mapping on $\Var(Z_3)$ which  is given by $h_{\gamma_1}$ on old variables, and by $h'$ on new variables. The node $(Z', \Inact')$ is obtained by removing the old variables and removing the prime on the new variables. This implies $\cup \bigcup_{p=1}^{p=\ell} M_p$ is present in $(Z', \Inact')$. But $\cup \bigcup_{p=1}^{p=\ell} M_p$  is simply $\gamma'$. This shows $\gamma' \in (Z', \Inact')$.
    
  \paragraph{To prove $(Z', \Inact') \incl S$} 
  Pick some $\gamma'$ such that $\gamma' \in (Z',\Inact')$ via a mapping $h'$. To prove that $\gamma' \in S$, we first prove that there is some $\gamma_{1} \in (Z_{1},\Inact_{1})$ such that $\gamma_{1} \xra{~a, (C'_1, C'_2, \dots, C'_{\ell})~} \gamma'$. Then, by correctness of the time successor computation in DBMs, we remark that $\gamma_{1} \in (Z_{1},\Inact_{1})$ only when there is some $\gamma \in (Z,\Inact)$ and $d \geq 0$ such that $\gamma + d = \gamma_{1}$. To prove the claim, we construct the required $\gamma_{1}$ and mapping $h_{\gamma_{1}}$ such that the following hold:
  \begin{enumerate}
    \item We start with $\gamma_{1} = \{(q_{1},v_{1}),\dots,(q_{m},v_{m})\}$ with some arbitrary values for $v_{1},\dots,v_{m}$. We will eventually fix values for these valuations, which may result in the number of states decreasing, but we will continue to refer to state $(q_{p},v_{p})$ for $1 \le p \le \ell$ that was initialized to $(q_{j},v_{j})$ in this step as $(q_{j},v_{j})$ for convenience. 
    \item If there is a variable $x_{q,i} \in \Var(Z', \Inact')$ that was added because of a $q$ conjunct in some $C_{j}$, we set $v_{j}$ to $val(h'(x_{q,i}))$ and $h_{\gamma_{1}}(x_{q_j, i_j}) = (q_{j},v_{j})$.
    \item If there is some variable $x_{q,i}  \in \Var(Z',\Inact')$ that was added because of a $x.q$ or a $\inact{x}.q$ conjunct in some $C_{j}$, we set the value of $v_{j}$ to be an arbitrary real such that it satisfies the constraints of $Z_{1}$ and $v_{j} \in I_{j}$ for any $I_{j}$ conjunct present in $C_{j}$. Such a value will exist because otherwise, in the construction of $(Z',\Inact')$, the zone $Z_{2}$ would have been empty. We set $h_{\gamma_{1}}(x_{q_j, i_j}) = (q_{j},v_{j})$.
  \end{enumerate} 
  Let the final $\gamma_{1}$ be $\{(q_{1},v'_{1}),\dots,(q_{\ell},v'_{\ell})\}$. Now, we see that because $\gamma_{1}$ will satisfy all constraints of $(Z_{1},\Inact_{1})$ by mapping $h_{\gamma_{1}}$, $\gamma_{1} \in (Z_{1},\Inact_{1})$. Also, if we define $C'_{1},\dots,C'_{\ell}$ such that $C'_{p} = C_{j}$ when $(q_{p},v'_{p}) = h_{\gamma_{1}}(x_{q_{j},i_{j}})$, by our construction of $\gamma_{1}$, $\gamma_{1} \xra{~a, (C'_1, C'_2, \dots, C'_{\ell})~} \gamma'$.
\end{proof}
    
\begin{thm}\label{thm:zonecas}
  The zone graph is sound and complete.
\end{thm}
\begin{proof}
  We need to prove that for every path $(Z_{0}, \Inact_0) \xra{a_{1},C^1} (Z_{1}, \Inact_1) \dots (Z_{n-1}, \Inact_{n-1}) \xra{a_{n},C^n} (Z_{n}, \Inact_n)$ in the zone graph, there exists a path in the configuration graph of the form $\gamma_{0} \xra{t_{1},a_{1},C^{'1}} \dots \xra{t_{n},a_{n},C^{'n}} \gamma_{n}$ such that $\gamma_{i} \in (Z_{i}, \Inact_i)$ for every $0 \leq i \leq n$, and vice versa. To prove the latter direction, we see that firstly, $\gamma_{0} \in (Z_{0}, \Inact_0)$ by definition. Also, by Lemma \ref{lem:succ-zone}, if $\gamma \in (Z, \Inact)$ and $\gamma \xra{d,a,C} \gamma'$ for some $d \ge 0$, then for the zone $Z \xra{a,C} (Z', \Inact')$, $\gamma' \in (Z', \Inact')$. This means that for some path $\gamma_{0} \xra{t_{1},a_{1},C^{'1}}\gamma_{1}\dots\gamma_{n-1}\xra{t_{n},a_{n},C^{'n}} \gamma_{n}$ of the configuration graph, we can always get a path $(Z_{0}, \Inact_0) \xra{a_{1},C^1} (Z_{1}, \Inact_1) \dots (Z_{n-1}, \Inact_{n-1}) \xra{a_{n},C^n} (Z_{n}, \Inact_n)$ in the zone graph with $\gamma_{i} \in (Z_{i}, \Inact_i)$ for each $0 \leq i \leq n$. Now to prove the soundness direction, looking at a path $(Z_{0}, \Inact_0) \xra{a_{1},C^1} (Z_{1}, \Inact_1) \dots (Z_{n-1}, \Inact_{n-1}) \xra{a_{n},C^n} (Z_{n}, \Inact_n)$ of the zone graph, because each of these zones are non-empty and using Lemma \ref{lem:succ-zone}, we can see that a path $\gamma_{0} \xra{t_{1},a_{1},C^{'1}}\gamma_{1}\dots\gamma_{n-1}\xra{t_{n},a_{n},C^{'n}} \gamma_{n}$ exists in the configuration graph.
\end{proof}


\section{The entailment relation}\label{sec:entailment}
  
Zone enumeration in 1-ATAs suffers from two sources of infinity: (1) the width of the zone (number of active states) can increase in an unbounded manner, and (2) the constants appearing in the zone constraints can be unbounded too. For timed automata, the number of clocks (and hence the width of the zones) is fixed. However, the second challenge does manifest and there is a long line of work coming up with better termination mechanisms that tackle the unbounded growth of constants. The termination mechanism is essentially a subsumption relation between zones, which allows to prune the search. 

For zone-based universality in 1-clock timed automata, an \emph{entailment} relation between zones was used as a termination mechanism~\cite{abdulla2007zone}. We provide an algorithm for this check that makes use of zone operations from the timed automata literature. We then prove that the entailment is $\NP$-hard. As the entailment check is an important operation in the zone graph, done each time a new zone is added, the $\NP$-hardness illuminates the difficulty caused due to point (1) above. Finally, for 1-ATA with bounded width, where (1) is not an issue any more, we provide a slight modification to the test, which makes it polynomial-time checkable, and yet ensures termination. 

The entailment check is based on an equivalence between configurations $\gamma \regeq_{\mc} \gamma'$, first proposed in~\cite{ouaknine2004language}. The equivalence can be adapted to our setting by considering the $\bot$ states by adding that there is a one-to-one correspondence between inactive states in $\gamma$ and $\gamma'$. Fixing a 1-ATA $\Aa$: 
\begin{defi}[Region Equivalence]\label{def:regeq}
  Let $\mc \in \mathbb{N}$ be the largest constant appearing in $\Aa$. For two configurations $\gamma$ and $\gamma'$, we say that $\gamma$ is region equivalent to $\gamma'$, or that $\gamma \regeq_{\mc} \gamma'$, if we can define a bijection $\pi : \gamma \to \gamma'$ such that for every $(q,v),(q_{1},v_{1}),(q_{2},v_{2}) \in \gamma$:
  \begin{itemize}
    \item $\loc(\pi(q, v)) = q$,
    \item $\val(\pi(q, v)) = \bot$ iff $v = \bot$, \quad and \quad $0 \le \val(\pi(q, v)) \le \mc$ iff $0 \le v \le \mc$,
    \item if $0 \le v \le \mc$, then $\lfloor v \rfloor = \lfloor \val(\pi(q,v)) \rfloor$, and $\fract(v) = 0$ iff $\fract(\val(\pi(q,v))) = 0$,
    \item if $0 \le v_1, v_2 \le \mc$, then $\fract(v_1) \le \fract(v_2)$ iff $\fract(\val(\pi(q_1,v_1))) \le \fract(\val(\pi(q_2, v_2)))$.
  \end{itemize}
\end{defi}

By definition, the empty configuration can only be equivalent to itself. 
We now observe that $\regeq_{\mc}$ is a time-abstract bisimulation on the configurations.
\begin{lem}\label{lem:configb}
  For configurations $\gamma_{1}$, $\gamma_{2}$, $\gamma_{3}$ such that $\gamma_{1} \regeq_{\mc} \gamma_{3}$ and $\gamma_{1} \xra{d,a,C} \gamma_{2}$, there exists a $d' \ge 0$ and configuration $\gamma_{4}$ such that $\gamma_{3} \xra{d',a,C} \gamma_{4}$ and $\gamma_{2} \regeq_{\mc} \gamma_{4}$.
\end{lem}
The proof for this is similar to the standard one for region equivalence in timed automata and hence is omitted. The equivalence $\regeq_{\mc}$ is defined on configurations with the same size. However, as noticed in Fig.~\ref{fig:overview-1}, the size of the configurations could be unbounded, and we need a way to relate configurations of different sizes.
  
\begin{defi}[Entailment relation]\label{def:entailment-relation}
 For configurations $\gamma,\gamma'$, we say that $\gamma$ is entailed by $\gamma'$, or $\gamma \entails_{\mc} \gamma'$, if there exists a subset $\gamma'' \incl \gamma'$ such that $\gamma \regeq_{\mc} \gamma''$.

 For nodes $(Z,\Inact), (Z',\Inact')$, we say $(Z,\Inact) \entails_{\mc} (Z',\Inact')$, or $(Z, \Inact)$ is entailed by $(Z', \Inact')$, if for all $\gamma' \in (Z',\Inact')$, there exists $\gamma \in (Z,\Inact)$ s.t. $\gamma \entails_{\mc} \gamma'$.

  When $\mc$ is clear from the context, we simply write $\entails$ instead of $\entails_{\mc}$. 
\end{defi}

For readers familiar with subsumption relations from timed automata literature, the entailment may seem confusing: in timed automata literature, we say $Z \sqsubseteq Z'$ if for all valuations of $Z$ (zone on the left), there exists an equivalent valuation in $Z'$ (zone on the right). In the entailment relation for 1-ATA zones as in Definition~\ref{def:entailment-relation}, it is the opposite. Notice that when $(Z,\Inact) \entails_{\mc} (Z',\Inact')$ the node $(Z', \Inact')$ on the right has more number of variables than $(Z, \Inact)$ the zone on the left. So it is a ``bigger'' zone in terms of the number of variables. Here is the intuition about how we make use of the entailment relation for 1-ATA zone graphs. 

The idea is that when $\gamma \entails \gamma'$, if $\gamma'$ reaches an accepting configuration, then so does $\gamma$. In other words, if $\gamma'$ (the bigger configuration) is able to discharge all its obligations, the small configuration $\gamma$ (with fewer obligations) is also able to discharge its own obligations. The same idea holds in the case of nodes. If $(Z, \Inact) \entails (Z', \Inact')$, then if $(Z', \Inact')$ reaches an accepting node, then so does $(Z,\Inact)$. Therefore, if we reach $(Z', \Inact')$ during the zone enumeration, while $(Z, \Inact)$ has already been visited, we can stop exploring $(Z', \Inact')$. We prove this formally in Lemma~\ref{lem:zoneent}. Before that, we provide an example to illustrate the definition of  the entailment relation.

\begin{exa}
  For $\Aa_1$ in Fig. \ref{fig:ata}, we have $M = 1$. If  $\gamma_{1} = \{(q_{0},1.2),(q_{1},0),(q_2,1.2)\}$ and $\gamma_{2} = \{(q_0,1.5),(q_1,0),(q_0,0.3),(q_2,1.5)\}$, then $\gamma_{1} \entails_1 \gamma_{2}$ because taking subset $\gamma'_{2} \incl \gamma_{2}$ where $\gamma'_{2} = \{(q_0,1.5),(q_1,0),(q_2,1.5)\}$, $\gamma'_{2} \regeq_1 \gamma_{1}$ by the mapping $\pi$ with $\pi((q_{0},1.2)) = (q_0,1.5)$, $\pi((q_{1},0)) = (q_1,0)$ and $\pi((q_2,1.2)) = (q_2,1.5)$.
\end{exa}

\begin{lem}\label{lem:coninvsim}
  For configurations $\gamma_{1}$, $\gamma_{2}$, $\gamma_{3}$ such that $\gamma_{3} \entails_{\mc} \gamma_{1}$ and $\gamma_{1} \xra{d,a,C} \gamma_{2}$, there exists a $d' \ge 0$ and a configuration $\gamma_{4}$ such that $\gamma_{3} \xra{d',a,C} \gamma_{4}$ and $\gamma_{4} \entails_{\mc} \gamma_{2}$.
\end{lem}
\begin{proof}
    If $\gamma_{3} \entails_{\mc} \gamma_{1}$, it means there is some $\gamma'_{1} \incl \gamma_{1}$ such that $\gamma_{3} \regeq_{\mc} \gamma'_{1}$. Let $\gamma'_{1} \xra{d,a,C} \gamma'_{2}$. Now, from Lemma \ref{lem:configb}, we know that there is some $d'$ and $\gamma_{4}$ such that $\gamma_{3} \xra{d',a,C} \gamma_{4}$ and $\gamma_{4} \regeq_{\mc} \gamma'_{2}$. But by definition, $\gamma'_{2} \incl \gamma_{2}$. This means $\gamma_{4} \entails_{\mc} \gamma_{2}$.
\end{proof}

\begin{lem}\label{lem:zoneent}
  For zones $(Z,\Inact)$ and $(Z',\Inact)$, if $(Z,\Inact) \entails_{\mc} (Z',\Inact)$ and $(Z',\Inact')$ has a path to an accepting node in the zone graph, then so does $(Z,\Inact)$.
\end{lem}
\begin{proof}
    Let $(Z,\Inact) \entails_{\mc} (Z',\Inact')$ and let $(Z',\Inact')$ have a path to an accepting zone $(Z'_{a},\Inact'_{a})$. 
    \[ (Z', \Inact') \xra{a_1, C^1} (Z'_1, \Inact'_1) \xra{a_2, C^2} \cdots (Z'_{n-1}, \Inact'_{n-1}) \xra{a_n, C^n} (Z'_a, \Inact'_a)\]
    Either $(Z'_a, \Inact'_a)$ is empty or it is non-empty and every configuration in $(Z'_a, \Inact'_a)$ is accepting. 
    
    Suppose $(Z'_a, \Inact'_a)$ is empty. This means that the atoms in each clause of $C^n$ are of the form $\true$ or $I \land \true$, and there exists a configuration $\gamma_{n-1}' \in (Z'_{n-1}, \Inact'_{n-1})$ which satisfies all the interval constraints and gets updated to the empty configuration. Notice that this is a special case of Lemma~\ref{lem:succ-zone}, but we have made the proof explicit for clarity. Now, from $\gamma'_{n-1}$, using Lemma~\ref{lem:succ-zone}, we can deduce the presence of a configuration $\gamma' \in (Z', \Inact')$ and a run on configurations of the following form:
     \begin{align*}
     \gamma' \xra{d_{1},a_{1},C^1} \gamma'_{1} \dots \gamma'_{n-1} \xra{d_{n},a_{n},C^n} \gamma'_a
     \end{align*}  
    where $\gamma_a'$ is the empty configuration. When $(Z'_a, \Inact'_a)$ is non-empty, we can pick a (non-empty) configuration $\gamma'_a \in (Z'_a, \Inact'_a)$ and get a run of the above shape.
    
    Now, as $(Z,\Inact) \entails_{\mc} (Z',\Inact')$, there is some $\gamma \in (Z,\Inact)$ such that $\gamma \entails_{\mc} \gamma'$. From Lemma \ref{lem:coninvsim}, we know that we can get $d'_{1},\dots d'_{n}$ and $\gamma_{1}, \dots ,\gamma_{n-1},\gamma_{a}$ such that $\gamma \xra{d'_{1},a_{1},C^1} \gamma_{1} \dots \gamma_{n-1} \xra{d'_{n},a_{n},C^n} \gamma_{a}$ and $\gamma_{i} \entails_{\mc} \gamma'_{i}$ for $1 \leq i \leq n-1$ and  $\gamma_{a} \entails_{\mc} \gamma'_{a}$. Hence, starting from $(Z, \Inact)$ and applying the sequence $(a_1, C^1) \dots (a_n, C^n)$ leads to a zone $(Z_a, \Inact_a)$ that contains $\gamma_a$. Moreover, every configuration in $(Z_a, \Inact_a)$ is accepting since the location signature of every configuration is the same in a zone, and we already know that $\gamma_a$ is accepting.
    Thus there is a path to an accepting zone from $(Z,\Inact)$.
\end{proof}

Lemma~\ref{lem:zoneent} proves that the zone graph obtained by not exploring entailed nodes is correct: there is an accepting node in this pruned zone graph iff the language of $\Aa$ is non-empty. 
We still need to show that this pruning results in a finite zone graph. To do so, we will use the following: a relation $\prec ~\incl A \times A$ for a set $A$ is a well-quasi order (WQO) if for every infinite sequence $a_{1} a_{2} \dots$ where $a_{1},a_{2},\dots \in A$, there exists some $i$ and $j$ such that $a_{i} \prec a_{j}$. It was shown in \cite{abdulla2007zone} that the entailment restricted to zones $Z \entails Z'$ is a well-quasi order. Since the $\Inact$ sets are finite, $IA \incl IA'$ is also a well-quasi order. Finite cartesian products of WQOs are WQOs~\cite{KRUSKAL1972297}. Now, we can observe since $\entails$ is a WQO on nodes, every sequence of nodes will hit a node which is bigger than an existing node w.r.t $\entails$ and hence the computation will terminate at this node, implying that every path in the zone graph pruned using $\entails$, is finite.
Our goal now is to study the algorithmic aspects of this entailment relation. Before doing so, we present an example.

\begin{figure}[t]
  \centering
  \begin{tikzpicture}[node distance = 1cm, thick, >=stealth]
    \node (q0) {$0$}; 
    \node [right = of q0] (q1) {$x_{q,1}$}; 
    \node [right = of q1] (q2) {$x_{q,2}$} ; 
    \path[->] (q1) edge [bend left=30,above] node {\scriptsize $\leq 0$} (q2); 
    \path[->] (q2) edge [bend left=30,below] node {\scriptsize $\leq 2$} (q1);
  \end{tikzpicture}
  \hspace*{1cm}
  \begin{tikzpicture}[node distance = 1cm, thick, >=stealth]
    \node (q0) {$0$}; 
    \node [right = of q0] (q1) {$x_{q,1}$}; 
    \node [right = of q1] (q2) {$x_{q,2}$}; 
    \node [right = of q2] (q3) {$x_{q,3}$}; 
    \path[->] (q1) edge [bend left=20,above] node {\scriptsize $\leq 0$} (q2); 
    \path[->] (q2) edge [bend left=20,below] node {\scriptsize $\leq 3$} (q1); 
    \path[->] (q2) edge [bend left=20,above] node {\scriptsize $\leq 0$} (q3); 
    \path[->] (q3) edge [bend left=20,below] node {\scriptsize $\leq 3$} (q2); 
    \path[->] (q1) edge [bend left=55,above] node {\scriptsize $\leq 0$} (q3); 
    \path[->] (q3) edge [bend left=55,below] node {\scriptsize $\leq 3$} (q1);
  \end{tikzpicture}
  \caption{Zone $Z_1$ on the left, and $Z_2$ on the right for Example \ref{eg:z2z3} represented in a graphical notation. An edge $x \xra{\lleq c} y$ stands for constraint $y - x \lleq c$. Edges $x \xra{\le 0} 0$ are all omitted, as well as edges $0 \xra{<\infty} x$.}\label{fig:nzc}
\end{figure}
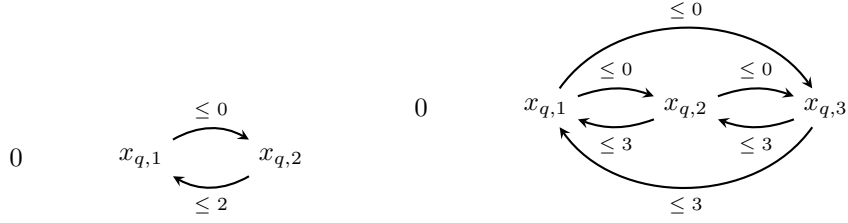
  
\begin{exa}\label{eg:z2z3}
  Let nodes $(Z_{1},\emptyset)$ and $(Z_{2},\emptyset)$ be defined by the constraints represented as a graph on Fig. \ref{fig:nzc}, where an edge $x \xra{\lleq~c_{xy}} y$ represents the constraint $y - x \lleq c_{xy}$. We take $\mc = 3$. Then, we claim that  $(Z_{1},\emptyset) \entails (Z_{2},\emptyset)$. We now explain why -- alongside, we illustrate a source of difficulty in the check. Firstly, $\emptyset \incl \emptyset$, and so the test on the inactive component trivially passes. Next, pick an arbitary $\gamma_{2} \in Z_{2}$ of the form $\gamma_{2} = \{(q,v_{1}),(q,v_{2}),(q,v_{3})\}$, with $0 \le v_1, v_2, v_3 \le 3$. Hence, with the value of $\mc = 3$, we can replace $\gamma \regeq_{\mc} \gamma_2$ with $\gamma = \gamma_2$. 
  
  Suppose  $v_{1} - v_{2} \leq 2$. Consider the projection $\gamma_2' = \{ (q, v_1), (q, v_2)\}$. By examining the constraints of $Z_1$, we conclude that $\gamma_2' \in Z_1$. Setting $\gamma_1 = \gamma_2'$, we have $\gamma_1 \in Z_1$, such that $\gamma_1 \entails \gamma_2$. Suppose $v_{1} - v_{2} > 2$. Then, $\gamma_2'$ as above, does not belong to $Z_1$. However, if $v_1 - v_2 > 2$ (equivalently $v_2 - v_1 < -2$), and since $\gamma_2 \in Z_2$, by the constraint $v_1 - v_3 \le 3$, we infer that $v_2 - v_3 < -1$. Now, we can take the subset $\gamma''_{2} = \{(q,v_{2}),(q,v_{3})\}$ and set $\gamma_1 = \gamma''_2$ to get $\gamma_1 \in Z_1$ (mapping $(q, v_2)$ to $x_{q, 1}$ and $(q, v_3)$ to $x_{q,2}$) such that $\gamma_1 \entails \gamma_2$.

  Notice that different configurations in $Z_2$ require different projections as witnesses for the entailment. As we will see, this makes the problem $\NP$-hard. 
\end{exa}

\subsection{Algorithm for the entailment check}
To get to an algorithm to check entailment on two zones, we look at the cases when the entailment will not hold, i.e. the conditions when $(Z,\Inact) \not\entails (Z',\Inact')$. We observe that $(Z,\Inact) \not \entails (Z',\Inact')$ if either $\Inact \not\incl \Inact'$, or 
\begin{align}\label{eq:1} 
  \text{ there is some $\gamma' \in Z'$ s.t. for every $\gamma \in Z$ and every $\gamma'' \incl \gamma'$, $\gamma \not\regeq_{\mc} \gamma''$}
\end{align}
As the first case is easy to check, we assume $\Inact \incl \Inact'$ and focus on the second case, which means that for this particular $\gamma' \in Z'$, for each of its subsets $\gamma''$, there is no configuration that is region equivalent to it in $Z$. We see that for $\gamma'' \incl \gamma'$ such that $\locsign(\gamma'') \neq \locsign(Z)$, there will trivially be no region equivalent configuration for $\gamma''$ in $Z$. Thus we look at the subsets $\gamma''$ such that $\locsign(\gamma'') = \locsign(Z)$ and investigate the entailment check \eqref{eq:1}.
  
Consider a one-to-one mapping $r: \Var(Z) \to \Var(Z')$ that preserves locations: that is, for every $x \in \Var(Z)$, we have $\loc(x) = \loc(r(x))$. Notice that $r$ is an injection (one-to-one) but not necessarily a surjection (onto), since $\Var(Z')$ can in general have more variables than $\Var(Z)$. Let $\range(r)$ be the set of variables of $\Var(Z')$ that are in the range of $r$. Rename variables $\Var(Z)$ to $y_1, y_2, \dots, y_n$, making $Z$ a zone over these fresh variables. Similarly, project $Z'$ to $\range(r)$ and rename $r(y_1), r(y_2), \dots, r(y_n)$ as $y_1, y_2, \dots, y_n$. Call the resulting zones $Z_r$ and $Z'_r$ respectively. These are zones on the same fixed set of variables and are hence amenable to techniques from timed automata literature. A valuation $v' \in Z'_r$ associates a real to each variable $y_i$. Define: $N_r := \{v' \in Z'_r \mid \forall v \in Z_r, v \not \regeq_{\mc} v' \}$. Given $Z_r, Z'_r$, there is a method to check if $N_r$ is non-empty~\cite{herbreteau2016better}. Using this  algorithm, we can in fact describe $N_r$ as a finite union of zones. We present the algorithm in Appendix~\ref{sec:append-ent}. 

The complexity of this method is quadratic in the number of variables used in $Z_r$ and $Z'_r$. Then, we translate $N_r$ back to $Z'$: let $N'_r = \{ \gamma' \in Z' \mid \gamma' \text{ restricted to $r$ belongs to } N_r\}$.
  
\begin{lem}\label{lem:entalgo}
  For two zones $(Z,\Inact)$ and $(Z',\Inact')$, $(Z,\Inact) \not\entails (Z',\Inact')$ iff $\Inact \not\incl \Inact'$ or $\bigcap_{r \in R_{loc}}N'_{r} \neq \emptyset$, where $R_{loc}$ denotes the set of all location preserving one-to-one mappings from $\Var(Z)$ to $\Var(Z')$ 
\end{lem}
\begin{proof}
  Suppose $(Z, \Inact) \not \entails (Z', \Inact')$ and $\Inact \not \incl \Inact'$. Then, there is a witness $\gamma' \in Z'$ satisfying the condition given in \eqref{eq:1}. Since $\gamma' \in Z'$, there is a surjective mapping $h': \Var(Z') \to \gamma'$ that preserves locations, and the tuple of values $val(h(x'))$ over the variables $x' \in Z'$ satisfies the constraints of $Z'$. We will now show that for every $r \in R_{loc}$, the valuation obtained by restricting $\gamma'$ to $r$ lies in $N'_r$, thereby completing the proof of this direction.
  
  Consider zone $Z$, and assume we have renamed variables of $\Var(Z)$ by $y_1, \dots, y_n$. For every mapping $r \in R_{loc}$, and variable $y_j$, consider the variable $r(y_j)$ and the value associated by $\gamma'$ to $r(y_j)$: that is, the value $val(h'(r(y_j)))$. This gives a valuation $v'_r$, giving a value for each variable $r(y_1), r(y_2), \dots, r(y_n)$, via $\gamma'$. We claim that $v'_r \in N_r$. Otherwise, we will be able to find a $\gamma \in Z$ for which $\gamma \regeq_{\mc} \gamma'_r$ where $\gamma'_r$ is the restriction of $\gamma'$ to $\range(r)$. This contradicts \eqref{eq:1}.
  
  For the other direction, pick an arbitrary $r \in R_{loc}$ and let $v_r \in N_r$. Then a $\gamma' \in Z'$ that corresponds to $v_r$ satisfies the following condition: for every $\gamma \in Z$  we have $\gamma \not \regeq_M \gamma'_r$. Since we now have a $v \in \bigcap_{r \in R_{loc}}N'_{r}$, for the corresponding $\gamma'$, the above condition is obtained with $\range(r)$ replaced with any arbitrary subset, giving~\eqref{eq:1}.
\end{proof}
  The above lemma leads to an algorithm: compute $N'_r$ for each location preserving mapping and check if the intersection is non-empty. 

\subsection{Complexity}
We first analyze the complexity of the algorithm outlined above to check non-entailment for two zones. It will take time $\mathcal{O}(|\Var(Z)|^2)$ to compute each $N'_{r}$ (detailed in Appendix \ref{sec:append-ent}). Also, the number of mappings $r$ could be exponential in $|\Var(Z)|$ as it is bounded by the number of subsets of size $|\Var(Z)|$ in $\Var(Z')$, which is given by ${|\Var(Z')|}\choose{|\Var(Z)|}$. The overall complexity is $\mathcal{O}({{|\Var(Z')|}\choose{|\Var(Z)|}} \cdot |\Var(Z)|^2)$. In practice, one can envisage an algorithm that enumerates each location preserving $r$, and keeps the intersection of $N'_r$ among all the $r$ enumerated so far. This intersection will be a union of zones. If at some point, the intersection becomes empty, the algorithm can stop. Otherwise, the algorithm continues until all location-preserving maps are considered. 

We now focus on getting a lower bound for the non-entailment check. Let $V = \{ p_1, p_2, \dots, p_n\}$ be a set of propositional variables, taking values either $\top$ (true) or $\bot$ (false). An assignment fixes a value to each of the variables. A positive literal is an element of $L_+ = V$ and a negative literal is an element of $L_- = \{ \bar{p}_1, \bar{p}_2, \dots, \bar{p}_n\}$. When $p_i = \top$, the literal $\bar{p}_i = \bot$ and vice-versa. We denote by $L := L_+ \cup L_-$, the set of all literals. A $3$-clause (hereafter called simply a clause) $C$ is a disjunction of \emph{three} literals. The clause is said to be monotone if either all its literals are positive, or all its literals are negative: $C \incl L_+$ or $C \incl L_-$. A boolean formula in $3$-CNF form is a conjunction of clauses: $\bigwedge_{i=1}^{i=m} C_i$. A $3$-CNF formula is monotone if each of its clauses is monotone. The \msat\ problem: given a monotone $3$-CNF formula, is it satisfiable? This problem is known to be $\NP$-complete~\cite{DBLP:books/fm/GareyJ79}.
  
Let \etmt\ be the decision problem which takes as input two nodes $(Z, \Inact)$, $(Z', \Inact')$, and a constant $\mc$, and checks if $(Z, \Inact) \not \entails (Z', \Inact')$, where $\entails$ makes use of the region equivalence w.r.t $\mc$.
  
\begin{thm}
  \etmt\ is $\NP$-hard.
\end{thm}
  
To prove the theorem, we present a reduction from \msat\ to \etmt. Given a monotone $3$-CNF formula $\varphi$, we construct zones $(Z_\varphi, \Inact_\varphi)$, $(Z'_\varphi, \Inact'_\varphi)$ together with the constant $\mc_\varphi$ such that $\varphi$ is satisfiable iff $(Z_\varphi, \Inact_\varphi) \not \entails (Z'_\varphi, \Inact'_\varphi)$. To start off with, we assume $\Inact_\varphi = \Inact'_\varphi = \emptyset$. This means $(Z_\varphi,\Inact_\varphi) \not \entails (Z'_\varphi, \Inact'_\varphi)$ iff $(Z_{\varphi}, \emptyset) \not\entails (Z'_{\varphi}, \emptyset)$. So for simplicity we will use the notation $Z_\varphi \not \entails Z'_\varphi$ in the rest of this section. We now define the zones constructed for the monotone $3$-CNF formula $\varphi = C_{1} \land C_{2} \land \dots \land C_{n}$, where we assume that $C_{1},\dots,C_{k}$ have all positive literals and $C_{k+1},\dots,C_{n}$ have all negative literals. 
  
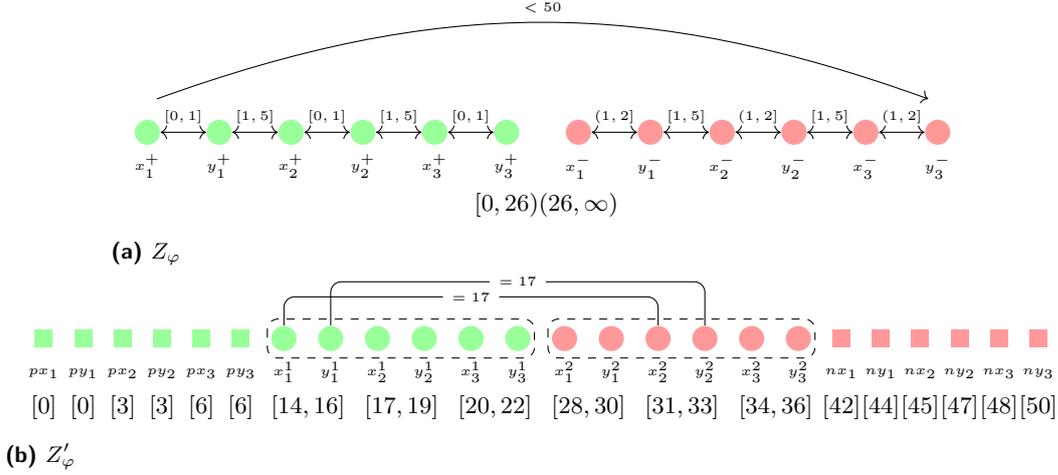
\begin{figure}[t]
  \centering
  \subfloat[ $Z_\varphi$ \label{fig:crz}]{
    \begin{tikzpicture}[node distance = 0.2cm and 0.6cm, scale=0.9]
      \node [circle, fill=green!40] (xp1) {}; \node [circle,
      fill=green!40,right = of xp1] (yp1) {}; \node [right = of
      yp1,circle,fill=green!40] (xp2) {}; \node [right = of
      xp2,circle,fill=green!40] (yp2) {}; \node [right = of
      yp2,circle,fill=green!40] (xp3) {}; \node [right = of
      xp3,circle,fill=green!40] (yp3) {}; \node [right = of
      yp3,circle,fill=red!40] (xn1) {}; \node [right = of
      xn1,circle,fill=red!40] (yn1) {}; \node [right = of
      yn1,circle,fill=red!40] (xn2) {}; \node [right = of
      xn2,circle,fill=red!40] (yn2) {}; \node [right = of
      yn2,circle,fill=red!40] (xn3) {}; \node [right = of
      xn3,circle,fill=red!40] (yn3) {};
  
      \path[<->] (xp1) edge [above=0.2cm,font = \tiny] node {$[0,1]$}
      (yp1); \path[<->] (xp2) edge [above=0.2cm,font = \tiny] node
      {$[0,1]$} (yp2); \path[<->] (xp3) edge [above=0.2cm,font =
      \tiny] node {$[0,1]$} (yp3); \path[<->] (xn1) edge
      [above=0.2cm,font = \tiny] node {$(1,2]$} (yn1); \path[<->]
      (xn2) edge [above=0.2cm,font = \tiny] node {$(1,2]$} (yn2);
      \path[<->] (xn3) edge [above=0.2cm,font = \tiny] node {$(1,2]$}
      (yn3); \path[<->] (yp1) edge [above=0.2cm,font = \tiny] node
      {$[1,5]$} (xp2); \path[<->] (yp2) edge [above=0.2cm,font =
      \tiny] node {$[1,5]$} (xp3); \path[<->] (yn1) edge
      [above=0.2cm,font = \tiny] node {$[1,5]$} (xn2); \path[<->]
      (yn2) edge [above=0.2cm,font = \tiny] node {$[1,5]$} (xn3);

      \node [below = 0.001cm of xp1,font=\tiny] (n1) {$x^{+}_{1}$};
      \node [below = 0.001cm of yp1,font=\tiny] (n2) {$y^{+}_{1}$};
      \node [below = 0.001cm of xp2,font=\tiny] (n3) {$x^{+}_{2}$};
      \node [below = 0.001cm of yp2,font=\tiny] (n4) {$y^{+}_{2}$};
      \node [below = 0.001cm of xp3,font=\tiny] (n5) {$x^{+}_{3}$};
      \node [below = 0.001cm of yp3,font=\tiny] (n6) {$y^{+}_{3}$};
      \node [below = 0.001cm of xn1,font=\tiny] (n7) {$x^{-}_{1}$};
      \node [below = 0.001cm of yn1,font=\tiny] (n8) {$y^{-}_{1}$};
      \node [below = 0.001cm of xn2,font=\tiny] (n9) {$x^{-}_{2}$};
      \node [below = 0.001cm of yn2,font=\tiny] (n10) {$y^{-}_{2}$};
      \node [below = 0.001cm of xn3,font=\tiny] (n11) {$x^{-}_{3}$};
      \node [below = 0.001cm of yn3,font=\tiny] (n12) {$y^{-}_{3}$};

      \node [below = 0.5cm of yp3,font=\small] (l1) {$[0,26)$}; \node
      [below = 0.5cm of xn1,font=\small] (l2) {$(26,\infty)$};

      \node [opacity=0,above = 0.1cm of xp1] (p1) {}; \node
      [opacity=0,above = 0.1cm of yn3] (p2) {};
        
      \path[->] (p1) edge [bend left=20,above,font = \tiny] node
      {$<50$} (p2);
  
      \end{tikzpicture}
    }\hfil
    \subfloat[ $Z'_\varphi$ \label{fig:crz'}]{
      \begin{tikzpicture}[node distance = 0.2cm and 0.27cm, scale=0.9]
        \node [fill=green!40] (dpx1) {}; \node [right = of
        dpx1,fill=green!40] (dpy1) {}; \node [right = of
        dpy1,fill=green!40] (dpx2) {}; \node [right = of
        dpx2,fill=green!40] (dpy2) {}; \node [right = of
        dpy2,fill=green!40] (dpx3) {}; \node [right = of
        dpx3,fill=green!40] (dpy3) {}; \node [right = of
        dpy3,circle,fill=green!40] (x11) {}; \node [right = 0.02cm of
        x11] (g1) {}; \node [right = of x11,circle,fill=green!40]
        (y11) {}; \node [right = of y11,circle,fill=green!40] (x12)
        {}; \node [right = 0.02cm of x12] (g2) {}; \node [right = of
        x12,circle,fill=green!40] (y12) {}; \node [right = of
        y12,circle,fill=green!40] (x13) {}; \node [right = 0.02cm of
        x13] (g3) {}; \node [right = of x13,circle,fill=green!40]
        (y13) {}; \node [right = of y13,circle,fill=red!40] (x21) {};
        \node [right = 0.02cm of x21] (g4) {}; \node [right = of
        x21,circle,fill=red!40] (y21) {}; \node [right = of
        y21,circle,fill=red!40] (x22) {}; \node [right = 0.02cm of
        x22] (g5) {}; \node [right = of x22,circle,fill=red!40] (y22)
        {}; \node [right = of y22,circle,fill=red!40] (x23) {}; \node
        [right = 0.02cm of x23] (g6) {}; \node [right = of
        x23,circle,fill=red!40] (y23) {}; \node [right = of
        y23,fill=red!40] (dnx1) {}; \node [right = of
        dnx1,fill=red!40] (dny1) {}; \node [right = of
        dny1,fill=red!40] (dnx2) {}; \node [right = of
        dnx2,fill=red!40] (dny2) {}; \node [right = of
        dny2,fill=red!40] (dnx3) {}; \node [right = of
        dnx3,fill=red!40] (dny3) {};

        \node [below = 0.15cm of dpx1,font=\tiny] (nn1) {$px_{1}$};
        \node [below = 0.15cm of dpy1,font=\tiny] (nn2) {$py_{1}$};
        \node [below = 0.15cm of dpx2,font=\tiny] (nn3) {$px_{2}$};
        \node [below = 0.15cm of dpy2,font=\tiny] (nn4) {$py_{2}$};
        \node [below = 0.15cm of dpx3,font=\tiny] (nn5) {$px_{3}$};
        \node [below = 0.15cm of dpy3,font=\tiny] (nn6) {$py_{3}$};
        \node [below = 0.01cm of x11,font=\tiny] (nn7) {$x^{1}_{1}$};
        \node [below = 0.01cm of y11,font=\tiny] (nn8) {$y^{1}_{1}$};
        \node [below = 0.01cm of x12,font=\tiny] (nn9) {$x^{1}_{2}$};
        \node [below = 0.01cm of y12,font=\tiny] (nn10) {$y^{1}_{2}$};
        \node [below = 0.01cm of x13,font=\tiny] (nn11) {$x^{1}_{3}$};
        \node [below = 0.01cm of y13,font=\tiny] (nn12) {$y^{1}_{3}$};
        \node [below = 0.01cm of x21,font=\tiny] (nn13) {$x^{2}_{1}$};
        \node [below = 0.01cm of y21,font=\tiny] (nn14) {$y^{2}_{1}$};
        \node [below = 0.01cm of x22,font=\tiny] (nn15) {$x^{2}_{2}$};
        \node [below = 0.01cm of y22,font=\tiny] (nn16) {$y^{2}_{2}$};
        \node [below = 0.01cm of x23,font=\tiny] (nn17) {$x^{2}_{3}$};
        \node [below = 0.01cm of y23,font=\tiny] (nn18) {$y^{2}_{3}$};
        \node [below = 0.15cm of dnx1,font=\tiny] (nn19) {$nx_{1}$};
        \node [below = 0.15cm of dny1,font=\tiny] (nn20) {$ny_{1}$};
        \node [below = 0.15cm of dnx2,font=\tiny] (nn21) {$nx_{2}$};
        \node [below = 0.15cm of dny2,font=\tiny] (nn22) {$ny_{2}$};
        \node [below = 0.15cm of dnx3,font=\tiny] (nn23) {$nx_{3}$};
        \node [below = 0.15cm of dny3,font=\tiny] (nn24) {$ny_{3}$};

        \node [below = 0.5cm of dpx1,font=\small] (lp1) {$[0]$}; \node
        [below = 0.5cm of dpy1,font=\small] (lp2) {$[0]$}; \node
        [below = 0.5cm of dpx2,font=\small] (lp3) {$[3]$}; \node
        [below = 0.5cm of dpy2,font=\small] (lp4) {$[3]$}; \node
        [below = 0.5cm of dpx3,font=\small] (lp5) {$[6]$}; \node
        [below = 0.5cm of dpy3,font=\small] (lp6) {$[6]$};

        \node [below = 0.5cm of g1,font=\small] (lc1) {$[14,16]$};
        \node [below = 0.5cm of g2,font=\small] (lc2) {$[17,19]$};
        \node [below = 0.5cm of g3,font=\small] (lc3) {$[20,22]$};

        \node [below = 0.5cm of g4,font=\small] (lc4) {$[28,30]$};
        \node [below = 0.5cm of g5,font=\small] (lc5) {$[31,33]$};
        \node [below = 0.5cm of g6,font=\small] (lc6) {$[34,36]$};

        \node [below = 0.5cm of dnx1,font=\small] (ln1) {$[42]$};
        \node [below = 0.5cm of dny1,font=\small] (ln2) {$[44]$};
        \node [below = 0.5cm of dnx2,font=\small] (ln3) {$[45]$};
        \node [below = 0.5cm of dny2,font=\small] (ln4) {$[47]$};
        \node [below = 0.5cm of dnx3,font=\small] (ln5) {$[48]$};
        \node [below = 0.5cm of dny3,font=\small] (ln6) {$[50]$};

        \node [above = 0.2cm of x13,font=\tiny] (e1) {$=17$}; \node
        [above = 0.4cm of y13,font=\tiny] (e2) {$=17$}; \draw[rounded
        corners] (x11) |- (e1) -| (x22); \draw[rounded corners] (y11)
        |- (e2) -| (y22);

        \node [right = 0.1cm of dpy3] (g1) {}; \node [above = 0.01cm
        of g1] (g2) {}; \node [left = 0.1cm of x21] (g3) {}; \node
        [below = 0.01cm of g3] (g4) {}; \draw[rounded corners,dashed]
        (g2) rectangle (g4);

        \node [right = 0.1cm of y13] (g5) {}; \node [above = 0.01cm of
        g5] (g6) {}; \node [left = 0.1cm of dnx1] (g7) {}; \node
        [below = 0.01cm of g7] (g8) {}; \draw[rounded corners,dashed]
        (g6) rectangle (g8);
  
      \end{tikzpicture}
    }
    \caption{Zones $Z_\varphi, Z'_\varphi$ for formula
      $\varphi = (p_{1} \lor p_{2} \lor p_{3}) \land (\bar{p}_{4} \lor
      \bar{p}_{1} \lor \bar{p}_{5})$.}
    \label{fig:complexity-example}
  \end{figure}

\paragraph*{The idea:} We see that $\varphi$ is satisfied iff \emph{there exists some assignment} of the variables of $\varphi$ such that \emph{for every clause} $C_{i}$ for $1 \leq i \leq n$, $C_{i}$ is true. Comparing this with the definition of non-entailment: $Z_{\varphi} \not\entails Z'_{\varphi}$ iff \emph{there exists} $\gamma' \in Z'_{\varphi}$ such that \emph{for every} $\gamma'' \incl \gamma'$ \emph{and} $\gamma \in Z_{\varphi}$, we have $\gamma'' \not\regeq_{\mc} \gamma$. In our construction, we will pick a large enough constant $M_{\varphi}$ so that $\gamma'' \regeq_{\mc} \gamma$ boils down to simply $\gamma'' = \gamma$. So, we see that we need to construct zones $Z_{\varphi}$ and $Z'_{\varphi}$ such that each $\gamma' \in Z'_{\varphi}$ corresponds to an assignment, and having no subset $\gamma'' \incl \gamma'$ with $\gamma'' \in Z_{\varphi}$ corresponds to every clause being true. To encode assignments through our zones, we use two variables, $x$ and $y$ and use the value of their difference, $y-x$, to decide the assignment of the variable they encode: $[0,1]$ as false and $(1,2]$ as true. We present the key ideas of our construction though an example $\varphi = (p_{1} \lor p_{2} \lor p_{3}) \land (\bar{p}_{4} \lor \bar{p}_{1} \lor \bar{p}_{5})$. Fig.~\ref{fig:complexity-example} illustrates the construction for $\varphi$. The details of the construction are presented in Appendix~\ref{sec:append-ent}.

\paragraph*{Zone $Z'_\varphi$:} We want to construct this zone to encode the possible assignments to the variables of $\varphi$. To do this, we consider variables $x_j^i, y_j^i$ for $1 \le j \le 3$ and $1 \le i \le m$ with constraints $0 \le y_j^i - x_j^i \le 2$. Each valuation to the $x, y$ variables assigns a value between $0$ to $2$ for the difference. For it to encode an assignment, the assigned values should be consistent across all occurrences of the same variable: for example, in Fig.~\ref{fig:complexity-example}, variable $p_1$ occurs in $C_1$ positively and in $C_2$ negatively. We add two constraints $x_2^2 - x_1^1 = 17$ and $y_2^2 - y_1^1 = 17$. These constraints ensure that $ y_1^1 - x_1^1 = y_2^2 - x_2^2$. Therefore, in all occurrences of the variable, we have the difference $y - x$ to be the same, and hence we can extract an assigment of the variables of $\varphi$ from a valuation of the $x, y$ variables in $Z'_\varphi$. Fig.~\ref{fig:complexity-example} also shows other variables $px_j, py_j$ and $nx_j, ny_j$, which we will explain later, after describing $Z_\varphi$.

\paragraph*{Zone $Z_{\varphi}$:} In the entailment check $Z_\varphi \entails Z'_\varphi$, we pick a $\gamma' \in Z_\varphi'$, select a subset $\gamma'' \incl \gamma'$ and check whether $\gamma'' \notin Z_\varphi$. The subset selection $\gamma''$ should correspond to a clause and $Z_\varphi$ intuitively encodes one clause and checks if all literals are false. Since negative and positive clauses need to be dealt with  differently, we instead encode two clauses, one positive (the green part of $Z_\varphi$ in Fig.~\ref{fig:complexity-example}) and one negative (the red part in Fig.~\ref{fig:complexity-example}). The constraints of $Z_\varphi$ ensure that all literals are false, hence the two selected clauses are false. The extra constraints $[1, 5]$ ensure that all three literals are picked from the same clause. Moreover, for the mapping to work correctly, we set $\loc(x_j^i) = q_x$ and $\loc(x_i^+) = \loc(x_i^-) = q_x$, and $\loc(y_j^i) = q_y$, $\loc(y_i^+) = \loc(y_i^-) = q_y$ for all relevant indices. 

While evaluating $Z_\varphi \not \entails Z'_\varphi$, we will consider each $\gamma' \in Z'_\varphi$, and find projections $\gamma'' \incl \gamma'$ that map to $Z_\varphi$. By our choice of the variable signatures, each projection will pick a positive clause and a negative clause.
\begin{itemize}
  \item when $\gamma'$ is a satisfying assignment, no projection $\gamma''$ belongs to $Z_\varphi$, since all clauses are true, and $Z_\varphi$ encodes a positive and a negative clause. 
  \item when $\gamma'$ is not a satisfying assignment (that is, some of the clauses are false), there should be some projection $\gamma''$ that should belong to $Z_\varphi$. However, with the construction mentioned so far, this may not be possible: suppose $\gamma'$ makes one positive clause false and all negative clauses true; any projection $\gamma''$ will have to include a negative clause, but then due to the constraints on $Z_\varphi$, $\gamma'' \notin Z_\varphi$. This is where we make use of dummy clauses.
\end{itemize}
We add $12$ extra variables $px_j, py_j, nx_j, ny_j$ for $j=1,2,3$ to correspond to a dummy positive and a negative clause. The values of $py_j - px_j$ and $ny_j - nx_j$ encode false for the corresponding literals. Therefore, for a $\gamma'$ that is not a satisfying assignment, we can pick a projection $\gamma''$ that chooses a clause that is made false by $\gamma'$ and a dummy clause of the opposite polarity. This takes care of the second point above. However, due to this addition, there is a projection $\gamma''$ consisting of both dummy clauses. One can always choose this pair of dummy clauses to belong to $Z_\varphi$, even when $\gamma'$ is a satisfying assignment. To eliminate this, we add a final constraint to $Z_\varphi$ which bounds the distance between the two extremes ($y_3^- - x_1^+ < 50$ in Fig.~\ref{fig:complexity-example}) preventing  choosing both the dummy clauses.

\subsection{Entailment for bounded width 1-ATAs} Suppose we start with a 1-ATA $\Aa$ with width $k$. Then in the zone graph computation as explained in Section~\ref{sec:zone-graph-1}, every variable name $x_{q, i}$ will have index $i \le k$. For each state $q$, there are at most $k$ variables in the zone (along with potentially an $x_{q, 0}$ in the inactive part). In such a scenario, we can use a modified entailment check that only compares zones with the same set of variable names and thereby avoids the costly identification of subsets.
  
Let $Z$ and $Z'$ be two zones such that $\Var(Z) = \Var(Z')$. Let $\iota:\Var(Z) \to \Var(Z')$ be the identity mapping. Let $\regeq_{\mc}^\iota$ be the equivalence of Definition~\ref{def:regeq} with $h$ being the identity function $\iota$. So, $\regeq_{\mc}^\iota$ is simply the classical region equivalence of \cite{alur1994theory}. This can be naturally lifted to zones: $(Z,\Inact) \entails^b_{\mc} (Z', \Inact')$ if $\Inact \incl \Inact'$ and for every $\gamma' \in Z'$ there is some $\gamma \in Z$ such that $\gamma \regeq^\iota_{\mc} \gamma'$. The check $\entails^b$ is identical to a simulation test $\preccurlyeq_{\mc}$ used in the timed automata literature~\cite{herbreteau2016better}, but with the direction reversed: $(Z, \Inact) \entails^b_{\mc} (Z', \Inact')$ iff $\Inact \incl \Inact'$ and $Z'  \preccurlyeq_{\mc} Z$. This test can be done with quadratic complexity and is known to be a well-quasi-order on zones with a fixed number of variables.

Notice that $\entails^b$ implies $\entails$, but not the other way around. Therefore, the zone graph obtained by $\entails^b$ is correct. Notice that using $\entails$ may lead to more pruning, in principle. However, the entailment check $\entails^b$ is efficient. Moreover, as there are at most $|Q| \times k$ number of variables generated, and the relation $\entails^b$ induces a well-quasi order for each $j \le |Q| \times k$, the overall graph will terminate. This is because if not, there is an infinite path in the zone graph with an infinite subsequence of zones over the same set of variables, which is non-increasing w.r.t $\entails^b_\mc$, which is a contradiction.


\section{Extending the zone graph for the model-checking problem}
\label{sec:zg-mc}

We come back to the model-checking problem: given a timed automaton model $\Aa$ and a specification given as a $1$-ATA $\Bb$, check  whether $L(\Aa) \incl L(\Bb)$. This problem is known to be decidable for finite words (\cite{ouaknine2007decidability}), and can be reduced to checking emptiness of $L(\Aa \times \Bb^{c})$, where $\Bb^{c}$ is the complement of $\Bb$, also a 1-ATA. In this section, we develop a zone graph computation for the product $\Aa \times \Bb$, where $\Aa$ is a timed automaton and $\Bb$ is a 1-ATA. Observe that this subsumes model-checking MTL specifications against timed automata models, since MTL specifications can be converted to 1-ATAs (Section~\ref{sec:improving-mtl}). The plan of this section is as follows.
\begin{enumerate}
\item We formally define the semantics of a product $\Aa \times \Bb$ where $\Aa$ is a timed automaton and $\Bb$ is a $1$-ATA. 
\item We define a zone graph for $\Aa \times \Bb$ and extend the entailment $\entails$ to this \emph{compound zone graph}.
\item When $\Bb$ has a bounded width, the number of variables generated in the zone graph of $\Aa \times \Bb$ is bounded and the modified entailment $\entails_b$ (with quadratic complexity) can be adapted for this compound zone graph.
\item Finally, we consider the case when $\Aa$ is strongly non-Zeno, that is, there is a number $k$ such that every $k$ consecutive actions of $\Aa$ incurs a duration of at least $1$ time unit. When $\Aa$ is strongly non-Zeno, we observe that the number of variables within the maximum constant $\mca{A}$ is bounded. Using this observation, we propose a new entailment relation for the compound zone graph of $\Aa \times \Bb$ that has a quadratic complexity -- essentially, the entailment uses $\entails_b$ for the variables within $\mca{A}$ and does a simple constant time check for the variables above $\mca{A}$.
\end{enumerate} 

We recall the definition and semantics of timed automata \cite{alur1994theory} that we will use in this section:
\begin{defi}[Timed Automata]
  A timed automaton (TA) is given by a tuple $\Aa = (Q,\Sigma,Y,q_{0},F,\delta)$, where $Q$ is a finite set of locations, $q_{0} \in Q$ and $F \incl Q$ are the initial and final locations respectively, $Y$ is a finite set of clocks, and the transition relation $\delta \incl Q \times \Sigma \times \Phi(Y) \times \mathcal{P}(Y) \times Q$, with $\Phi(Y)$ being the set of constraints generated by $\varphi := y \sim c \mid \varphi \land \varphi$ for $\sim \mathord{\in} \{<,\leq,>,\geq\}$, $c \in \N$, and $y \in Y$, and $\mathcal{P}(Y)$ is a set of clocks. A transition (which we sometimes call an \emph{edge}) of $\Aa$ is of the form $e = (q,a,g,R,q')$ where $g \in \Phi(Y)$ is called a guard and $R \incl Y$ is the set of clocks that are reset. 
\end{defi}

Given a TA $\Aa = (Q,\Sigma,Y,q_{0},F,\delta)$ with $Y = \{y_{1},\dots,y_{m}\}$, a clock valuation on $\Aa$ is a vector $\bar{\nu} = (\nu_{1},\nu_{2},\dots,\nu_{m})$ where each $\nu_i \in \Rpos$. Given $d \in \Rpos$ and $R \incl Y$, $\bar{\nu} + d = (\nu_{1}+d,\dots,\nu_{m}+d)$ and $[R]\bar{\nu} = \bar{\nu}'$ such that $\nu'_{i} = 0$ if $y_{i} \in R$, $\nu'_{i} = \nu_{i}$ otherwise. Given a clock valuation $\bar{\nu} = (\nu_{1},\dots,\nu_{m})$ and guard $g$, $\bar{\nu} \sat g$ if replacing $y_{i}$ with $\nu_{i}$ satisfies the constraints of $g$.

We fix a TA $\Aa = (Q_{A},\Sigma,Y,q_{A},F_{A},\delta_{A})$ with $m$ clocks and maximum constant $\mca{A}$, and a 1-ATA $\Bb = (Q_{B}, \Sigma, q_B, F_{B}, \delta_{B})$ with maximum constant $\mca{B}$ for the rest of this section. We say the maximum constant of $\Aa$ and $\Bb$ combined is $\mc = \max(\mca{A},\mca{B})$.

\begin{defi}[Compound Configurations]
  A compound configuration is of the form $\Gamma = (p,\bar{\nu},\gamma)$, where $p\in Q_{A}$ is a location of $\Aa$, $\bar{\nu}$ is a clock valuation of $\Aa$, and $\gamma = \{ (q_1, v_1), \dots, (q_n, v_n)\}$ is a configuration of $\Bb$. A compound configuration is accepting if $p \in F_{A}$ and $\gamma$ is an accepting configuration of $\Bb$.
\end{defi}
We will define two kinds of transitions for $\Gamma$:
\begin{enumerate}
  \item Timed Transitions: for $d \in \Rpos$, we define $\Gamma + d$ as $(p,\bar{\nu}+d,\gamma+d)$, and we add edge $\Gamma \xra{d} \Gamma + d$ for all $d \in \Rpos$.
  \item Discrete Transitions: for $a \in \Sigma$, an edge $e = (p,a,g,R,p')$ in $\Aa$, and a combination $C = (C_{1},\dots,C_{n})$ where each $C_{i}$ is a disjunct of $\delta_{B}(q_{i},a)$, we add an edge $\Gamma \xra{a,e,C} \Gamma'$, where $\Gamma' = \{p',\bar{\nu}',\gamma'\}$, if $\gamma \xra{a,C} \gamma'$, $\bar{\nu} \sat g$, and $\bar{\nu}' = [R]\bar{y}$.
\end{enumerate}

The semantics is given by a \emph{compound configuration graph} where the initial node is $\Gamma_{0} = (q_{A},\bar{\nu_{0}},\gamma_{0})$ where $\bar{\nu_{0}} = (0,\dots,0)$ and $\gamma_{0} = \{(q_{B},0)\}$ and the successor computation is as defined above. A run on this graph for a timed word $w = (d_{1},a_{1})\dots (d_{n},a_{n})$ is a sequence of transitions of the form $\Gamma_{0} \xra{d_{1}} \Gamma'_{0} \xra{a_{1},e_{1},C^{1}} \dots \xra{a_{n},e_{n},C^{n}} \Gamma_{n}$, where it is an accepting run if $\Gamma_{n}$ is an accepting compound configuration. By our transitions, it is easy to see that a word will have an accepting run on both TA $\Aa$ and 1-ATA $\Bb$ if and only if the word has an accepting run in the compound configuration graph. Checking emptiness of $\Aa \times \Bb$ reduces to checking whether the compound configuration graph as an accepting run.

\subsection{Compound zone graph} 
We can now define the extended zone graph. We again assume that we have an infinite supply $\Var$ of variable names $x_{q,1},\dots,x_{q,i}$ for each $q \in Q_{B}$, and for a variable $x_{q,i}$, define $\loc(x_{q,i}) = q$. In the rest of the section, we will use both $\Var$ (the set of variables) and $Y$ (the set of clocks of $\Aa$) in the definitions. Let us remark that we will refer to the elements of $\Var$ as variables and the elements of $Y$ as clocks.
\begin{defi}
  A compound zone $Z_{AB}$ is a conjunction of constraints of the form $(x \sim k)$ or $(x-z \sim k)$ where $\sim \mathord{\in} \{<,\leq,\geq,>\}$, $k \in \Z$, and $x,z \in \Var ~\cup~ Y$. We will define $\Var(Z_{AB}) \incl \Var$ as the set of variables appearing in $Z_{AB}$ and $\locsign(Z_{AB}) = \locsign(\Var(Z_{AB}))$.
\end{defi}

A node in the zone graph of $\Aa \times \Bb$ is of the form $N = (p, Z_{AB},\Inact)$ where $p \in Q_{A}$ is a location of $\Aa$, $Z_{AB}$ is a compound zone, and $\Inact \incl \{x_{q,0} \mid q \text{ is a location of }\Bb\}$ is the set of inactive variables. For a node $N = (p,Z_{AB},\Inact)$, we say $\Var(N) = \Var(Z_{AB}) \cup \Inact$ and $\locsign(N) = \locsign(Z_{AB}) \cup \locsign(\Inact)$.
A node $N = (p, Z_{AB},\Inact)$ is accepting if $p\in F_{A}$ and for every $x \in \Var(N)$, $\loc(x) \in F_{B}$.

We can now naturally extend the definition of a compound configuration satisfying a node. Given a node $N = (p,Z_{AB},\Inact)$, we say that a compound configuration $\Gamma = (p',\bar{\nu},\gamma)$ satisfies $N$, written as $\Gamma \sat N$, if $p = p'$ and there exists a surjection $h: \Var(N) \mapsto \gamma$ such that (i) for every $x \in \Var(N)$, $\loc(x) = \loc(h(x))$, (ii) for every $x_{q,0} \in \Inact$, $\val(h(x_{q,0})) = \bot$, and (iii) replacing every clock in $Y$ with the respective clock value in $\bar{\nu}$ and every variable $x \in \Var(Z_{AB})$ with $\val(h(x))$ satisfies all the constraints of $Z_{AB}$. We will define $\llbracket N \rrbracket$ as the set of compound configurations $\{\Gamma \mid \Gamma \sat N\}$.

\paragraph*{Computing Successors.} 
To compute successors of such compound nodes, we build on the successor computation of Section \ref{sec:zone-graph-1}. We need to now include a target outgoing edge for the TA $\Aa$ along with the set of disjuncts we select as target for $\Bb$. Assuming $N = (p,Z_{AB},\Inact)$ is a non-empty node with $\Var(Z_{AB}) = \{x_{q_{1},i_{1}},\dots,x_{q_{k},i_{k}}\}$ (indices 1 to $k$) and $\Inact = \{x_{q_{k+1},0},\dots,x_{q_n, 0}\}$ (indices $k+1$ to $n$), we extend the definition of the target set for an action $a \in \Sigma$ as follows, where $e$ is an edge of $\Aa$ of the form $(p,a,g,R,p')$:
\[target(N,a) = \{(e,(C_{1},\dots,C_{n}) \mid C_{j} \text{ is a disjunct in }\delta_{B}(q_{j},a) \text{ for }1 \leq j \leq n)\}\]
Picking one $(e,(C_{1},\dots,C_{n}))$ from this set where no $C_{j}$ contains $\false$, the time elapse computation remains the same as before. For the guard intersection, we first compute $Z_{AB} \cap g$, and on the resulting node, we perform the guard intersection step in the successor computation of Section~\ref{sec:zone-graph-1} using the target $(C_1, \dots, C_n)$, to get a node $(p,Z^{2}_{AB},\Inact_{2})$.
The computation to reset and move to new variables will remain unchanged on the $\Bb$ variables, and additional constraints $\bigwedge_{y' \in R} (y' = 0)$ will be added to the computed zone. After the canonicalization and removing old variables to compute $Z'_{AB}$ and $\Inact'$, the successor node will finally be $(p',Z'_{AB},\Inact')$. 

We can now define the compound zone graph. The initial node will now be $N_{0} = (q_{A},Z^{0}_{AB},\emptyset)$ where $Z^{0}_{AB}$ is a zone with constraints $\bigwedge_{y \in Y}(y=0) \land (x_{q_{B},1} = 0)$, and the successor nodes are computed as defined above. We can see the following:
\begin{thm}
  The compound zone graph is sound and complete.
\end{thm}
\begin{proof}
  The proof for this will be very similar to the 1-ATA zone graph proofs for Lemma \ref{lem:succ-zone} and Theorem \ref{thm:zonecas}, and therefore the full proof is omitted here. The key difference is that for this case, the zone graph computation also has to incorporate the transition edges taken by the TA $\Aa$, i.e. that the guard constraints and resets are incorporated in the successor node. Since we ensure the guard constraints are satisfied by all variables and clocks in a successor node, the successor node will contain exactly those compound configurations that can, after some time delay, satisfy the guard of an edge along with the intervals involving the transitions of $\Bb$. 
\end{proof}
\paragraph*{Extending the entailment relation} 
Similar to the 1-ATA zone graph, this zone graph can also be potentially infinite, so we still require a way to prune the zone graph. We will do so by extending the entailment relation for this case.

\begin{defi}[Flattening compound configurations]
Let $\Gamma = (p, \bar{\nu}, \gamma)$ be a compound configuration with $\bar{\nu} = (\nu_1, \dots, \nu_m)$. We define $\Gamma_{\flat}$ to be a configuration $\gamma \cup \{ (p^i, \nu_i) \mid 1 \le i \le m\}$ where $p^i$ is a fictitious location, created as a copy of location $p$ for clock $i$.
\end{defi}

\begin{defi}\label{def:entext}
For compound configurations $\Gamma = (p, \bar{\nu}, \gamma)$ and $\Gamma' = (p', \bar{\nu}', \gamma')$, we say $\Gamma \regeq_{\mc} \Gamma'$ if their flattened configurations are equivalent, that is $\Gamma_{\flat} \regeq_{\mc} \Gamma'_{\flat}$ according to Definition~\ref{def:regeq}. Secondly, we say $\Gamma \entails_{\mc} \Gamma'$ if $\Gamma_{\flat} \entails_{\mc} \Gamma'_{\flat}$.
\end{defi}

This definition can be extended to compound nodes: we say that node $(p, Z_{AB},\Inact) \entails_{\mc} (p', Z'_{AB}, \Inact')$ if for every $\Gamma' \in (p,Z'_{AB},\Inact')$, there is some $\Gamma \in (p,Z_{AB},\Inact)$ such that $\Gamma \entails_{\mc} \Gamma'$. The algorithm from Section~\ref{sec:entailment} can be directly adapted to compute the current entailment for the compound zone graph.

\subsection{Bounded cases for model checking}
We now look at restrictions on $\Aa$ and $\Bb$ that will give us a compound zone graph with a global bound on the variables (and clocks) used in every reachable node. We can notice straightaway that because there is a constant number of clocks $Y$, if we restrict $\Bb$ to be a 1-ATA with width $k$, we will get a compound zone graph with a bounded number of variables.  
\begin{lem}
Let $\Aa$ be a timed automaton with $m$ clocks and $\Bb$ a 1-ATA with width $k$. Then, for every reachable node $(p, Z_{AB}, \Inact)$ in the compound zone graph of $\Aa \times \Bb$, $Z_{AB}$ is a zone over atmost $k + m$ variables plus clocks. 
\end{lem}

We can use the same reasoning as in Section \ref{sec:entailment} to argue that the more efficient entailment check $\entails^{b}$ is sufficient here too. We now look at a restriction of $\Aa$ that will lead to a similar situation. We recall the definition of strongly non-zeno timed automata below.

\begin{defi}
  A timed automaton $\Aa$ is strongly non-zeno if there exists a constant $k$ such that for any run of $\Aa$ with length $\ge k$, at least 1 time unit is elapsed. 
\end{defi} 
Restricting the TA $\Aa$ to be strongly non-zeno does not directly ensure that the zone graph for $\Aa \times \Bb$ has a bounded number of variables, but we can notice the following. 

Consider a node $N = (p, Z_{AB}, \Inact)$ and a variable $x \in \Var(Z_{AB})$. We say that $x$ is \emph{irrelevant} in $Z_{AB}$ if in \emph{every} configuration $\Gamma \in N$, the value of $x$ is strictly bigger than $M$. Variable $x$ is said to be \emph{relevant} otherwise.

\begin{lem}\label{lem:nzbdd}
Let $\Aa$ be a TA which is strongly non-Zeno with a constant $k$. Let $\Bb$ be an arbitrary 1-ATA. Then, in any reachable node $N = (p, Z_{AB}, \Inact)$ in the compound zone graph of $\Aa \times \Bb$, the number of relevant variables is atmost $|Q_\Bb| \times k \times (\mc + 1)$ where $Q_{\Bb}$ is the set of states of $\Bb$.
\end{lem}
\begin{proof}
  Let $Y$ be the set of clocks of $\Aa$. 
We first remark that every reachable zone $N = (p, Z_{AB}, \Inact)$ is \emph{totally ordered}: for every $x, z \in \Var(Z_{AB}) \cup Y$, the constraints of $Z_{AB}$ implies $x \le z$ or $z \le x$. The ordering reflects the order of resets along the path leading to $N$. The remark can be proved by an induction on the length of the path. The initial node is easily seen to satisfy this property. Suppose $N_1 = (p_1, Z_1, \Inact_1)$ is totally ordered and let $N_1 \xra{a, e, C} N_2 = (p_2, Z_2, \Inact_2)$. In $Z_2$, all new variables created due to an $x.q$ atom in $C$, and all clocks of $Y$ that were reset in $e$ will have value $0$ in every configuration of $Z_2$. For the rest of the variables, the ordering is maintained. This shows that $(p_2, Z_2, \Inact_2)$ is totally ordered.

We will show that for any reachable node $N = (p,Z_{AB},\Inact)$ of the zone graph, there are at most $k \times (\mc +1) \times |Q_{B}|$ relevant variables in $Z_{AB}$. For convenience, we will assume that the indices of variables of $Z_{AB}$ are ordered according to the ordering of their values: $x_{q,i} \leq x_{q,j}$ if $i \leq j$. We can ensure this naming convention by a small modification to the successor computation where the new variables that are created and assigned indices according to the ordering of their values, and thus we can make this assumption without loss of generality. 
  
  Now, given a $q \in Q_{B}$, we first consider the variables of $Z_{AB}$ of the form $x_{q,i}$ for $i \in \{1,2,\dots\}$. Consider $x_{q, n}$ with $n > k \times (\mc + 1)$. This variable was created sometime in the past in a node $N'$ along the path from the initial node $N_0$ to the current node $N$. When it was created, $x_{q, n}$ had value $0$. Moreover, after it was created, at least, $k \times (M + 1)$ edges appear along the path. Since $\Aa$ is strongly non-Zeno, this implies that at least $k \times (\mc + 1)$ time has elapsed in every run encoded by the path from $N'$ to $N$. Hence, in every configuration of $N$, the value of $x_{q, n}$ is at least  $(\mc + 1)$, hence $> \mc$. This shows that $x_{q, n}$ is irrelevant for every $n \ge k \times (\mc + 1)$.
\end{proof}

Using the above lemma, we see that in the zone graph, however the number of variables might grow in the graph, there are only a bounded number of them relevant to us with respect to the entailment relation. Given a compound zone $Z_{AB}$, we can define $\bd(Z_{AB})$ as $Z_{AB}$ projected onto the first $k.(\mc + 1)$ variables of the form $x_{q,i}$ for each $q \in Q_{B}$, along with the clocks in $Y$; to deal with the rest of the variables which have become irrelevant, we define the notion $\ubd(Z_{AB})$ as the multiset $\{q \mid x_{q,i} \in \Var(Z_{AB}), i > k.\mc\}$. Given a node $N = (p,Z_{AB},\Inact)$, we will similarly define $\bd(N) = (p,\bd(Z_{AB}),\Inact)$ and $\ubd(N) = \ubd(Z_{AB})$. We can now say the following.
\begin{lem}
  Given two nodes $N = (p,Z_{AB},\Inact)$ and $N' = (p',Z'_{AB},\Inact')$, $N \entails_{\mc} N'$ iff $\bd(N) \entails_{\mc} \bd(N')$ and $\ubd(N) \incl \ubd(N')$.
\end{lem} 
\begin{proof}
Let $\Gamma \in N$ and $\Gamma' \in N'$. We inherit the definition of $\bd$ and $\ubd$ at the configuration level in a natural way. We will now show that:
\begin{align} \label{eq:snz-ent}
\Gamma \entails_{\mc} \Gamma' \text{ iff } \bd(\Gamma) \entails_{\mc} \bd(\Gamma') \text{ and } \ubd(\Gamma) \incl \ubd(\Gamma')
\end{align}
Notice that proving the above statement completes the proof of the lemma.

Let us now prove \eqref{eq:snz-ent}.
From Definition~\ref{def:entext}, we have 
\begin{align*}
\Gamma \entails_{\mc} \Gamma' \text{ iff } \Gamma_{\flat} \entails_{\mc} \Gamma'_{\flat}
\end{align*}
This means there exists a subset $\Gamma'' \incl \Gamma'_{\flat}$ such that $\Gamma_{\flat} \regeq_{\mc} \Gamma''$. According to the region equivalence (Definition~\ref{def:regeq}), states of $\Gamma_{\flat}$ that correspond to $\ubd(N)$ are mapped to states of $\Gamma''$ that come from $\ubd(N')$. In other words:
\begin{align*}
\Gamma_{\flat} \regeq_\mc \Gamma'' \text{ iff } \bd(\Gamma_{\flat}) \regeq_{\mc} \bd(\Gamma'') \text{ and } \ubd(\Gamma_{\flat}) \incl \ubd(\Gamma'')
\end{align*}
This proves \eqref{eq:snz-ent}.
\end{proof}

As the final observation, as the number of relevant variables in every reachable node is bounded (Lemma~\ref{lem:nzbdd}), we can replace $\bd(N) \entails_M \bd(N')$ in the above lemma with the modified entailment $\entails^b$ of Section~\ref{sec:entailment}. This shows that for the compound zone graph of $\Aa \times \Bb$, we have an efficient entailment check when $\Aa$ is strongly non-Zeno. 


\section{Conclusion}
In timed automata literature, 1-ATAs have played the role of an excellent technical device for logic-to-automata translations. Perhaps, owing to the high general complexity, there have been no attempts at aligning the analysis of 1-ATAs to the well developed timed automata algorithms. Typically, 1-ATAs are converted to equivalent network of timed automata, with a blowup in the control states. Our aim in this paper is to pass on the message that these conversions of 1-ATAs to TAs may not really be needed and instead one could embed the analysis of 1-ATAs into the current timed automata algorithms. To substantiate this claim, we have given a zone-based emptiness algorithm for 1-ATAs. We have also shown how we can lift it to a model-checking algorithm, by providing a zone graph for the product of a timed automaton $\Aa$ with a 1-ATA $\Bb$. When $\Aa$ is strongly non-Zeno or when $\Bb$ has a bounded-width, the zone graph computation uses efficient operations, with the same complexity as in timed automata. We have also demonstrated a logical fragment of MTL which induces bounded width 1-ATAs. One important idea that we wish to highlight is the addition of an explicit deactivation operation for 1-ATAs. This is a simple trick that can have a substantial impact on the analysis.
  
We believe that the theoretical foundations we lay here pave the way for studying further optimizations, implementing the ideas and understanding the practical impact. As part of future work, it would be interesting to study whether the best known simulation techniques for timed automata~\cite{DBLP:conf/formats/BouyerGHSS22} can be incorporated into the 1-ATA zone graphs.  What about ATAs with multiple clocks? They are undecidable, of course. But, are there good restrictions that make the modeling more succinct and still enable a zone-based analysis? It would be interesting to extend the notion of boundedness on multi-clock ATAs and find out if there are fragments of MTL (MITL for instance) that are expressible by it. This study also motivates the question of understanding zone-based liveness algorithms for ATAs, in particular, bounded-width ATAs. 
Another direction is to explore the extensive literature based on antichain algorithms studied for alternating finite automata emptiness, in the  real-time setting. Here is another natural problem that we leave open: given a 1-ATA $\Aa$ and a constant $k$, decide whether $\Aa$ has width $k$?  

\bibliography{ATA-zones}

@article{alur1994theory,
  title={A theory of timed automata},
  author={Alur, Rajeev and Dill, David L},
  journal={Theoretical computer science},
  volume={126},
  number={2},
  pages={183--235},
  year={1994},
  publisher={Elsevier}
}

@inproceedings{abdulla2007zone,
  title={Zone-based universality analysis for single-clock timed automata},
  author={Abdulla, Parosh Aziz and Ouaknine, Jo{\"e}l and Quaas, Karin and Worrell, James},
  booktitle={International Conference on Fundamentals of Software Engineering},
  pages={98--112},
  year={2007},
  organization={Springer}
}

@article{alur1999event,
  title={Event-clock automata: A determinizable class of timed automata},
  author={Alur, Rajeev and Fix, Limor and Henzinger, Thomas A},
  journal={Theoretical Computer Science},
  volume={211},
  number={1-2},
  pages={253--273},
  year={1999},
  publisher={Elsevier}
}

@inproceedings{ouaknine2004language,
  title={On the language inclusion problem for timed automata: Closing a decidability gap},
  author={Ouaknine, Jo{\"e}l and Worrell, James},
  booktitle={Proceedings of the 19th Annual IEEE Symposium on Logic in Computer Science, 2004.},
  pages={54--63},
  year={2004},
  organization={IEEE}
}

@inproceedings{lasota2005alternating,
  title={Alternating timed automata},
  author={Lasota, S{\l}awomir and Walukiewicz, Igor},
  booktitle={International Conference on Foundations of Software Science and Computation Structures},
  pages={250--265},
  year={2005},
  organization={Springer}
}

@article{herbreteau2016better,
  title={Better abstractions for timed automata},
  author={Herbreteau, Fr{\'e}d{\'e}ric and Srivathsan, B and Walukiewicz, Igor},
  journal={Information and Computation},
  volume={251},
  pages={67--90},
  year={2016},
  publisher={Elsevier}
}

@InProceedings{gastin_et_al2018diag,
  author =	{Paul Gastin and Sayan Mukherjee and B. Srivathsan},
  title =	{{Reachability in Timed Automata with Diagonal Constraints}},
  booktitle =	{29th International Conference on Concurrency Theory  (CONCUR 2018)},
  pages =	{28:1--28:17},
  series =	{Leibniz International Proceedings in Informatics (LIPIcs)},
  ISBN =	{978-3-95977-087-3},
  ISSN =	{1868-8969},
  year =	{2018},
  volume =	{118},
  editor =	{Sven Schewe and Lijun Zhang},
  publisher =	{Schloss Dagstuhl--Leibniz-Zentrum fuer Informatik},
  address =	{Dagstuhl, Germany},
  URL =		{http://drops.dagstuhl.de/opus/volltexte/2018/9566},
  URN =		{urn:nbn:de:0030-drops-95660},
  doi =		{10.4230/LIPIcs.CONCUR.2018.28}
}

@inproceedings{baier2009timed,
  title={When are timed automata determinizable?},
  author={Baier, Christel and Bertrand, Nathalie and Bouyer, Patricia and Brihaye, Thomas},
  booktitle={Automata, Languages and Programming: 36th Internatilonal Collogquium, ICALP 2009, Rhodes, greece, July 5-12, 2009, Proceedings, Part II 36},
  pages={43--54},
  year={2009},
  organization={Springer}
}

@article{ouaknine2007decidability,
  title={On the decidability and complexity of metric temporal logic over finite words},
  author={Ouaknine, Jo{\"e}l and Worrell, James},
  journal={Logical Methods in Computer Science},
  volume={3},
  year={2007},
  publisher={Episciences. org}
}

@inproceedings{brihaye2013mitl,
  title={On MITL and alternating timed automata},
  author={Brihaye, Thomas and Esti{\'e}venart, Morgane and Geeraerts, Gilles},
  booktitle={Formal Modeling and Analysis of Timed Systems: 11th International Conference, FORMATS 2013, Buenos Aires, Argentina, August 29-31, 2013. Proceedings 11},
  pages={47--61},
  year={2013},
  organization={Springer}
}

@inproceedings{bouyer2007cost,
  title={The cost of punctuality},
  author={Bouyer, Patricia and Markey, Nicolas and Ouaknine, Jo{\"e}l and Worrell, James},
  booktitle={22nd Annual IEEE Symposium on Logic in Computer Science (LICS 2007)},
  pages={109--120},
  year={2007},
  organization={IEEE}
}

@article{alur1996benefits,
  title={The benefits of relaxing punctuality},
  author={Alur, Rajeev and Feder, Tom{\'a}s and Henzinger, Thomas A},
  journal={Journal of the ACM (JACM)},
  volume={43},
  number={1},
  pages={116--146},
  year={1996},
  publisher={ACM New York, NY, USA}
}

@inproceedings{DBLP:conf/avmfss/Dill89,
  author       = {David L. Dill},
  editor       = {Joseph Sifakis},
  title        = {Timing Assumptions and Verification of Finite-State Concurrent Systems},
  booktitle    = {Automatic Verification Methods for Finite State Systems, International
                  Workshop, Grenoble, France, June 12-14, 1989, Proceedings},
  series       = {Lecture Notes in Computer Science},
  volume       = {407},
  pages        = {197--212},
  publisher    = {Springer},
  year         = {1989},
  url          = {https://doi.org/10.1007/3-540-52148-8\_17},
  doi          = {10.1007/3-540-52148-8\_17},
  bibsource    = {dblp computer science bibliography, https://dblp.org}
}

@inproceedings{DBLP:conf/sfm/BehrmannDL04,
  author       = {Gerd Behrmann and
                  Alexandre David and
                  Kim Guldstrand Larsen},
  editor       = {Marco Bernardo and
                  Flavio Corradini},
  title        = {A Tutorial on Uppaal},
  booktitle    = {Formal Methods for the Design of Real-Time Systems, International
                  School on Formal Methods for the Design of Computer, Communication
                  and Software Systems, {SFM-RT} 2004, Bertinoro, Italy, September 13-18,
                  2004, Revised Lectures},
  series       = {Lecture Notes in Computer Science},
  volume       = {3185},
  pages        = {200--236},
  publisher    = {Springer},
  year         = {2004},
  url          = {https://doi.org/10.1007/978-3-540-30080-9\_7},
  doi          = {10.1007/978-3-540-30080-9\_7},
  bibsource    = {dblp computer science bibliography, https://dblp.org}
}

@inproceedings{DBLP:conf/formats/BouyerGHSS22,
  author       = {Patricia Bouyer and
                  Paul Gastin and
                  Fr{\'{e}}d{\'{e}}ric Herbreteau and
                  Ocan Sankur and
                  B. Srivathsan},
  editor       = {Sergiy Bogomolov and
                  David Parker},
  title        = {Zone-Based Verification of Timed Automata: Extrapolations, Simulations
                  and What Next?},
  booktitle    = {Formal Modeling and Analysis of Timed Systems - 20th International
                  Conference, {FORMATS} 2022, Warsaw, Poland, September 13-15, 2022,
                  Proceedings},
  series       = {Lecture Notes in Computer Science},
  volume       = {13465},
  pages        = {16--42},
  publisher    = {Springer},
  year         = {2022},
  url          = {https://doi.org/10.1007/978-3-031-15839-1\_2},
  doi          = {10.1007/978-3-031-15839-1\_2},
  bibsource    = {dblp computer science bibliography, https://dblp.org}
}

@inproceedings{DBLP:conf/tacas/DawsT98,
  author       = {Conrado Daws and
                  Stavros Tripakis},
  editor       = {Bernhard Steffen},
  title        = {Model Checking of Real-Time Reachability Properties Using Abstractions},
  booktitle    = {Tools and Algorithms for Construction and Analysis of Systems, 4th
                  International Conference, {TACAS} '98, Held as Part of the European
                  Joint Conferences on the Theory and Practice of Software, ETAPS'98,
                  Lisbon, Portugal, March 28 - April 4, 1998, Proceedings},
  series       = {Lecture Notes in Computer Science},
  volume       = {1384},
  pages        = {313--329},
  publisher    = {Springer},
  year         = {1998},
  url          = {https://doi.org/10.1007/BFb0054180},
  doi          = {10.1007/BFB0054180},
  bibsource    = {dblp computer science bibliography, https://dblp.org}
}

@book{DBLP:books/fm/GareyJ79,
  author       = {M. R. Garey and
                  David S. Johnson},
  title        = {Computers and Intractability: {A} Guide to the Theory of NP-Completeness},
  publisher    = {W. H. Freeman},
  year         = {1979},
  isbn         = {0-7167-1044-7},
  bibsource    = {dblp computer science bibliography, https://dblp.org}
}

@article{DBLP:journals/fmsd/Bouyer04,
  author       = {Patricia Bouyer},
  title        = {Forward Analysis of Updatable Timed Automata},
  journal      = {Formal Methods Syst. Des.},
  volume       = {24},
  number       = {3},
  pages        = {281--320},
  year         = {2004},
  url          = {https://doi.org/10.1023/B:FORM.0000026093.21513.31},
  doi          = {10.1023/B:FORM.0000026093.21513.31},
  bibsource    = {dblp computer science bibliography, https://dblp.org}
}

@article{DBLP:journals/sttt/BehrmannBLP06,
  author       = {Gerd Behrmann and
                  Patricia Bouyer and
                  Kim Guldstrand Larsen and
                  Radek Pel{\'{a}}nek},
  title        = {Lower and upper bounds in zone-based abstractions of timed automata},
  journal      = {Int. J. Softw. Tools Technol. Transf.},
  volume       = {8},
  number       = {3},
  pages        = {204--215},
  year         = {2006},
  url          = {https://doi.org/10.1007/s10009-005-0190-0},
  doi          = {10.1007/S10009-005-0190-0},
  bibsource    = {dblp computer science bibliography, https://dblp.org}
}

@article{DBLP:journals/siglog/Srivathsan22,
  author       = {B. Srivathsan},
  title        = {Reachability in timed automata},
  journal      = {{ACM} {SIGLOG} News},
  volume       = {9},
  number       = {3},
  pages        = {6--28},
  year         = {2022},
  url          = {https://doi.org/10.1145/3559736.3559738},
  doi          = {10.1145/3559736.3559738},
  bibsource    = {dblp computer science bibliography, https://dblp.org}
}

@inproceedings{DBLP:conf/cav/BrihayeGHM17,
  author       = {Thomas Brihaye and
                  Gilles Geeraerts and
                  Hsi{-}Ming Ho and
                  Benjamin Monmege},
  editor       = {Rupak Majumdar and
                  Viktor Kuncak},
  title        = {MightyL: {A} Compositional Translation from {MITL} to Timed Automata},
  booktitle    = {Computer Aided Verification - 29th International Conference, {CAV}
                  2017, Heidelberg, Germany, July 24-28, 2017, Proceedings, Part {I}},
  series       = {Lecture Notes in Computer Science},
  volume       = {10426},
  pages        = {421--440},
  publisher    = {Springer},
  year         = {2017},
  url          = {https://doi.org/10.1007/978-3-319-63387-9\_21},
  doi          = {10.1007/978-3-319-63387-9\_21},
  bibsource    = {dblp computer science bibliography, https://dblp.org}
}

@inproceedings{DBLP:conf/concur/KrishnaMP18,
  author       = {Shankara Narayanan Krishna and
                  Khushraj Madnani and
                  Paritosh K. Pandya},
  editor       = {Sven Schewe and
                  Lijun Zhang},
  title        = {Logics Meet 1-Clock Alternating Timed Automata},
  booktitle    = {29th International Conference on Concurrency Theory, {CONCUR} 2018,
                  September 4-7, 2018, Beijing, China},
  series       = {LIPIcs},
  volume       = {118},
  pages        = {39:1--39:17},
  publisher    = {Schloss Dagstuhl - Leibniz-Zentrum f{\"{u}}r Informatik},
  year         = {2018},
  url          = {https://doi.org/10.4230/LIPIcs.CONCUR.2018.39},
  doi          = {10.4230/LIPICS.CONCUR.2018.39},
  bibsource    = {dblp computer science bibliography, https://dblp.org}
}

@misc{TChecker,
  title        = {{TChecker}},
  author       = {F. Herbreteau and G. Point},
  howpublished = {\url{https://github.com/fredher/tchecker}},
  year         = {v0.2 - April 2019}
}

@article{Larsen:1997:UPPAAL,
  author  = {Kim Guldstrand Larsen and
	Paul Pettersson and
	Wang Yi},
  title   = {{UPPAAL} in a Nutshell},
  journal = {{STTT}},
  volume  = {1},
  number  = {1-2},
  pages   = {134--152},
  year    = {1997}
}

@inproceedings{LTSmin,
  author    = {Gijs Kant and
               Alfons Laarman and
               Jeroen Meijer and
               Jaco van de Pol and
               Stefan Blom and
               Tom van Dijk},
  title     = {{LTS}min: High-Performance Language-Independent Model Checking},
  booktitle = {{TACAS}},
  series    = {Lecture Notes in Computer Science},
  volume    = {9035},
  pages     = {692--707},
  publisher = {Springer},
  year      = {2015}
}

@inproceedings{PAT,
  author    = {Jun Sun and Yang Liu and Jin Song Dong and Jun Pang},
  title     = {{PAT}: Towards Flexible Verification under Fairness},
  journal   = {Proceedings of the 21th International Conference on Computer Aided Verification (CAV'09)},
  year      = {2009},
   publisher = {Springer},
  pages     = {709-714},
  series    = {Lecture Notes in Computer Science},
  volume    = {5643},
}

@inproceedings{Theta,
    author     = {T\'oth, Tam\'as and Hajdu, \'{A}kos and V\"or\"os, Andr\'as and Micskei, Zolt\'an and Majzik, Istv\'an},
    year       = {2017},
    title      = {Theta: a Framework for Abstraction Refinement-Based Model Checking},
    booktitle  = {Proceedings of the 17th Conference on Formal Methods in Computer-Aided Design},
    isbn       = {978-0-9835678-7-5},
    editor     = {Stewart, Daryl and Weissenbacher, Georg},
    pages      = {176--179},
    doi        = {10.23919/FMCAD.2017.8102257},
}

@inproceedings{DBLP:conf/lics/OuaknineW05,
  author       = {Jo{\"{e}}l Ouaknine and
                  James Worrell},
  title        = {On the Decidability of Metric Temporal Logic},
  booktitle    = {20th {IEEE} Symposium on Logic in Computer Science {(LICS} 2005),
                  26-29 June 2005, Chicago, IL, USA, Proceedings},
  pages        = {188--197},
  publisher    = {{IEEE} Computer Society},
  year         = {2005},
  url          = {https://doi.org/10.1109/LICS.2005.33},
  doi          = {10.1109/LICS.2005.33},
  bibsource    = {dblp computer science bibliography, https://dblp.org}
}

@article{DBLP:journals/tocl/LasotaW08,
  author       = {Slawomir Lasota and
                  Igor Walukiewicz},
  title        = {Alternating timed automata},
  journal      = {{ACM} Trans. Comput. Log.},
  volume       = {9},
  number       = {2},
  pages        = {10:1--10:27},
  year         = {2008},
  url          = {https://doi.org/10.1145/1342991.1342994},
  doi          = {10.1145/1342991.1342994},
  bibsource    = {dblp computer science bibliography, https://dblp.org}
}

@inproceedings{DBLP:conf/cav/AkshayGGJS23,
  author       = {S. Akshay and
                  Paul Gastin and
                  R. Govind and
                  Aniruddha R. Joshi and
                  B. Srivathsan},
  editor       = {Constantin Enea and
                  Akash Lal},
  title        = {A Unified Model for Real-Time Systems: Symbolic Techniques and Implementation},
  booktitle    = {Computer Aided Verification - 35th International Conference, {CAV}
                  2023, Paris, France, July 17-22, 2023, Proceedings, Part {I}},
  series       = {Lecture Notes in Computer Science},
  volume       = {13964},
  pages        = {266--288},
  publisher    = {Springer},
  year         = {2023},
  url          = {https://doi.org/10.1007/978-3-031-37706-8\_14},
  doi          = {10.1007/978-3-031-37706-8\_14},
  bibsource    = {dblp computer science bibliography, https://dblp.org}
}

@inproceedings{DBLP:conf/cav/WulfDHR06,
  author       = {Martin De Wulf and
                  Laurent Doyen and
                  Thomas A. Henzinger and
                  Jean{-}Fran{\c{c}}ois Raskin},
  editor       = {Thomas Ball and
                  Robert B. Jones},
  title        = {Antichains: {A} New Algorithm for Checking Universality of Finite
                  Automata},
  booktitle    = {Computer Aided Verification, 18th International Conference, {CAV}
                  2006, Seattle, WA, USA, August 17-20, 2006, Proceedings},
  series       = {Lecture Notes in Computer Science},
  volume       = {4144},
  pages        = {17--30},
  publisher    = {Springer},
  year         = {2006},
  url          = {https://doi.org/10.1007/11817963\_5},
  doi          = {10.1007/11817963\_5},
  bibsource    = {dblp computer science bibliography, https://dblp.org}
}

@article{DBLP:journals/corr/abs-0902-3958,
  author       = {Laurent Doyen and
                  Jean{-}Fran{\c{c}}ois Raskin},
  title        = {Antichains for the Automata-Based Approach to Model-Checking},
  journal      = {Log. Methods Comput. Sci.},
  volume       = {5},
  number       = {1},
  year         = {2009},
  url          = {http://arxiv.org/abs/0902.3958},
  bibsource    = {dblp computer science bibliography, https://dblp.org}
}

@inproceedings{DBLP:conf/tacas/DoyenR10,
  author       = {Laurent Doyen and
                  Jean{-}Fran{\c{c}}ois Raskin},
  editor       = {Javier Esparza and
                  Rupak Majumdar},
  title        = {Antichain Algorithms for Finite Automata},
  booktitle    = {Tools and Algorithms for the Construction and Analysis of Systems,
                  16th International Conference, {TACAS} 2010, Held as Part of the Joint
                  European Conferences on Theory and Practice of Software, {ETAPS} 2010,
                  Paphos, Cyprus, March 20-28, 2010. Proceedings},
  series       = {Lecture Notes in Computer Science},
  volume       = {6015},
  pages        = {2--22},
  publisher    = {Springer},
  year         = {2010},
  url          = {https://doi.org/10.1007/978-3-642-12002-2\_2},
  doi          = {10.1007/978-3-642-12002-2\_2},
  bibsource    = {dblp computer science bibliography, https://dblp.org}
}

@inproceedings{DBLP:conf/cade/FiedorHHRSV23,
  author       = {Tom{\'{a}}s Fiedor and
                  Luk{\'{a}}s Hol{\'{\i}}k and
                  Martin Hruska and
                  Adam Rogalewicz and
                  Juraj S{\'{\i}}c and
                  Pavol Vargovc{\'{\i}}k},
  editor       = {Brigitte Pientka and
                  Cesare Tinelli},
  title        = {Reasoning About Regular Properties: {A} Comparative Study},
  booktitle    = {Automated Deduction - {CADE} 29 - 29th International Conference on
                  Automated Deduction, Rome, Italy, July 1-4, 2023, Proceedings},
  series       = {Lecture Notes in Computer Science},
  volume       = {14132},
  pages        = {286--306},
  publisher    = {Springer},
  year         = {2023},
  url          = {https://doi.org/10.1007/978-3-031-38499-8\_17},
  doi          = {10.1007/978-3-031-38499-8\_17},
  bibsource    = {dblp computer science bibliography, https://dblp.org}
}

@inproceedings{DBLP:conf/sat/HolikV24,
  author       = {Luk{\'{a}}s Hol{\'{\i}}k and
                  Pavol Vargovc{\'{\i}}k},
  editor       = {Supratik Chakraborty and
                  Jie{-}Hong Roland Jiang},
  title        = {Antichain with {SAT} and Tries},
  booktitle    = {27th International Conference on Theory and Applications of Satisfiability
                  Testing, {SAT} 2024, August 21-24, 2024, Pune, India},
  series       = {LIPIcs},
  volume       = {305},
  pages        = {15:1--15:24},
  publisher    = {Schloss Dagstuhl - Leibniz-Zentrum f{\"{u}}r Informatik},
  year         = {2024},
  url          = {https://doi.org/10.4230/LIPIcs.SAT.2024.15},
  doi          = {10.4230/LIPICS.SAT.2024.15},
  bibsource    = {dblp computer science bibliography, https://dblp.org}
}

@inproceedings{DBLP:conf/aplas/VargovcikH21,
  author       = {Pavol Vargovc{\'{\i}}k and
                  Luk{\'{a}}s Hol{\'{\i}}k},
  editor       = {Hakjoo Oh},
  title        = {Simplifying Alternating Automata for Emptiness Testing},
  booktitle    = {Programming Languages and Systems - 19th Asian Symposium, {APLAS}
                  2021, Chicago, IL, USA, October 17-18, 2021, Proceedings},
  series       = {Lecture Notes in Computer Science},
  volume       = {13008},
  pages        = {243--264},
  publisher    = {Springer},
  year         = {2021},
  url          = {https://doi.org/10.1007/978-3-030-89051-3\_14},
  doi          = {10.1007/978-3-030-89051-3\_14},
  bibsource    = {dblp computer science bibliography, https://dblp.org}
}

@inproceedings{DBLP:conf/concur/0001G0S24,
  author       = {S. Akshay and
                  Paul Gastin and
                  R. Govind and
                  B. Srivathsan},
  editor       = {Rupak Majumdar and
                  Alexandra Silva},
  title        = {{MITL} Model Checking via Generalized Timed Automata and a New Liveness
                  Algorithm},
  booktitle    = {35th International Conference on Concurrency Theory, {CONCUR} 2024,
                  September 9-13, 2024, Calgary, Canada},
  series       = {LIPIcs},
  volume       = {311},
  pages        = {5:1--5:19},
  publisher    = {Schloss Dagstuhl - Leibniz-Zentrum f{\"{u}}r Informatik},
  year         = {2024},
  url          = {https://doi.org/10.4230/LIPIcs.CONCUR.2024.5},
  doi          = {10.4230/LIPICS.CONCUR.2024.5},
  bibsource    = {dblp computer science bibliography, https://dblp.org}
}

@article{KRUSKAL1972297,
title = {The theory of well-quasi-ordering: A frequently discovered concept},
journal = {Journal of Combinatorial Theory, Series A},
volume = {13},
number = {3},
pages = {297-305},
year = {1972},
issn = {0097-3165},
doi = {https://doi.org/10.1016/0097-3165(72)90063-5},
url = {https://www.sciencedirect.com/science/article/pii/0097316572900635},
author = {Joseph B Kruskal}
}

@inproceedings{BouyerSV:Fossacs25,
  author       = {Patricia Bouyer and
                  B. Srivathsan and
                  Vaishnavi Vishwanath},
  editor       = {Parosh Aziz Abdulla and
                  Delia Kesner},
  title        = {Model-Checking Real-Time Systems: Revisiting the Alternating Automaton
                  Route},
  booktitle    = {Foundations of Software Science and Computation Structures - 28th
                  International Conference, FoSSaCS 2025, Held as Part of the International
                  Joint Conferences on Theory and Practice of Software, {ETAPS} 2025,
                  Hamilton, ON, Canada, May 3-8, 2025, Proceedings},
  series       = {Lecture Notes in Computer Science},
  volume       = {15691},
  pages        = {399--421},
  publisher    = {Springer},
  year         = {2025},
  url          = {https://doi.org/10.1007/978-3-031-90897-2\_19},
  doi          = {10.1007/978-3-031-90897-2\_19},
  bibsource    = {dblp computer science bibliography, https://dblp.org}
}

@article{ASARIN1998447,
title = {Controller Synthesis for Timed Automata1},
journal = {IFAC Proceedings Volumes},
volume = {31},
number = {18},
pages = {447-452},
year = {1998},
note = {5th IFAC Conference on System Structure and Control 1998 (SSC'98), Nantes, France, 8-10 July},
issn = {1474-6670},
doi = {https://doi.org/10.1016/S1474-6670(17)42032-5},
url = {https://www.sciencedirect.com/science/article/pii/S1474667017420325},
author = {Eugene Asarin and Oded Maler and Amir Pnueli and Joseph Sifakis}
}

@InProceedings{Tripakis:progress:99,
author="Tripakis, Stavros",
editor="Katoen, Joost-Pieter",
title="Verifying Progress in Timed Systems",
booktitle="Formal Methods for Real-Time and Probabilistic Systems",
year="1999",
publisher="Springer Berlin Heidelberg",
address="Berlin, Heidelberg",
pages="299--314",
isbn="978-3-540-48778-4"
}

@article{DBLP:journals/fac/BowmanG06,
  author       = {Howard Bowman and
                  Rodolfo G{\'{o}}mez},
  title        = {How to stop time stopping},
  journal      = {Formal Aspects Comput.},
  volume       = {18},
  number       = {4},
  pages        = {459--493},
  year         = {2006},
  url          = {https://doi.org/10.1007/s00165-006-0010-7},
  doi          = {10.1007/S00165-006-0010-7},
  bibsource    = {dblp computer science bibliography, https://dblp.org}
}

@inproceedings{Bansal:ATVA23,
  author       = {Suguman Bansal and
                  Yong Li and
                  Lucas M. Tabajara and
                  Moshe Y. Vardi and
                  Andrew M. Wells},
  editor       = {{\'{E}}tienne Andr{\'{e}} and
                  Jun Sun},
  title        = {Model Checking Strategies from Synthesis over Finite Traces},
  booktitle    = {Automated Technology for Verification and Analysis - 21st International
                  Symposium, {ATVA} 2023, Singapore, October 24-27, 2023, Proceedings,
                  Part {I}},
  series       = {Lecture Notes in Computer Science},
  volume       = {14215},
  pages        = {227--247},
  publisher    = {Springer},
  year         = {2023},
  url          = {https://doi.org/10.1007/978-3-031-45329-8\_11},
  doi          = {10.1007/978-3-031-45329-8\_11},
  bibsource    = {dblp computer science bibliography, https://dblp.org}
}

@article{Tempora,
  author       = {S. Akshay and
                  Prerak Contractor and
                  Paul Gastin and
                  R. Govind and
                  B. Srivathsan},
  title        = {Efficient Verification of Metric Temporal Properties with Past in
                  Pointwise Semantics},
  journal      = {CoRR},
  volume       = {abs/2510.14699},
  year         = {2025},
  url          = {https://doi.org/10.48550/arXiv.2510.14699},
  doi          = {10.48550/ARXIV.2510.14699},
  eprinttype    = {arXiv},
  eprint       = {2510.14699},
  bibsource    = {dblp computer science bibliography, https://dblp.org}
}
\appendix
\section{Appendix for Section \ref{sec:entailment}}\label{sec:append-ent}
\subsection{Algorithm for Entailment Check}
\paragraph*{Computing $N_{r}$}
We use the notion of Distance Graphs~\cite{herbreteau2016better} for zones and regions in the computation of $N_{r}$ and the following result:

\begin{lemC}[\cite{herbreteau2016better}]\label{lem:nvecycle}
  A distance graph $G$ has some negative cycle iff $\llbracket G \rrbracket = \emptyset$.
\end{lemC}

Coming back to $N_{r}$ for a given $r: \Var(Z) \mapsto \Var(Z')$, we know that some $v' \in Z'_{r}$ is in $N_{r}$ when $v \not\regeq_M v'$ for every $v \in Z$. In other words, looking at the region of $v'$, $reg(v')$, then $reg(v') \not\in Reg(Z)$, where $Reg(Z)$ is the set of regions present in $Z$. Also, by this definition, if $v' \in N_{r}$ for some $v' \in Z'_{r}$, then for any $v'' \in Z'_{r}$ with $v'' \regeq_M v'$, even $v'' \in N_{r}$. So it is enough to check a representative valuation from each region in $Z'_{r}$.
As we have renamed the variables of $Z$ and $Z'_{r}$, we can directly talk about the distance graphs for $Z$ and regions of $Z'_{r}$. This means if $G_{Z}$ and $G_{reg(v')}$ are the distance graphs for $Z$ and the region of $v' \in Z_{r}$ respectively, then $v' \in N_{r}$ iff $G_{Z} \cap G_{reg(v')} = \emptyset$. From \ref{lem:nvecycle}, we know that it means $v' \in N_{r}$ iff $G_{Z} \cap \gamma_{reg(v')}$ has a negative cycle. Now, we use another result from \cite{herbreteau2016better}, modifying it from the context of LU-simulations to the maximum constant $M$ we use here.
\begin{lemC}[\cite{herbreteau2016better}]\label{lem:2cyc}
  $G_{Z} \cap G_{reg(v')}$ has a negative cycle iff there are two variables $y,y'$ such that:
  \begin{enumerate}
    \item $v'(y) \leq M$
    \item If $v'(y') > M$, then $(<,-M) + (\leq, \lceil v'(y) \rceil) + (\lessdot_{yy'},c_{yy'}) < (\leq,0)$, and if $v'(y') \leq M$, then $(\leq,\lceil v'(y) \rceil - \lfloor v'(y') \rfloor) + (\lessdot_{yy'},c_{yy'}) < (\leq,0)$, where $y \xra{(\lessdot_{yy'},c_{yy'})} y'$ is the edge between $y,y'$ in $G_{Z}$.
  \end{enumerate}
\end{lemC}

This means if there is a negative cycle in $G_{Z} \cap G_{reg(v')}$, there is a specific kind of negative cycle involving just 2 variables. This simplifies the way to compute $N_{r}$, as we can break it down into computing the valuations pairwise on all possible $y, y'$, then taking their union. If we define $N_{r,y,y'} = \{v' \in Z'_{r} \mid G_{Z} \cap G_{reg(v')} \text{ has a -ve cycle on }y,y'\}$, then we see that
\[N_{r} = \bigcup_{\forall y,y'} N_{r,y,y'} \text{ where }y,y' \in \Var(Z)\cup \{0\} \]
Finally, we see that computing $N_{r,y,y'}$ is straightforward as we see from Lemma \ref{lem:2cyc} that $N_{r,y,y'}$ can be described as a zone. 
To compute this zone, we first build, for each $y,y'$, a distance graph $G_{y,y'}$ with vertices $\Var(Z'_{r}) \cup \{0\}$ as follows (where $y \xra{(\lessdot_{yy'},c_{yy'})} y'$ is an edge in $G_{Z}$):
\begin{enumerate}
    \item We add edge $0 \xrightarrow{(\leq,\, M)} y$ to $G_{y,y'}$.
    \item We add edge $y \xrightarrow{(<,\, c'_{yy'})} y'$ to $G_{y,y'}$.
    \item If $(\leq,\, M) > (<, c'_{yy'} - M)$, we replace the previously added edge to $0 \xrightarrow{(<,\,c'_{yy'} - M)} y$ in $G_{y,y'}$.
    \item We add edge $(<,\, \infty)$ between every other pair of variables in $G_{y,y'}$.
\end{enumerate}
The required $N_{r,y,y'} = G_{Z'_{r}} \cap G_{y,y'}$.

\subsection{Complexity}
  We define $M_{\varphi}$ to be a value more than $14(m+2)-6$, say
  $14(m+2)$. Hence, every value in $Z'_\varphi$ and $Z_\varphi$ is
  below the maximum constant. Due to this, checking if
  $\gamma'' \regeq_M \gamma$ for some $\gamma \in Z_\varphi$ simply
  becomes checking $\gamma'' \in Z_\varphi$. 

\paragraph*{Details in the construction of $Z_\varphi$.}

 We want to construct this zone to encode
  the process of selecting one clause and falsifying it. 
  We thus define the variables of this zone to have a set
  representing the literals of a positive clause, as
  $x^{+}_{j},y^{+}_{j}$ for $j = 1,2,3$ and another set for the
  literals of a negative clause as $x^{-}_{j},y^{-}_{j}$ for
  $j = 1,2,3$. We say that $loc(x^{+}_{j}) = loc(x^{-}_{j}) = q_{x}$
  for $j = 1,2,3$ and $loc(y^{+}_{j}) = loc(y^{-}_{j}) = q_{y}$ for
  $j = 1,2,3$. We can now define the constraints of this zone as
  follows:
  \begin{enumerate}
  \item We add constraints $(0 \leq y^{+}_{j} - x^{+}_{j} \leq 1)$ for
    $j=1,2,3$. This is to ensure that the positive literals get
    assigned the value false by our definition.
  \item We add constraints $(1 < y^{-}_{j} - x^{-}_{j} \leq 2)$ for
    $j=1,2,3$. This is to ensure that the negative literals get
    assigned the value true by our definition.
  \item We add constraints
    $(1 \leq x^{\sim}_{j+1} - y^{\sim}_{j} \leq 5)$ for
    $\sim \in \{+,-\}$ and $j = 1,2$. This is to ensure that each
    mapping from $Z'_{\varphi}$ to $Z_{\varphi}$ picks 3 literals from
    the same clause.
  \item We add constraints $y^{+}_{3} < 14(k+1) - 2$ and
    $14(k+1) - 2 < x^{-}_{1}$ (recall that $k$ is the number of
    positive clauses in the formula). This is to ensure that the
    positive clause variables of $Z'_\varphi$ are mapped only to the
    positive variables of $Z_\varphi$, and similarly for the negative
    variables.
  \item We add constraint $y^{-}_{3} - x^{+}_{1} < 14(m+2)-6$. This is
    to ensure that the mapping from $Z'_\varphi$ to $Z_\varphi$
    selects only one among the positive and negative ``dummy
    clauses''. 
  \end{enumerate}
  
\paragraph*{Details in the construction of $Z'_\varphi$.}

We want to construct this zone to encode the possible assignments to the variables of $\varphi$.  To do this, we consider variables $x_j^i, y_j^i$ for $1 \le j \le 3$ and $1 \le i \le m$. 
  We also add $12$ extra variables $px_j, py_j, nx_j, ny_j$ for $j=1,2,3$
  to correspond to a dummy positive and a negative clause. Similar to $Z_\varphi$, we define $\loc(x_j^i) = q_x$ and
  $\loc(y_j^i) = q_y$. We define the constraints of $Z'_\varphi$ as follows:
\begin{enumerate}[label=(\arabic*')]
    \item We add constraints $px_{j} = py_{j} = 3(j-1)$ for each $j=1,2,3$. This ensures that the dummy positive clause variables get assigned value false by our definition.
    \item We add constraints $nx_{j} = 14(m+1) + 3(j-1)$ and $ny_{j} =14(m+1) +3(j-1) + 2$ for each $j=1,2,3$. This ensures that the dummy negative clause variables get assigned value true by our definition.
    \item We add constraints $14i +3(j-1) \leq x^{i}_{j} \leq 14i+3(j-1)+2$, $14i +3(j-1) \leq y^{i}_{j} \leq 14i+3(j-1)+2$, and $y^{i}_{j} - x^{i}_{j} \geq 0$ for $1 \leq i \leq m$ and $1 \leq j \leq 3$. This explicitly sets the values of all the clause variables (s.t. $y-x$ values are within 0 to 2), and ensures that there is a minimum gap between values of variables in different clauses, and there is a global bound on the values of the positive clause variables.
    \item For each $l^{i}_{j}$, $l^{i'}_{j'}$ such that $var(l^{i}_{j}) = var(l^{i'}_{j'})$, we add constraints $x^{i'}_{j'} - x^{i}_{j} =y^{i'}_{j'} - y^{i}_{j} =  14(i'-i) + 3(j'-j)$. This ensures that the literals on the same variable of the formula are assigned the same value (true or false) by the assignment we define.
\end{enumerate}

Now, we define the way zone $Z'_{\varphi}$ encodes an assignment. For a configuration $\gamma' \in Z'_{\varphi}$, we say that $\alpha_{\gamma'}$ is the corresponding assignment if for every literal $l^{i}_{j}$,
\[\alpha_{\gamma'}(var(l^{i}_{j})) = \begin{cases}
    \top & \text{ if }y^{i}_{j} - x^{i}_{j} > 1\\
    \bot & \text{ if }y^{i}_{j} - x^{i}_{j}  \leq 1
  \end{cases}\]

\begin{lem}\label{lem:assignval}
  For every $\gamma' \in Z'_{\varphi}$, $\alpha_{\gamma'}$ is a valid assignment of the variables of $\varphi$
\end{lem}
\begin{proof}
  By the construction of $Z'_{\varphi}$, for any $x^{i}_{j},y^{i}_{j}$, $0 \leq y^{i}_{j} - x^{i}_{j} \leq 2$. This means every variable is assigned some value by the definition of $\alpha_{\gamma'}$ above. Also, if there are two literals $l^{i}_{j},l^{i'}_{j'}$ on the same variable, the constraint 4 on $Z'_{\varphi}$ ensures that $y^{i}_{j} - x^{i}_{j} > 1$ iff $y^{i'}_{j'} - x^{i'}_{j'} > 1$, and so the asisgnment of values to the variables is consistent, and so $\alpha_{\gamma'}$ is a valid assignment.
\end{proof}

\begin{lem}\label{lem:oneclause}
  For every $\gamma'$ such that $\gamma' \in Z'_{\varphi}$ by some mapping $h'$, if there is some $\gamma'' \incl \gamma'$ such that $\gamma'' \in Z_{\varphi}$, then there is some clause $C_{i}$ such that $\gamma''$ contains $h'(x^{i}_{j}),\,h'(y^{i}_{j})$, for each $j=1,2,3$.
\end{lem}
\begin{proof}
  Let $h'$ and $\gamma'' \in Z_{\varphi}$ by mapping $h$. We observe the following:
  \begin{itemize}
      \item Due to the locations of the variables, $h(x^{\sim}_{c})$ is only among $h'(px_{i}),h'(x^{i'}_{j'}),h'(nx_{i''})$, and $h(y^{\sim}_{c})$ is only among $h'(py_{i}),h'(y^{i'}_{j'}),h'(ny_{i''})$ for $c,i,i',j',i''=1,2,3$.

      \item By constraints 1 and 2, if $h(x^{\sim}_{i}) = h'(px_{j})$ or $h'(nx_{i})$ for $i,j \in \{1,2,3\}$, then $h(y^{\sim}_{i}) = h'(py_{j})$ or $h'(ny_{i})$ respectively. Similarly, if $h(x^{\sim}_{i}) = h'(x^{i}_{j})$ for $i,j \in \{1,2,3\}$, then $h(y^{\sim}_{i}) = h'(y^{i}_{j})$.

      \item If for some $i = 1,2$, $h(y^{\sim}_{i}) = h'(y^{i_{1}}_{j_{1}})$ and $h(x^{\sim}_{i+1}) = h'(x^{i_{2}}_{j_{2}})$ for $i_{1},i_{2},j_{1},j_{2} \in \{1,2,3\}$, then because of constraints on $y^{i_{1}}_{j_{1}}$ and $x^{i_{2}}_{j_{2}}$ in $Z'_{\varphi}$, $14(i_{2}-i_{1}) +3(j_{2}-j_{1}) - 2 \leq h'(x^{i_{1}}_{j_{1}}) - h'(y^{i_{2}}_{j_{2}}) \leq 14(i_{2}-i_{1}) +3(j_{2}-j_{1}) +2$. Also, by constraint $1 \leq x^{\sim}_{i+1} - y^{\sim}_{i} \leq 5$ in $Z_{\varphi}$, $1 \leq h'(x^{i_{1}}_{j_{1}}) - h'(y^{i_{2}}_{j_{2}}) \leq 5$. This means $i_{1} = i_{2}$, and $j_{2} = j_{1} + 1$, i.e. $x^{i_{i}}_{j_{1}},y^{i_{2}}_{j_{2}}$ are variables on literals in the same clause, and occur adjacent to each other in $\varphi$. As this is true for $i=1,2$, we see that if $h(y^{\sim}_{1}) = h'(y^{i'}_{j'})$, then $h(x^{\sim}_{2}) = h'(x^{i'}_{j'+1})$, $j'=1$, and so the 6 variables $x^{\sim}_{j},y^{\sim}_{j}$ for $j=1,2,3$ will be mapped to $h'(x^{i'}_{j})$, $h'(y^{i'}_{j})$ respectively. We can similarly see that if $h(y^{+}_{i}) = h'(py_{j_{1}})$ or $h(y^{-}_{i}) = h'(ny_{j_{1}})$, then the 6 variables $x^{\sim}_{j},y^{\sim}_{j}$ for $j=1,2,3$ will be mapped to $h'(px_{j})$, $h'(py_{j})$ or $h'(nx_{j}),h'(ny_{j})$ respectively.

      \item If some $h(x^{+}_{1}) = h'(px^{1})$, then because of constraint 5 on $x^{+}_{1}$ and $y^{-}_{3}$ in $Z_{\varphi}$, $h(y^{-}_{3}) \neq h'(ny_{3})$. Similarly, if $h(y^{-}_{3}) = h'(ny_{3})$, then $h(x^{+}_{1}) \neq h'(px_{1})$. Meaning there is at least one $h'(x^{i}_{j})$ in $\gamma''$.
  \end{itemize}
This means there will be at least one clause $C_{i}$ such that $h'(x^{i}_{j}),h'(y^{i}_{j}) \in \gamma''$ for each $j=1,2,3$.
\end{proof}

\begin{lem}\label{lem:zfalcl}
  For each $\gamma' \in Z'_{\varphi}$, there is some $\gamma'' \incl \gamma'$ such that $\gamma'' \in Z_{\varphi}$ iff $\alpha_{\gamma'}$ falsifies at least one clause in $\varphi$.
\end{lem}
\begin{proof}
  Let $\gamma' \in Z'_{\varphi}$ by mapping $h'$. To prove one direction, assuming there is some $\gamma'' \incl \gamma'$ such that $\gamma'' \in Z_{\varphi}$ by mapping $h$, we see that by Lemma \ref{lem:oneclause}, there is some clause $C_{i}$ such that $\gamma''$ contains $h'(x^{i}_{j}),h'(x^{i}_{j})$ for each $j=1,2,3$. By constraints of $Z_{\varphi}$, 
   we observe that in the assignment $\alpha_{\gamma'}$, each literal of $C_{i}$ will be falsified. For the other direction, if $\alpha_{\gamma'}$ falsifies at least one clause $C_{i}$ in $\varphi$, then we define a subset $\gamma'' \incl \gamma'$ and a mapping $h: \Var(Z_{\varphi}) \mapsto \gamma''$ as follows:
  \begin{description}
    \item[Case 1 ]If $C_{i}$ has all positive literals: for each literal $l^{i}_{j}$ in $C_{i}$, we add $h'(x^{i}_{j})$ and $h'(y^{i}_{j})$ to $\gamma''$ and define $h(x^{+}_{j}) = h'(x^{i}_{j})$, $h(x^{-}_{j}) = h'(y^{i}_{j})$. Also, we add $h'(nx_{j}),h'(ny_{j})$ for $j=1,2,3$ to $\gamma''$ and define $h(x^{-}_{j}) = h'(nx_{j})$, $h(y^{-}_{j}) = h'(ny_{j})$.
    \item[Case 2 ]If $C_{i}$ has all negative literals: for each literal $l^{i}_{j}$ in $C_{i}$, we add $h'(x^{i}_{j})$ and $h'(y^{i}_{j})$ to $\gamma''$ and define $h(x^{-}_{j}) = h'(x^{i}_{j})$, $h(x^{+}_{j}) = h'(y^{i}_{j})$. Also, we add $h'(px_{j}),h'(py_{j})$ for $j=1,2,3$ to $\gamma''$ and define $h(x^{+}_{j}) = h'(nx_{j})$, $h(y^{+}_{j}) = h'(ny_{j})$.
  \end{description}  
  To see that $\gamma'' \in Z_{\varphi}$ by mapping $h$, we see that as $\alpha_{\gamma'}$ falsifies $C_{i}$, it means that $\alpha_{\gamma'}$ assigns all literal variables $\bot$ or $\top$ depending on whether $C_{i}$ has all positive or negative literals. This means the 6 states corresponding to the clause $C_{i}$ we added to $\gamma''$ will satisfy the constraints 1 or 2 respectively. Also, by constraints 1 and 2 of $Z'_{\varphi}$, the 6 states we added corresponding to either $px^{j},py_{j}$ or $nx_{j},ny_{j}$ will satisfy constraints 1 or 2 of $Z_{\varphi}$ respectively. It is easy to see that the constraints 2-5 of $Z_{\varphi}$ are also satisfied. 
\end{proof}

\begin{lem}\label{lem:z'assign}
  For every assignment $\alpha$ of the variables of $\varphi$, there is some $\gamma' \in Z'_{\varphi}$ such that $\alpha_{\gamma'} = \alpha$.
\end{lem}
\begin{proof}
  Given an assignment $\alpha$, we construct the required $\gamma'$ and the mapping $h'$ such that $\gamma' \in Z'_{\varphi}$ as follows:
  \begin{itemize}
      \item For each variable $px_{i},py_{i}$ in $Z'_{\varphi}$, we add $(q_{x},3(i-1))$ and $(q_{y},3(i-1))$ to $\gamma'$ and define $h'(px_{i}) = (q_{x},3(i-1))$, $h'(py_{i}) = (q_{y},3(i-1))$.
      \item For each variable $x^{i}_{j},y^{i}_{j}$ in $Z'_{\varphi}$, if $\alpha(var(l^{i}_{j})) = \true$, we add $(q_{x},14i +3(j-1))$ and $(q_{y},14i +3(j-1)+2)$ to $\gamma'$ and define $h'(x^{i}_{j}) = (q_{x},14i +3(j-1))$, $h'(y^{i}_{j}) = (q_{y},14i +3(j-1)+2)$. Otherwise, if $\alpha(var(l^{i}_{j})) = \false$, we add $(q_{x},14i +3(j-1))$ and $(q_{y},14i +3(j-1))$ to $\gamma'$ and define $h'(x^{i}_{j}) = (q_{x},14i +3(j-1))$, $h'(y^{i}_{j}) = (q_{y},14i +3(j-1))$.
      \item For each variable $nx_{i},ny_{i}$ in $Z'_{\varphi}$, we add $(q_{x},14(m+1))$ and $(q_{y},14(m+1)+2)$ to $\gamma'$ and define $h'(px_{i}) = (q_{x},14(m+1))$, $h'(py_{i}) = (q_{y},14(m+1)+2)$.
  \end{itemize}
  Now, we have ensured by this definition and mapping $h'$ that constraints 1-4 of $Z'_{\varphi}$ are satisfied by $\gamma'$. Also, $\alpha_{\gamma'}$ defined for $\gamma'$ will be such that $\alpha = \alpha_{\gamma'}$. 
\end{proof}

\begin{lem}
$\varphi$ is satisfiable iff
    $Z_{\varphi} \not\entails Z'_{\varphi}$.
\end{lem}
\begin{proof}
  To prove one direction, we assume that $\varphi$ is satisfiable, and $\alpha$ is a satisfying assignment of the variables of $\varphi$. From Lemma \ref{lem:z'assign}, we know that there is some $\gamma' \in Z'_{\varphi}$ such that $\alpha_{\gamma'} = \alpha$. Now, if there was some $\gamma'' \incl \gamma'$ such that $\gamma'' \in Z_{\varphi}$, then from Lemma \ref{lem:zfalcl}, we get that $\alpha_{\gamma'}$ falsifies at least one clause in $\varphi$, which is a contradiction to the fact that $\alpha$ was a satisfying assignment. Thus no such $\gamma''$ exists, and for every $\gamma'' \incl \gamma'$, $\gamma'' \not\in Z_{\varphi}$, meaning $Z_{\varphi}\not\entails Z'_{\varphi}$. To prove the other direction, we assume that $Z_{\varphi} \not\entails Z'_{\varphi}$. This means there is some $\gamma' \in Z'_{\varphi}$ such that for every $\gamma'' \incl \gamma'$, $\gamma'' \not\in Z_{\varphi}$. Now, we claim that the corresponding $\alpha_{\gamma'}$ is in fact a satisfying assignment of $\varphi$. To prove this, we first see that by Lemma \ref{lem:assignval}, $\alpha_{\gamma'}$ is a valid assignment. To prove it is a satisfying assignment, we assume on the contrary that $\alpha_{\gamma'}$ is not a satisfying assignment, meaning it falsifies at least one clause of $\varphi$. This means by Lemma \ref{lem:zfalcl}, there is some $\gamma'' \incl \gamma'$ such that $\gamma'' \in Z_{\varphi}$, which is a contradiction. Hence $\alpha_{\gamma'}$ is a satisfying assignment, and so $\varphi$ is satisfiable.
\end{proof}

\end{document}